\documentclass[12pt, a4paper]{article}
\usepackage[absolute]{textpos}
\usepackage[top=3cm,bottom=3cm,left=2cm,right=2cm]{geometry}
\usepackage{amsmath,amssymb,mathtools,dsfont,emptypage,graphicx,dcolumn,bm,booktabs,colortbl,collcell,enumerate,float,verbatim,multirow,xparse,slashed,mathrsfs,color}
\usepackage[normalem]{ulem}
\usepackage{cite}
\usepackage{braket}
\usepackage{tabularx}
\usepackage{multirow}
\usepackage{graphicx}
\usepackage{subcaption}
\usepackage{appendix}
\usepackage{booktabs} % Enhance the tables
\usepackage{colortbl}
\usepackage{collcell}
\usepackage{xparse}
\usepackage[labelfont=bf, labelsep=period]{caption}
\usepackage{sectsty}
\usepackage{pifont}
\usepackage{relsize}
\usepackage{pifont}
\usepackage{bbold}
\sectionfont{\fontsize{15}{12}\selectfont}
\subsectionfont{\fontsize{13}{12}\selectfont}

\usepackage[compat=1.1.0]{tikz-feynman}
\usepackage{contour}%FeynamnDiagrams e contorni

\newcommand{\virg}[1]{``#1''}
\newcommand{\mnulight}{m_{\nu}^{\rm lightest}}

\DeclareMathOperator{\myRe}{Re}

\def\Mav{{M}_\text{av}}
\def\wMav{\tilde{M}_\text{av}}
\def\ch{\text{ch}}
\def\sh{\text{sh}}
\def\cNt{c_N^{(3)}}
\def\sNt{s_N^{(3)}}
\def\cNd{c_N^{(2)}}
\def\sNd{s_N^{(2)}}
\def\cNu{c_N^{(1)}}
\def\sNu{s_N^{(1)}}
\def\cnud{c_\nu^{(2)}}
\def\snud{s_\nu^{(2)}}
\def\cnuu{c_\nu^{(1)}}
\def\snuu{s_\nu^{(1)}}
\def\tNt{x^N_3}
\def\tNd{x^N_2}
\def\tNu{x^N_1}
\def\tnud{x^\nu_2}
\def\tnuu{x^\nu_1}

\def\nubb{0\nu\beta\beta}
\def\wTheta{\widetilde{\Theta}}

\def\sph{\text{sph}}
\def\LNC{\text{LNC}}
\def\LNV{\text{LNV}}

\newcommand{\smallerbullet}{\,\begin{picture}(-1,1)(-1,-3)\circle*{2}\end{picture}\ }

\definecolor{myred}{cmyk}{0,1,1,0.55}
\definecolor{mygold}{rgb}{0.8, 0.54, 0.08}
\definecolor{mygray}{gray}{.95}
\usepackage[linktoc = page, colorlinks=true,urlcolor=mygold,linkcolor=mygold,citecolor=mygold]{hyperref}

\begin{document}

\begin{center}
{\bf \large Dirac-Phase CP-Violation in the Low-Scale Type-I Seesaw with\\ Three Right-Handed Neutrinos}\\
[5mm]
\renewcommand*{\thefootnote}{\fnsymbol{footnote}}
Alessandro Granelli$^{\,a,\,}$\footnote{\href{mailto:alessandro.granelli@ific.uv.es}{alessandro.granelli@ific.uv.es}},
Juraj Klari\'c$^{\,b, c, d,\,}$\footnote{\href{mailto:jklaric@phy.hr}{jklaric@phy.hr}} and 
S.~T.~Petcov$^{\,e, f,\,}$\footnote{\href{mailto:petcov@sissa.it}{petcov@sissa.it}. Also at: Institute of Nuclear Research and Nuclear Energy, Bulgarian Academy of Sciences, 1784 Sofia, Bulgaria.}
\\
\vspace{2mm}
$^{a}$\,{\it Instituto de Física Corpuscular (IFIC), CSIC-Universitat de València, Parc Científic UV,\\ c/ Catedrático José Beltrán 2, E-46980 Paterna, Spain;}\\
$^{b}${\it Institute for Theoretical Physics Amsterdam and Delta Institute for Theoretical Physics,
University of Amsterdam, Science Park 904, 1098 XH Amsterdam, The Netherlands;}\\
$^{c}${\it Nikhef, Theory Group, Science Park 105, 1098 XG, Amsterdam, The Netherlands;}\\
$^{d}${\it University of Zagreb, Faculty of Science, Department of Physics, 10000 Zagreb, Croatia;}\\
$^{e}${\it INFN/SISSA, via Bonomea 265, 34136, Trieste, Italy;}\\
$^{f}${\it Kavli IPMU (WPI), UTIAS, University of Tokyo, Kashiwa, Chiba 277-8583, Japan.}\\

\end{center}

%%%%%%%%%%%%%%%%ABSTRACT%%%%%%%%%%%%%%%%%%%55
\begin{center}
    {\bf Abstract}
\end{center}
We study the low-scale type-I seesaw with three right-handed neutrinos (i.e.~heavy Majorana neutrinos) when the CP-violation arises solely from the low-energy Dirac phase $\delta$ of the Pontecorvo-Maki-Nakagawa-Sakata (PMNS) neutrino mixing matrix and the heavy neutrinos have testable mixings. We derive a CP-conserving and non-real structure of the $3\times 3$ orthogonal matrix entering the Casas-Ibarra parametrisation in terms of two real angles and one single imaginary parameter, ensuring that the only CP-violating phases in the neutrino Yukawa couplings are those of the PMNS matrix. We then focus on the case of CP-violation from $\delta$ alone and discuss the phenomenological implications of this hypothesis. We concentrate on quasi-degenerate heavy Majorana neutrinos with masses within $\sim (0.1-100)\,\text{GeV}$, as relevant for low-scale leptogenesis. Only certain subregions of the full ternary space defined by the ratios $\Theta^2_e:\Theta^2_\mu:\Theta^2_\tau$ -- where $\Theta^2_\alpha$ denotes the squared coupling of the heavy neutrinos to leptons of flavour $\alpha = e,\,\mu,\,\tau$ -- are compatible with Dirac-phase CP-violation while being testable at collider experiments. Our assumption also implies specific forms of the effective Majorana mass parameter that can be tested at neutrinoless double-beta decay searches. Finally, low-scale leptogenesis under this restriction can reproduce the baryon asymmetry of the Universe (BAU) and be tested through the combination of direct heavy neutral lepton searches and neutrino experiments. The BAU vanishes in the exact limit of CP-conserving values of the Dirac phase $\delta = 0,\,\pi,\,2\pi$, but the observed BAU can be reproduced within the testable region even if $\delta$ deviates from these values by a factor as small as $\mathcal{O}(10^{-5})$, with important implications for ultraviolet completions with approximate CP-symmetry.

\renewcommand*{\thefootnote}{\arabic{footnote}}
\setcounter{footnote}{0}

\newpage
%%%TABLE OF CONTENTS%%%
\tableofcontents

%%%%%%%%%%%%%%%%%%%%%%%%%%%%%%%%%%%%%%%%%%

\newpage
\section{Introduction}
That of the type-I seesaw \cite{Minkowski:1977sc,Yanagida:1979as,Gell-Mann:1979ijt,GellMann:1980vs,Glashow:1979nm,Mohapatra:1979ia} is arguably
the most natural extension of the SM that can account for the generation and smallness of non-zero
light neutrino masses.
It consists of adding to the SM (at least two) %$n_R\geq 2$ 
right-handed (RH) neutrino SM gauge singlet fields.
Without the additional assumption of total lepton charge conservation,
    the RH neutrino fields admit a Majorana mass term in the Lagrangian
    which respects the SM gauge symmetries.
    These fields also couple to the Higgs and the three left-handed (LH) lepton
doublets via gauge-invariant Yukawa interaction.
Following the spontaneous breaking of the electroweak symmetry, 
the LH and RH neutrino fields form a Dirac neutrino mass term.
The interplay between the Dirac and the RH neutrino Majorana mass terms
leads to a set of %light 
seesaw-suppressed masses for the observed light neutrinos,
and to a small but important correction to the yet-to-be-discovered heavy neutrino states,
with both sets of states being Majorana particles.
In this setting the LH flavour (RH) neutrino fields are linear combinations of the massive light and heavy Majorana neutrino fields.

When the heavy Majorana neutrino mass scale is close to that of Grand Unification,
being, say, $\sim 10^{14}\,\text{GeV}$, the seesaw suppression that generates and explains the smallness of the light neutrino masses can take place with $\mathcal{O}(1)$ Yukawa couplings, rendering rather natural this seesaw realisation. However, it is extremely challenging, if not impossible, to test the physics associated with the type-I seesaw mechanism at these scales. The alternative low-scale seesaw with sub-TeV heavy Majorana neutrino masses is, in principle, an experimentally testable scenario in which the seesaw suppression of the light neutrino masses can be made technically natural using symmetry arguments \cite{Wyler:1982dd,Mohapatra:1986aw,Malinsky:2005bi}   (see also \cite{Leung:1983ti}).
In such scenarios, the light neutrino masses are generated through approximate symmetry-protected cancellations between the contributions of the heavy Majorana neutrinos in the seesaw relation \cite{Shaposhnikov:2006nn,Kersten:2007vk,Gavela:2009cd,Ibarra:2010xw,Ibarra:2011xn,Antusch:2015mia} (see also \cite{Granelli:2025lds} for a modular-symmetry-protected low-scale seesaw), or at the loop level \cite{Pilaftsis_1992, Grimus_2002,Aristizabal_Sierra_2011, Lopez_Pavon_2013}. Then, at sub-TeV scales, the phenomenology of the type-I seesaw associated to the existence of heavy Majorana neutrinos and their weak couplings to the SM leptons, generated through the mixing of the RH and flavour neutrinos, is widely testable. For instance, the creation of heavy Majorana neutrinos becomes kinematically accessible at colliders and extracted beam-line facilities, and their mixing can be sufficiently sizeable to enable copious production through particle collisions, or via decays of vector bosons, the $\tau$-lepton and mesons (see e.g.~\cite{Abdullahi:2022jlv, Antel:2023hkf} and references therein).

Another remarkable feature of the type-I seesaw extension is that it provides a mechanism \cite{Kuzmin:1985mm,Fukugita:1986hr}
to generate the observed baryon asymmetry of the Universe (BAU) \cite{Planck2018, Cooke_2018}.
The existence of Yukawa interactions between the RH neutrinos, the Higgs and the LH lepton doublets allows for SM lepton-number (L), charge-conjugation (C) and charge-conjugation-plus-parity (CP) violating processes involving these particles. With these processes occurring out-of-equilibrium in the expanding Universe, all the three Sakharov’s conditions for a dynamical generation of a lepton asymmetry get fulfilled \cite{Sakharov:1967dj}. Once a lepton asymmetry is generated in the early Universe, it can be converted into the present BAU by the non-perturbative (B+L)-violating, but (B-L)-conserving, SM sphaleron processes \cite{Kuzmin:1985mm} (B is the SM baryon number), while these are in thermal equilibrium above the electroweak scale \cite{DOnofrio:2014rug}. This mechanism of BAU generation is referred to as \textit{leptogenesis} \cite{Fukugita:1986hr}  and there exist several different realisations of it depending on the values and the hierarchy of masses of the heavy Majorana neutrinos\cite{Pilaftsis:1997jf, Pilaftsis:2003gt, Akhmedov:1998qx, Asaka:2005pn, Racker:2012vw} (see \cite{Bodeker:2020ghk} for a review).
At the low scales accessible to collider experiments, two conceptually different mechanisms of leptogenesis are active: the \textit{freeze-out} version in which the out-of-equilibrium Sakharov condition
is caused by an overabundance of heavy Majorana neutrinos
\cite{Hambye:2016sby, Hambye:2017elz,Granelli:2020ysj};
and the \textit{freeze-in} scenario in which the heavy Majorana neutrinos are being produced from the SM plasma (i.e.~they are underabundant)
compared to the equilibrium density \cite{Akhmedov:1998qx, Asaka:2005pn}. 
Both mechanisms result from an interplay of decays, inverse decays, scatterings and CP-violating oscillations of the Majorana neutrinos,
and can be understood as \virg{two sides of the same coin}, described by the same density matrix formalism,
leading to a largely overlapping parameter space \cite{Klaric:2020phc, Klaric:2021cpi, Drewes:2021nqr}.

The leptonic CP-violation (CPV), as is well known,  plays a crucial role in 
the generation of BAU via leptogenesis.
It can be provided by the charged lepton mass term, and/or  neutrino Dirac
and/or RH neutrino Majorana mass terms.
In the charged lepton and light and heavy
Majorana neutrino mass eigenstate basis,
it can originate either from the Casas-Ibarra (CI) matrix
(the orthogonal matrix entering the CI parameterisation \cite{Casas:2001sr}), which accounts for the mixing originating from the RH neutrino Majorana mass term, or the Pontecorvo-Maki-Nakagawa-Sakata (PMNS) neutrino mixing matrix,
or from both these matrices, present in the neutrino Yukawa couplings.
A particularly interesting and appealing possibility is that the requisite
CPV in leptogenesis is induced exclusively by the Dirac CP-violating phase $\delta$
present in the PMNS matrix, i.e., the possibility of existence of
a link between the generation of BAU in the early Universe and
CPV in the process of three-flavour neutrino oscillations at low energies.
Searches for CP-violating effects in flavour neutrino oscillations are currently
conducted in the T2K \cite{T2K:2023smv} and NO$\nu$A \cite{NOvA:2025tmb}
experiments with accelerator neutrinos and
these studies will continue with higher precision
in the upcoming T2HK \cite{Hyper-Kamiokande:2025fci} and
DUNE \cite{DUNE:2021mtg}  experiments.

High-scale leptogenesis based on the type-I seesaw, in which the CI matrix is CP-conserving and the only source of CPV is, more generally, the Dirac and/or the Majorana phases \cite{Bilenky:1980cx} of the PMNS matrix was considered in \cite{Pascoli:2006ie,Pascoli:2006ci,Blanchet:2006be,Branco:2006hz,Branco:2006ce,Uhlig:2006xf,Anisimov:2007mw,Molinaro:2009lud,Molinaro:2008cw,Dolan:2018qpy, Moffat:2018smo,Xing:2020erm,Xing:2020ghj,Granelli:2021fyc,Granelli:2023tcj,Granelli:2025cho}.
CP-conserving CI matrix appears in many theoretical frameworks
(see, e.g., \cite{ Meroni:2012ze,Mohapatra:2015gwa,Samanta:2018efa,Nishi:2018vlz,Chen:2016ptr,Hagedorn:2016lva,Li:2017zmk}).

Type-I seesaw leptogenesis  (LG)
  with low-energy Dirac-phase CPV was investigated in \cite{Pascoli:2006ie,Pascoli:2006ci,Moffat:2018smo,Granelli:2021fyc,Granelli:2023tcj}. In \cite{Pascoli:2006ci} the general CP-invariance constraints on the neutrino Yukawa couplings,
CI and PMNS matrices, were derived and the  conditions under which the
requisite CPV is provided exclusively by the Dirac and/or Majorana phases
in the PMNS matrix in the case of high-scale LG with two RH neutrinos
were identified. Successful flavoured LG with purely Dirac-phase CPV
was shown to be possible for a normal hierarchical (NH) light
neutrino mass spectrum  at a scale of $\sim 10^{11}$ GeV
for $\delta \sim \pi/2,3\pi/2$ \cite{Pascoli:2006ci}.
The study of LG with three RH neutrinos \cite{Moffat:2018smo}
demonstrated that the observed BAU can be reproduced
at a scale of $\sim 10^{10}$ GeV and mildly hierarchical heavy Majorana
neutrino masses for a much larger set of values of $\delta$,
e.g., $40^\circ \lesssim \delta \lesssim 160^\circ$ ($2\sigma$ C.L.).
Moreover, the flavour effects were shown in \cite{Moffat:2018smo} to persist
beyond the scale of $10^{12}$ GeV in the case of CP-conserving CI matrix.
The analysis with two RH neutrinos revealed \cite{Granelli:2021fyc}
that the BAU can be reproduced at scales $\sim (2.7 - 10.0)\times 10^{12}$ GeV
in the case of NH (hierarchical) light (heavy) Majorana neutrino
mass spectrum for  $202.4^\circ \lesssim \delta \lesssim 337.6^\circ$.

Viable low-scale freeze-in Dirac-phase
LG with two quasi-degenerate heavy Majorana neutrinos having
masses $\sim (0.1-100)$ GeV was shown also to be possible
in the case, e.g., of NH light neutrino mass spectrum,
for $195^\circ \lesssim \delta \lesssim 345^\circ$
and a range of heavy Majorana neutrino 
couplings to charged leptons, part of which is testable
at future experiments (such as, e.g., SHiP \cite{SHiP:2018xqw} and
the discussed future circular colliders FCC-ee/CEPC
\cite{Blondel:2022qqo, CEPCStudyGroup:2018ghi}) \cite{Granelli:2023tcj}.
Moreover, the parameter space of the viable Dirac-phase LG
differs from those associated with  CI- and/or the Majorana- phase CPV,
opening up the possibility to distinguish experimentally between
the different sources of LG CPV.
In the limit of two RH neutrinos exactly degenerate in mass,
in which the CI matrix is CP-conserving \cite{Pascoli:2006ci},
successful LG can still take place \cite{Antusch:2017pkq},
with the low-energy CPV phases being the only source
of CP violation~\cite{Sandner:2023tcg}.

In all the cases of Dirac-phase leptogenesis 
studied so far and briefly discussed above relatively large CP-violating values of the Dirac phase $\delta$ are required to have a viable leptogenesis.

The present article extends the studies on type-I seesaw Dirac-phase leptogenesis to the low-scale case with three RH neutrinos with testable mixings, that has not been investigated so far. We derive the conditions that a non-real $3\times 3$ CI  matrix and the
Majorana phases should satisfy to ensure that the only leptonic source of CPV is the Dirac phase of the PMNS 
matrix. Examples of CP-conserving non-real $3\times 3$ CI matrices were given in
\cite{Moffat:2018smo}. Here we derive the most general form 
a non-real $3\times 3$ CP-conserving CI matrix can have. The choice of a non-real CI matrix ensures that the mixings of the heavy neutrinos can be sufficiently large to be testable at low-energy experiments.

Focusing on heavy neutrino mass scales of $0.1-100\,\text{GeV}$, we analyse the phenomenological implications of these
conditions for the heavy Majorana neutrino searches,
for neutrinoless double-beta decay experiments, and perform a detailed study of leptogenesis in the considered framework. We showcase the potential to falsify the hypothesis of Dirac-phase CPV starting from a non-real CI matrix.

The paper is organised as follows. In Sec.~\ref{sec:framework} we introduce the type-I seesaw framework to set the overall notation. We review the seesaw mechanism up to second order in the associated perturbative expansion.
We  review the differences between the heavy neutrino mass eigenstates before and after the electroweak symmetry breaking, and the loop corrections to light neutrino masses;
we summarise the current knowledge of neutrino masses and mixing to fix some of the seesaw parameters according to oscillation data;
we introduce the CI parametrisation used throughout this work. In Sec.~\ref{sec:LECPV} we discuss the condition of low-energy CPV, presenting the general CP-conserving form of the non-real CI matrix and analysing implications of purely Dirac-phase CPV on heavy neutrino searches and neutrinoless double-beta decay.
Low-scale leptogenesis under the assumption of Dirac-phase CPV is studied in Sec.~\ref{sec:LG}, where we introduce the density matrix equations, and present the results of our scans in the parameter spaces relevant to heavy neutrino searches.
We also discuss there, for selected benchmarks, the ranges of values of $\delta$ necessary to generate the BAU in our scenario. Additional derivations and complementary results are collected in the Appendices. We summarise the results and conclude in Sec.~\ref{sec:conclusions}.

%%%%%%%%%

\section{The type-I seesaw framework}\label{sec:framework}
We briefly review here the type-I seesaw mechanism of neutrino mass generation in the case of three right-handed (RH) neutrinos and up to the second order in the seesaw perturbative expansion. This allows us to set the notation for our subsequent analysis.

\subsection{The type-I seesaw mechanism to second order}\label{sec:type-I}
We write below the Lagrangian in the basis in which the charged lepton Yukawa matrix is diagonal 
(sum over equal indices is implicit and the space-time variable dependence of the fields is omitted):
%%%%%%%%%%%%%%%%%%%%%%%%%%%%%%%
\begin{equation}
\label{eq:Lseesaw}
{\cal L}_{\rm Type-I} ={\cal L}_{\rm SM} + i \,\overline{\nu_{\kappa R}}\slashed{\partial}\nu_{\kappa R} -\left[
\,\lambda_{\alpha \kappa} \overline{\psi_{\alpha L}}\,i\sigma_2\,\Phi^*\,\nu_{\kappa R}
 + \,\frac{1}{2}\,\overline{\nu_{\kappa L}^c}(M_R)_{\kappa\rho} \nu_{\rho R}+ \hbox{h.c.}\right]\,,
\end{equation}
%%%%%%%%%%%%%%%%%%%%%%%%%%%%%%%%%%%
%
where $\lambda_{\alpha \kappa}$ are the entries of the neutrino Yukawa 
matrix $\lambda$ in the considered basis and  
$\sigma_2$ is the second Pauli matrix. The fields $\psi_{\alpha L}(x) = (\nu_{\alpha L}^T(x)\,\alpha_L^T(x))^T $ and $\Phi(x) = (\Phi^+(x)\,\Phi^{(0)}(x))^T$ describe respectively the 
SM left-handed (LH) lepton and Higgs doublets, $\nu_{\alpha L}(x)$ and $\alpha_L(x)$ being the LH flavour neutrino and
charged lepton fields, 
while $\nu_{\kappa L}^c (x) = C (\overline{\nu_{\kappa R}}(x))^T$,
where $\nu_{\kappa(\rho) R}(x)$ is 
the field of the RH 
neutrino $\nu_{\kappa(\rho) R}$, $\kappa,\rho = e,\mu,\tau$. 
The complex RH neutrino mass matrix $M_R$ is symmetric. The matrix $M_R$ can be diagonalised by the 
Autonne-Takagi transformation \cite{Autonne, Takagi} (see also, e.g., \cite{Haber_2021}):
$\tilde{M}_N^d = \tilde{V}^\dagger M_R \tilde{V}^* = \text{diag}(\tilde{M}_1,\,\tilde{M}_2,\,\tilde{M}_3)$, where $\tilde{V}$ is a unitary matrix and $\tilde{M}_{j} >0$ is the mass of the corresponding 
Majorana neutrino 
$\tilde{N}_{j}$ described by the Majorana field 
$\tilde{N}_j(x) = \tilde{N}_{j R}(x) + \tilde{N}_{j L}(x)$, where 
$\tilde{N}_{j R}(x) = \tilde{V}^T_{j \kappa }\nu_{\kappa R}(x)$, 
$\tilde{N}_{j L}(x) = \tilde{V}^\dagger_{j \kappa} \nu^c_{\kappa L}(x)$ 
and 
$\tilde{N}_j^c(x) \equiv C(\overline{\tilde{N}_j}(x))^T = \tilde{N}_j(x)$, $j = 1,\,2,\,3$.
After such diagonalisation, the Lagrangian in Eq.~\eqref{eq:Lseesaw} becomes
\begin{equation}\label{eq:Lseesaw2}
{\cal L}_{\rm Type-I} ={\cal L}_{\rm SM} +  \frac{i}{2} \,\overline{\tilde{N}_{j}}\slashed{\partial}\tilde{N}_{j} - \,\frac{1}{2}\,\tilde{M}_{j}\overline{\tilde{N}_{j}} \tilde{N}_{j}-\left[
    \,Y_{\alpha j} \overline{\psi_{\alpha L}}\,i\sigma_2\,\Phi^*\,\tilde{N}_{j R}
    + \hbox{h.c.}\right],   
\end{equation}
with $Y_{\alpha j} \equiv \tilde{V}^*_{\kappa j}\lambda_{\alpha \kappa}$.
The masses of the Majorana neutrinos $\tilde{N}_j$ are typically much larger than the eV masses of the light massive neutrinos, and thus we will refer to $\tilde{N}_j$  as \textit{heavy Majorana neutrinos} or simply \textit{heavy neutrinos}.

After the spontaneous breaking of the electroweak symmetry, with the neutral component of the Higgs doublet acquiring
a non-vanishing vacuum expectation value $v/\sqrt{2} \simeq 174$ GeV, 
the terms in the square brackets  in Eq.~\eqref{eq:Lseesaw} can be re-arranged as follows:
\begin{equation}\label{eq:Lmnu}
{\cal L}_m^{\nu} = -\frac{1}{2} 
\begin{pmatrix}
    \overline{\nu_{\alpha L}}& \overline{\nu_{\kappa L}^c}
\end{pmatrix}
\begin{pmatrix}
    \mathbb{O}_{\alpha \beta}&(M_D)_{\alpha \rho}\\(M_D)_{\beta \kappa}& (M_R)_{\kappa \rho}
\end{pmatrix}
\begin{pmatrix}
    \nu_{\beta R}^c\\ \nu_{\rho R}
\end{pmatrix}
+ {\rm h.c.}\,.,
\end{equation}
with $M_D \equiv (v/\sqrt{2})\lambda$. The complex matrix 
\begin{equation}\label{eq:calM}
\mathcal{M} =
\begin{pmatrix}
     \mathbb{O}&M_D\\M_D^T& M_R
\end{pmatrix}
    \end{equation}
    represents the full neutrino mass matrix in the non-diagonal basis, at tree level. It is symmetric, and can thus be diagonalised as \cite{Autonne, Takagi}
%%%%%%%%%%%%%%%%%%%%%%%%%%
\begin{equation}
\label{eq:AT_diag0}
\mathcal{M} = T \begin{pmatrix}\hat{m}_\nu &\mathbb{O}\\ \mathbb{O}&\hat{M}_N\end{pmatrix}T^T\,,
\end{equation}
%%%%%%%%%%%%%%%%%%%%%%%%
%
where $T^\dagger = T^{-1}$, $\hat{m}_\nu=\text{diag}(m_1,\,m_2,\,m_3)$ and
$\hat{M}_N = \text{diag}(M_1,\,M_2,\,M_3)$. It proves convenient to recast $\mathcal{M}$ in the form:
%%%%%%%%%%%%%%%%%%%%%%%%%%%%
\begin{eqnarray}
\label{eq:AT_diag}
 \mathcal{M} 
&=& \,\text{exp}
\begin{pmatrix}
\mathbb{O} & R\\ 
-R^\dagger& \mathbb{O}
\end{pmatrix}
\begin{pmatrix}
 m^{(0)}_\nu & \mathbb{O}\\
\mathbb{O}& M_N 
\end{pmatrix}
\text{exp}
\begin{pmatrix}
\mathbb{O} &-R^*\\ 
R^T & \mathbb{O}
\end{pmatrix} 
\nonumber
\\
& = &\,\text{exp}
\begin{pmatrix}
\mathbb{O} & R\\ 
-R^\dagger& \mathbb{O}
\end{pmatrix}
\begin{pmatrix}
U & \mathbb{O} \\
\mathbb{O} & V
\end{pmatrix}
\begin{pmatrix}
\hat m_\nu & \mathbb{O}\\
\mathbb{O} & \hat{M}_N 
\end{pmatrix}
\begin{pmatrix}
U^T & \mathbb{O} \\
\mathbb{O} & V^T
\end{pmatrix}
\text{exp}
\begin{pmatrix}
\mathbb{O} & -R^*\\ 
R^T & \mathbb{O}
\end{pmatrix}\,.
\label{eq:TRUtV}
\end{eqnarray}
%%%%%%%%%%%%%%%%%%%%%%%%
%
Here $R$ is a generic $3\times3$ complex matrix, 
 $U$ and $V$ are unitary matrices, 
$m^{(0)}_\nu \equiv U\hat m_{\nu} U^T$ 
and $M_N \equiv V \hat{M}_N V^T$.      
Assuming that the matrix 
$R$ has relatively small entries, which will be justified below, we expand the exponential in the above expression up to second-order in $R$, and we rewrite $T$ as
%%%%%%%%%%%%%%%%%%%%%%%%%%%%%%%%%%%%%%
\begin{equation}\label{eq:T}
 T = 
\begin{pmatrix}
\left(\mathbb{1} - \frac{1}{2}RR^\dagger\right)U & R V\\ 
-R^\dagger U  & \left(\mathbb{1} - \frac{1}{2}R^\dagger R\right)V
\end{pmatrix} + \mathcal{O}(R^3)\,,
\end{equation}
with $T\,T^\dagger = T^\dagger\, T =  
\mathbb{1} + \mathcal{O}(R^3)$.
%%%%%%%%%%%%%%%%%%%%%%%%%%
%
Using the above expression for $T$ in Eq.~\eqref{eq:AT_diag0}, % for $T$,
we find, up to terms $\mathcal{O}(R^3)$,
%%%%%%%%%%%%%%%%%%%%%%%%%%%%%%%%%%%%
\begin{eqnarray}
\left(\mathbb{1} - \frac{1}{2}RR^\dagger\right)U \hat{m}_\nu U^T
\left(\mathbb{1} - \frac{1}{2}RR^\dagger\right)^T + R\,M_N\,R^T &=& 0\,,
\label{eq:seesaw1}
\\
-\, U \hat{m}_\nu U^T\,R^* + R\,M_N &=& M_D\,,
\label{eq:seesaw2}
\\
R^\dagger U \hat{m}_\nu U^T\,R^* + 
\left(\mathbb{1} - \frac{1}{2}R^\dagger R\right) M_N  
\left(\mathbb{1} - \frac{1}{2}R^\dagger R\right)^T  &=& M_R \,.
\label{eq:seesaw3}
\end{eqnarray}
%%%%%%%%%%%%%%%%%%%%%%%%%%%%%%%%%%%
%
It follows from Eq.~\eqref{eq:seesaw1} that 
$m^{(0)}_\nu = U \hat{m}_\nu U^T$ can be expressed as a linear combination of terms
all of which are $\mathcal{O}(R^2)$.
Thus, the terms involving 
$ U \hat{m}_\nu U^T R^*$ in 
Eqs.~\eqref{eq:seesaw2} and \eqref{eq:seesaw3} are contributions of order $\mathcal{O}(R^3)$  and $\mathcal{O}(R^4)$,
which, in the approximation with which we are 
working, can be neglected. Thus, we get:
%%%%%%%%%%%%%%%%%%%%%%%%%%%%%%%%%5
\begin{eqnarray}
  R\,M_N &=& M_D\,,
\label{eq:seesaw4}
\\
\left(\mathbb{1} - \frac{1}{2}R^\dagger R\right) M_N  
\left(\mathbb{1} - \frac{1}{2}R^\dagger R\right)^T  &=& M_R \,.
\label{eq:seesaw5}
\end{eqnarray}
%%%%%%%%%%%%%%%%%%%%%%%%%%%%%%%%%%%
%
Eq.~\eqref{eq:seesaw5} implies that 
$M_R =  M_N + \mathcal{O}(R^2)$. This, combined 
with Eq.~\eqref{eq:seesaw4}, leads to the relation
$R M_N \cong R M_R + O(R^3)  = M_D$, and consequently to: 
%%%%%%%%%%%%%%%%%%%%%%%%%
\begin{eqnarray}
 R &\simeq& M_D M_R^{-1}\, ,
\label{eq:RMDtMR}
\end{eqnarray}
%%%%%%%%%%%%%%%%%%%%%%%%
 
Within the seesaw approach, the matrix  
$M_D M^{-1}_R$
has elements whose magnitude is considered to be much smaller than unity. This justifies the approximation of neglecting terms $\mathcal{O}(R^n)$, $n\geq 3$, that we have adopted. 
Inserting the expression in Eq.~\eqref{eq:RMDtMR} for $R$ 
into Eq.~\eqref{eq:seesaw1}, and taking into account that 
$M_R^T = M_R$, we obtain the well-known seesaw relation, valid at tree level:
%%%%%%%%%%%%%%%%%%%%%%%%%%%%
\begin{eqnarray}
m_\nu = \left(\mathbb{1} - \frac{1}{2}RR^\dagger\right)U \hat m_{\nu} U^T\, 
\left(\mathbb{1} - \frac{1}{2}RR^\dagger\right)^T \simeq -\, 
M_D\, M^{-1}_R\,M_D^T\,,
\label{eq:seesaw6}
\end{eqnarray}
%%%%%%%%%%%%%%%%%%%%%%%%%%%%%%%%%%%%
%
where $m_\nu$ is the Majorana mass matrix of the 
 LH flavour fields $\nu_{\alpha L}(x)$. 
 
Inserting the expression for $\mathcal{M}$ given 
by Eq.~\eqref{eq:AT_diag0} in Eq.~\eqref{eq:Lmnu}, we find 
the expressions for the Majorana mass eigenstate fields  
$\chi_a(x) $ and $N_j(x) $ which  describe respectively the 
light and heavy Majorana neutrinos $\chi_a$ and $N_j$ with masses $m_a\geq 0$ and 
$M_j\gg m_a$, $a,j = 1,2,3$ in terms of the LH and RH neutrino  
fields $\nu_{\alpha L}(x) $, $\nu^c_{\kappa L}(x) $ and
$\nu^c_{\alpha R}$, $\nu_{\kappa R}$,
$\alpha,\kappa = e,\,\mu,\,\tau$.
Introducing the column matrices of fields,  
 $\nu_L = (\nu^T_{eL}\, \nu^T_{\mu L}\, \nu^T_{\tau L})^T$,
$\nu_R = (\nu^T_{e R}\, \nu^T_{\mu R}\, \nu^T_{\tau R})^T$, 
$\nu^c_R =  ( (\nu^c_{e R})^T\, (\nu^c_{\mu R})^T\, (\nu^c_{\tau R})^T)^T$,
$\nu^c_L =  ( (\nu^c_{e L})^T\, (\nu^c_{\mu L})^T\, (\nu^c_{\tau L})^T)^T$, where we have omitted the spacetime variable, 
we have:
%%%%%%%%%%%%%%%%%%%%%%%%%%%%%%%
\begin{equation}
\begin{pmatrix}
\chi_L \\ 
N_L 
\end{pmatrix}
+ 
\begin{pmatrix}
\chi_R \\ 
N_R 
\end{pmatrix}
= 
\begin{pmatrix}
\chi \\ 
N 
\end{pmatrix}
=
T^\dagger 
\begin{pmatrix}
\nu_L \\ 
\nu^c_L 
\end{pmatrix}
+ T^T
 \begin{pmatrix}
\nu^c_R \\ 
\nu_R 
\end{pmatrix}
= 
\begin{pmatrix}
C \bar{\chi}^T \\ 
C \bar{N}^T 
\end{pmatrix}
\equiv 
\begin{pmatrix}
\chi^c \\ 
N^c 
\end{pmatrix}
\,,
\label{eq:chitN}
\end{equation}
%%%%%%%%%%%%%%%%%%%%%%%%%%%%%%%
where $\chi_{L(R)} = (\chi^T_{1L(R)}\, \chi^T_{2L(R)}\, \chi^T_{3L(R)})^T$,
$\chi = (\chi^T_{1}\, \chi^T_{2}\, \chi^T_{3})^T$,
$N_{L(R)}=(N^T_{1L(R)}\,N^T_{2L(R)}\,N^T_{3L(R)})^T$,
$N = (N^T_{1}\,N^T_{2}\,N^T_{3})^T$, and
$C \bar{\chi}^T$ and $C \bar{N}^T$ are given by analogous 
expressions. Taking into account the explicit form of the matrix $T$ given in Eq.~\eqref{eq:T} we get%, in terms of $\Theta$:
%%%%%%%%%%%%%%%%%%%%%%%%%%%%%%
\begin{eqnarray}
\chi_L(x) 
&=&  U_\text{PMNS}^\dagger\,\nu_L(x)  -\,U^\dagger\,\Theta\,V^\dagger\,\nu^c_L(x) \,,
\\
N_L(x)  
&=& \Theta^\dagger\,\nu_L(x)  + 
\left(\mathbb{1} - \frac{1}{2}\Theta^\dagger \Theta\right)V^\dagger\,\nu^c_L(x) \,,
\label{eq:chLtNL}
\end{eqnarray}
%%%%%%%%%%%%%%%%%%%%%%%%%%%%
where we have identified the PMNS neutrino mixing matrix as
%%%%%%%%%%%%%%%%%%%%%%%%%%%
\begin{equation}
 U_\text{PMNS} = (\mathbb{1} + \eta)\,U\,,
\label{eq:UPMNS}
\end{equation}
%%%%%%%%%%%%%%%%%%%%%%
%
with
%%%%%%%%%%%%%%%%%%%%%%
\begin{equation}
 \eta = -\,\frac{1}{2}\,RR^\dagger = 
-\,\frac{1}{2}\,(R\,V)\,(R\,V)^\dagger  
= -\,\frac{1}{2}\Theta\Theta^\dagger\,, 
\label{eq:eta}
\end{equation}
%%%%%%%%%%%%%%%%%%%%%%%%%%%%
%
and we have employed the parameter $\Theta \equiv RV$, 
which will play a major role in our subsequent discussion.

By recalling that $\nu^c_L(x) = \tilde{V} \tilde{N}_L(x)$, it follows from Eq.~\eqref{eq:chLtNL}
that the fields $N_L(x)$ ($N(x)$) 
differ from the fields $\tilde{N}_L(x)$  ($\tilde{N}(x)$) 
by the presence of the LH flavour neutrino field $\nu_L(x)$ 
(of the fields $\nu_L(x)$ and $\nu^c_R(x)$) at order $\Theta$ 
in the expression for $N_L(x)$.
Inverting the relations in Eq.~\eqref{eq:chLtNL} we find:
%%%%%%%%%%%%%%%%%%%%%%%%%%%%%%%%%%%%%
\begin{eqnarray}
 \nu_L(x) 
&=&  U_\text{PMNS}\,\chi_L(x)  + \Theta\,N_L(x) \,,  
\label{eq:nuL}
\\
 \nu^c_L(x) 
&=& -\, V\,\Theta^\dagger\,U\, \chi_L(x)  + 
V\,\left(\mathbb{1} - \frac{1}{2}\Theta^\dagger \Theta\right)\,N_L(x) \,.    
\label{eq:nucL}
\end{eqnarray}
%%%%%%%%%%%%%%%%%%%%%%%%%%%%%%%
%
From Eq.~\eqref{eq:chitN} or \eqref{eq:nucL} we also get:
%%%%%%%%%%%%%%%%%%%%%%%%%%%%%%%%%%%%
\begin{equation}
\nu_R(x) = -\,V^*\,\Theta^T\,U^*\,\chi_R(x) + 
V^*\,\left(\mathbb{1} - \frac{1}{2}\Theta^T \Theta^*\right)\,N_R(x)\,. 
\label{eq:nuR}
\end{equation}
%%%%%%%%%%%%%%%%%%%%%%%%%%%%%%%%
%

If in Eq.~\eqref{eq:nuL} we substitute $N_L(x) $ 
with the expression given in Eq.~\eqref{eq:chLtNL} 
and use $\nu^c_L(x) = \tilde{V} \tilde{N}_L(x)$  
we obtain:
%%%%%%%%%%%%%%%%%%%%%%%
\begin{equation}
\nu_L(x) =  U_\text{PMNS}\,\chi_L(x)  + \Theta \Theta^\dagger \nu_L(x)  +  
R\tilde{V} \,\tilde{N}_L(x)  + \mathcal{O}(R^3)\,.  
\label{eq:nuL2}
\end{equation}
%%%%%%%%%%%%%%%%%%%%%%%%
%
where we have used the fact that $R =\Theta\,V^\dagger$. In order for the expression in Eq.~\eqref{eq:nuL} for $\nu_L(x)$ to formally 
coincide with that in Eq.~\eqref{eq:nuL2} with just $N_L(x)$
replaced by $\tilde{N}_L(x)$, we have to neglect the term 
$\Theta \Theta^\dagger \nu_L(x)$ in Eq.~\eqref{eq:nuL2}, i.e., 
neglect a term of $\mathcal{O}(R^2)$, and approximate 
$V^\dagger\,\tilde{V} \simeq \mathbb{1}$ so that $\Theta \simeq R \tilde{V}$,  
but even in this case it follows 
from Eq.~\eqref{eq:chLtNL} that 
$N_L(x)  \simeq \Theta^\dagger\,\nu_L (x) + \tilde{N}_L(x) $.
The approximation $V^\dagger\,\tilde{V} \simeq \mathbb{1}$, however, depends on the size of the mass splittings between the $\tilde{N}_j$ compared to the corrections of order $\mathcal{O}(R^2)$ in Eq.~\eqref{eq:seesaw5}.
\footnote{Although $M_N = M_R + \mathcal{O}(R^2)$, it does not necessarily follow that $V = \tilde{V} + \mathcal{O}(R^2)$. Consider for simplicity a $2\times 2$ sub-matrix $M_N$ with $\mathcal{O}(R^2)$ real corrections parameterised as
$M_N = \text{diag}(\tilde{M}_1, \tilde{M}_2) + \tilde{M}_1\begin{pmatrix}\delta_1 & \epsilon\\ \epsilon & (\tilde{M}_2/\tilde{M}_1)\delta_2\end{pmatrix}$,
where $\epsilon, \delta_{1,2}$ are $\sim \mathcal{O}(R^2)$ and real. 
The matrix $M_N$ is diagonalised by a rotation of an angle that satisfies
$\tan 2\theta \simeq 2\epsilon/[(\tilde{M}_2 -\tilde{M}_1) /\tilde{M}_1 + (\delta_2 - \delta_1)] = \mathcal{O}(R^2)/[(\tilde{M}_2 -\tilde{M}_1) /\tilde{M}_1 + \mathcal{O}(R^2)]$.
Thus, when $(\tilde{M}_2 -\tilde{M}_1) /\tilde{M}_1 \sim \mathcal{O}(R)$, we have $\theta \sim \mathcal{O}(R)$, leading to $V = \mathbb{1} + \mathcal{O}(R)$. In the extreme situation for which $(\tilde{M}_2 -\tilde{M}_1) /\tilde{M}_1 \ll \mathcal{O}(R^2)$, then $\theta \sim \pi/4$ and the correction to $V$ would be maximal.
}
It follows from this discussion that, in general, the heavy Majorana neutrinos $\tilde{N}_j$ differ in mass and mixing from the eigenstates $N_j$.

\subsubsection{Differences between the heavy neutrino high-energy and low-energy mass eigenstates}\label{sec:MassSplits}
The differences in the masses of the high-energy heavy states $\tilde{N}_j$ and the low-energy ones $N_j$ can be evaluated starting from Eq.~\eqref{eq:seesaw5}, which can be inverted following a perturbative approach to get 
\begin{equation}\label{eq:tMNMN_2}
    V \hat{M}_N V^T = M_R + \frac{1}{2} R^\dagger R M_R + \frac{1}{2} M_R R^T R^* + \mathcal{O}(R^3).
\end{equation}
The above equation gets simplified in the basis for which $M_R$ is diagonal, that is for $M_R \equiv \tilde{M}_N^d$ and $\tilde{V} \equiv 1$, so that $R = (v/\sqrt{2})Y(\tilde{M}_N^d)^{-1} = (v/\sqrt{2})\lambda (\tilde{M}_N^d)^{-1}$.
In such a basis, in order to find the masses $M_j$ in terms of $\tilde{M}_j$, and the corresponding mass differences, up to terms of order $\mathcal{O}(R^3)$, it is sufficient to perform the diagonalisation of the symmetric matrix $A\equiv \tilde{M}_N^d + (1/2) R^\dagger R \tilde{M}_N^d + (1/2)\tilde{M}_N^d R^T R^*$.  Note also that in the limit of vanishing splittings among the heavy neutrinos $\tilde{N}_j$, the matrix $A$ is also real, and thus $V$ is a real orthogonal matrix. This procedure allows us to obtain the mass of the heavy states $N_j$ from the masses of $\tilde{N}_j$ and the $R$-matrix, an expression for the unitary matrix $V$ at the second order in the seesaw expansion and to compute the matrix of mixing of the $N_j$ states as $\Theta = RV$. In the analysis that follows, we perform such factorisation numerically.
Perhaps surprisingly, in the minimal scenario with two heavy neutrinos,
this correction only causes a splitting of $\mathcal{O}(m_\nu)$,
which can in general be $\mathcal{O}(m_\nu)/\tilde{M} \ll R^2$ (see e.g.~\cite{Drewes:2019byd}), only affecting already highly mass-degenerate heavy neutrinos.
In the case with three heavy neutrinos this correction follows a similar pattern,
causing a $\mathcal{O}(R^2 \tilde{M})$ correction  between the mass of one of the heavy neutrino states and the masses of the other two, and leaving the remaining pair degenerate up to corrections of $\mathcal{O}(m_\nu)$~\cite{Drewes:2024pad}.
The indicated corrections are of the same order as the corrections one obtains in the scenario with approximate lepton number conservation broken by neutrino Yukawa coupling or by a term in the RH neutrino Majorana mass matrix (see, e.g., 
\cite{Ibarra:2010xw,Granelli:2025lds}).

\subsubsection{Radiative corrections to light neutrino masses}\label{sec:loop}
So far, for the sake of simplicity, we have not considered the radiative corrections to light neutrino masses, $m_\nu^{\rm loop}$, which arise from the flavour neutrino self-energy diagrams involving the Higgs doublet and the Z boson \cite{Pilaftsis_1992, Grimus_2002, Aristizabal_Sierra_2011, Lopez_Pavon_2013}.
Such loop corrections modify the full neutrino mass matrix as follows
\begin{equation}\label{eq:calM_loop}
\mathcal{M} = \begin{pmatrix}
     \mathbb{O}&M_D\\M_D^T& M_R
\end{pmatrix}
\to \begin{pmatrix}m_\nu^{\rm loop}&M_D\\M_D^T&M_R\end{pmatrix},
    \end{equation}
The term $m_\nu^{\rm loop}$ then enters the right-hand side of Eq.~\eqref{eq:seesaw1} and its dominant contribution is of the same order in the seesaw expansion as the tree level contribution, $m_\nu^{\rm tree} \equiv - M_D M_R^{-1} M_D^T$. As such, the inclusion of the loop corrections in Eq.~\eqref{eq:seesaw1}, within the approximation of neglecting terms of order $\mathcal{O}(R^3)$, only affects the seesaw relation in Eq.~\eqref{eq:seesaw6}. This,
in the basis in which $M_R$ is diagonal, becomes
\begin{equation}\label{eq:sessaw_loop}
     m_\nu = m_\nu^{\rm tree} + m_\nu^{\rm loop} \simeq -\, \frac{v^2}{2}Y~\tilde{M}_{\rm loop}^{-1}Y^T,
\end{equation}
where $\tilde{M}_{\rm loop} = \text{diag}(\tilde{M}^{\rm loop}_1, \tilde{M}^{\rm loop}_2, \tilde{M}^{\rm loop}_3)$ and \cite{Pilaftsis_1992, Grimus_2002, Aristizabal_Sierra_2011, Lopez_Pavon_2013}
\begin{equation}
\tilde{M}_j^{\rm loop} \equiv \tilde{M}_j\left\{1 - \frac{\tilde{M}_j^2}{16\pi^2v^2}\left[\frac{\log{(\tilde{M}_j^2/m_H^2)}}{\tilde{M}_j^2/m_H^2 - 1}+3\frac{\log{(\tilde{M}_j^2/m_Z^2)}}{\tilde{M}_j^2/m_Z^2 - 1}\right]\right\}^{-1},
~j=1,2,3\,,
\label{eq:Mloop}
\end{equation}
where $m_H = 125$ GeV and 
$m_Z = 91.2$ GeV are the Higgs and $Z$ boson 
masses, respectively.\\

\subsection{The mixing and oscillation data, and the absolute light neutrino mass scale}\label{sec:nu_osc_data}

It has been largely proved at atmospheric, solar, reactor and accelerator neutrino experiments that flavour neutrinos oscillate \cite{PhysRevD.110.030001}. In the type-I seesaw framework of interest here, this phenomenon is explained by Eq.~\eqref{eq:nuL}, which describes the mixing of the light neutrinos among themselves and with the heavy ones. The matrix $\eta$ that enters the definition of the PMNS light neutrino mixing matrix $U_\text{PMNS}$ in Eq.~\eqref{eq:UPMNS} quantifies
the deviations of $U_\text{PMNS}$ from unitarity.
Electroweak precision data and data
on flavour observables set upper limits on its matrix entries \cite{Fernandez-Martinez:2015hxa,Blennow:2016jkn, Blennow:2023mqx}. 
These limits, in the type-I seesaw scenario considered in this work with three quasi-degenerate heavy Majorana neutrinos, 
range between $1.2\times 10^{-5} - 1.4\times 10^{-3}$ (at $95\%$ C.~L.) \cite{Blennow:2023mqx}, 
depending on the element of $\eta$ and the ordering. 
Given such stringent upper bounds on $\eta$, 
we have $U_{\text{PMNS}} \cong U$ to a very good approximation. In this work, we adopt the standard numbering of the light neutrinos, having a spectrum with either normal ordering (NO) $m_1 < m_2 < m_3$ or inverted ordering (IO) $m_3<m_1<m_2$, and use the following parameterisation
for $U$ \cite{Tanabashi:2018oca}:
%%%%%%%%%%%%%%%%%%%%%%%%%%%%
\begin{equation}
\label{PMNS}
U = \begin{pmatrix}
c_{12}c_{13}&s_{12}c_{13}&s_{13}\text{e}^{-i\delta}\\
-s_{12}c_{23}-c_{12}s_{23}s_{13}\text{e}^{i\delta}&c_{12}c_{23}-s_{12}s_{23}s_{13}\text{e}^{i\delta}&s_{23}c_{13}\\
s_{12}s_{23}-c_{12}c_{23}s_{13}\text{e}^{i\delta}&-c_{12}s_{23}-s_{12}c_{23}s_{13}\text{e}^{i\delta}&c_{23}c_{13}
\end{pmatrix}
\begin{pmatrix}
1&0&0\\
0&\text{e}^{i\alpha_{21}/2}&0\\
0&0& \text{e}^{i\alpha_{31}/2}
\end{pmatrix},
\end{equation}
%%%%%%%%%%%%%%%%%%
%
where $c_{ij} \equiv \cos\theta_{ij}$, $s_{ij} \equiv \sin\theta_{ij}$,
$0\leq \delta < 2\pi$ is the Dirac phase,
while $0 \leq \alpha_{21}, \alpha_{31} < 2\pi$ are the two Majorana phases \cite{Bilenky:1980cx}.
In the numerical analysis that follows,
we consider the best-fit values and $3\sigma$ ranges of the three neutrino mixing angles 
$\theta_{12}$, $\theta_{23}$ and $\theta_{13}$, 
and the two neutrino mass squared differences -- $\Delta m_{21}^2 \equiv m_2^2-m_1^2$ and $\Delta m_{31(32)}^2 \equiv m_3^2-m_{1(2)}^2$, with $\Delta m_{21}^2> 0 $ and $\Delta m_{31(32)}^2 >(<) \,0$ for NO (IO) -- as obtained in the \texttt{NuFit 6.0} 
global fit analysis \cite{%nufit, 
Esteban:2024eli}.
%%%
\footnote{The latest JUNO results \cite{JUNO:2025gmd} have measured $\theta_{12}$ and $\Delta m_{21}^2$ with leading precision after only 59.1 days of data collection, giving as best-fit $\pm 1 \sigma$ results $\theta_{12} = 33.78^\circ \pm 0.54^\circ$ and $\Delta m_{21}^2 = (7.50 \pm 0.12) \times 10^{-5}\,\text{eV}^2$. The result of the \texttt{NuFit 6.0} global analysis that we are using here are previous to such measurements and thus do not include such refinements. The \texttt{NuFit~6.1} release incorporates the JUNO data and provides more precise determinations of
$\theta_{12}$ and $\Delta m_{21}^2$. We do not update our numerical inputs according to \texttt{NuFit~6.1}
which would lead only to minimal changes in our results.}
%%%%
The version of the analysis with the inclusion of the Super-Kamiokande dataset gives, in particular, the values reported in Table \ref{tab:nu_params}.
The uncertainty in the determination of the Dirac phase is relatively large. In particular, for the NO case, CP-conserving values are allowed within the $3\sigma$ range. This justifies our approach of treating the Dirac phase as a free parameter in our subsequent analysis. The phenomenon of neutrino oscillation is not sensitive to the Majorana phases $\alpha_{21}$ and $\alpha_{31}$ \cite{Bilenky:1980cx, Langacker:1986jv}, which remain undetermined at present, while they play a role in the processes where the Majorana nature of the neutrinos is manifest, like the neutrinoless double-beta decay (we discuss it in Sec.~\ref{sec:nubb}).

\newcommand{\thickhline}{\noalign{\hrule height 3pt}} 

\setlength{\arrayrulewidth}{.8pt}

\setlength{\tabcolsep}{0pt}
\newcolumntype{C}{@{}>{\centering\arraybackslash}X}
\begin{table}
\centering
\begin{tabularx}{\linewidth}{|@{}C||C|C@{}|}
    \hline
    \multicolumn{3}{|c|}{\rule{0pt}{2.5ex} \cellcolor{gray!20}\bf Neutrino Oscillation Parameters from \texttt{NuFit 6.0}}\\
    \thickhline
    \rule{0pt}{2.5ex}
{\bf Parameter} & {\bf Normal Ordering} & {\bf Inverted Ordering} \\
\hline
    \rule{0pt}{2.5ex}
$\theta_{12}\,(^\circ)$ & $33.68^{+2.27}_{-2.05}$ & $33.68^{+2.27}_{-2.05}$ \\
$\theta_{13}\,(^\circ)$ & $8.56^{+0.33}_{-0.37}$ & $8.59^{+0.34}_{-0.34}$ \\
$\theta_{23}\,(^\circ)$ & $43.3^{+6.6}_{-2.0}$ & $47.9^{+1.9}_{-6.4}$ \\
$\Delta m_{21}^2\,(10^{-5}\,\text{eV}^2)$ & $7.49^{+0.56}_{-0.57}$ & $7.49^{+0.56}_{-0.57}$ \\
$\Delta m_{31(32)}^2\,(10^{-3}\,\text{eV}^2)$ & $2.513^{+0.065}_{-0.062}$ & $-2.484^{+0.063}_{-0.063}$ \\
$\delta\,(^\circ)$ & $212^{+152}_{-88}$ & $274^{+61}_{-73}$ \\
\hline
\end{tabularx}
\caption{Best-fit values and $3\sigma$ uncertainties for the neutrino oscillation parameters used throughout this work. These are the values obtained in the \texttt{NuFit 6.0} 
global fit analysis \cite{%nufit, 
Esteban:2024eli}, with the inclusion of Super-Kamiokande data. The results in the second (third) column are for NO (IO).}
\label{tab:nu_params}
\end{table}

 The presently available data still allow for the possibility that the lightest SM neutrino mass, $m_\nu^{\text{lightest}}\equiv m_{1(3)}$ in the NO (IO) case, is zero. That is, the light neutrino mass spectrum can be either with a normal hierarchy (NH) or inverted hierarchy (IH). In general, using the standard numbering of light massive neutrinos,
  in the NO case we have $m_1 < m_2 < m_3$ and 
  $m_2 = \sqrt{m_1^2 + \Delta m^2_{21}}$, $m_3 = \sqrt{m_1^2 + \Delta m^2_{31}}$.
  The analogous expressions for the IO spectrum read:
  $m_3 < m_1 < m_2$ and  $m_2 = \sqrt{m^2_3 - \Delta m^2_{32}}$,
  $m_1 = \sqrt{m^2_3 - \Delta m^2_{32} - \Delta m^2_{21}}$. Lastly, a quasi-degenerate (QD) spectrum, where \( m_1 \simeq m_2 \simeq m_3 \), 
is also not ruled out by current data.

 Current experiments and cosmological observations provide constraints on the absolute light neutrino mass scale. The latest results from the KATRIN experiment, designed to precisely measure the endpoint of tritium $\beta$-decay spectrum, set a constraint of $m_{1,2,3} < 0.45\,{\rm eV}$ at 90$\%$ C.L.~\cite{KATRIN:2024cdt}. The most stringent limit on neutrinoless double-beta ($0\nu\beta\beta$-)decay lifetime of ${}^{136}{\rm Xe}$ gives $\mnulight\leq (0.084-0.353)\,{\rm eV}$ at $90\%$ C.L.~\cite{KamLAND-Zen:2024eml} (see also \cite{Penedo:2018kpc,Tanabashi:2018oca, Penedo:2026fpv}), valid only under the assumption that massive light neutrinos are Majorana particles and neglecting the contributions from heavy neutrinos. Finally, cosmological probes set stringent bounds on the sum of neutrino masses, see, e.g.~\cite{DESI:2024mwx, DESI:2025zgx, DES:2026fyc}, which translate into upper bounds on $\mnulight$ once the two light neutrino mass splittings are fixed according to the available oscillation data. 
However, cosmological bounds can vary substantially depending on the model assumptions, statistical uncertainties, and input data, and should therefore be interpreted with caution. 
Here we prefer to remain somewhat conservative by choosing $ \mnulight\lesssim 0.1\,\text{eV}$ as an illustrative benchmark in our subsequent analysis.

\subsection{The Casas-Ibarra parameterisation}\label{sec:Casas_Ibarra}
The seesaw relation in Eq.~\eqref{eq:sessaw_loop}
allows to express the matrix of the neutrino Yukawa matrix
$Y$ in terms of the low-energy observables via the CI parameterisation \cite{Casas:2001sr, Lopez-Pavon:2015cga} 
%%%%%%%%%%%%%%%%%%%%%%%%%%%%%%%%%%%%%
\begin{equation}
Y =
 \pm i\, \dfrac{\sqrt{2}}{v}\,U\, \sqrt{\hat{m}_\nu}\,O\sqrt{\tilde{M}_{\rm loop}}\,,
\label{eq:Casas-Ibarra}
\end{equation}
%%%%%%%%%%%%%%%%%%%%%%%%%%%%%%%%%%%%%%
%
where $O$ is a $3\times 3$ complex orthogonal matrix, $O^T\,O = O\,O^T = 1$.
The CI matrix $O$ is often parameterised as
\begin{equation}\label{eq:Euler_O}
O = R^{(23)}_1R^{(13)}_2R^{(12)}_3,
\end{equation}
% %%%%%%%%%%%%%%%%%%%%%%%%%%%%%%%%%%
% %
where $R^{(jk)}_l\equiv R^{(jk)}(z_l)$, $j,k,l = 1,2,3$, $j\neq k$, $j\neq l$, $k\neq l$, $R^{(jk)}$ being a $3\times 3$ complex orthogonal matrix that describes a rotation in the $j$-$k$ plane by a given complex angle: 
\begin{equation}\label{eq:complex_rotations}
R^{(23)}(z_1)=\begin{pmatrix}
 1&0&0\\
 0&c_1&s_1\\
 0&-s_1&c_1
 \end{pmatrix},\,
R^{(13)}(z_2)= \begin{pmatrix}
 c_2&0&s_2\\
 0&1&0\\
 -s_2&0&c_2
 \end{pmatrix},\,
 R^{(12)}(z_3)=\begin{pmatrix}
 c_3&s_3&0\\
 -s_3&c_3&0\\
 0&0&1
 \end{pmatrix},
 \end{equation}
 where $c_l \equiv \cos z_l$ and $s_l \equiv \sin z_l$, $z_l = x_l+i y_l$, $l=1,2,3$, with $x_1$, $x_2$, $x_3$, $y_1$, $y_2$ and $y_3$ being six free real parameters \footnote{We note that each of the complex matrices defined in Eq.~\eqref{eq:complex_rotations} can be decomposed into its real and purely imaginary associated rotations, namely $R^{(jk)}(z_l) = R^{(jk)}(x_l) R^{(jk)}(iy_l) = R^{(jk)}(i y_l) R^{(jk)}(x_l)$. We also recall that $\cos(iy_l) = \cosh(y_l)$ and $\sin(i y_l) = i\sinh(y_l)$.}.
However, an alternative convenient parameterisation in terms of one single complex angle is \cite{Drewes:2021nqr}
%%%%%%%%%%%%%%%%%%%%%%%%%%%5
\begin{equation}
\label{eq:alt_O}
O = R^{(13)}(x_2^\nu)R^{(23)}(x_1^\nu) R^{(12)}(z)R^{(23)}(x_1^N)R^{(13)}(x_2^N),
\end{equation}
%%%%%%%%%%%%%%%%%%%%%%%%%
%
where $z = x + i y$, with $x,\,y$ real,  $x_1^\nu$, $x_2^\nu$, $x_1^N$, $x_2^N$ are four real angles. We demonstrate in Appendix \ref{app:ConnectingParams} that for any $O$-matrix in the form given in Eq.~\eqref{eq:Euler_O} there exists an equivalent  parameterisation as in Eq.~\eqref{eq:alt_O}, and vice versa. The latter parameterisation is practical in the framework with three heavy Majorana neutrinos as it involves only one complex angle rather than three as in Eq.~\eqref{eq:Euler_O} and the angles $x_1^N$, $x_2^N$, $x_1^\nu$ and $x_2^\nu$ have a more direct interpretation and connection to observables than the angles $x_1$, $x_2$ and $x_3$.  The imaginary part of $z$, $\text{Im}(z) = y$, entering the neutrino Yukawa matrix via hyperbolic functions, controls the overall magnitude of the squared mixing $\Theta^2$. Specifically, 
 values of $|y| \gg 1$
 lead to sizeable mixings, allowing for a potential production at colliders and beam-line facilities.
 
\section{Low-energy CP-violation}\label{sec:LECPV}
If CP-symmetry were conserved by the seesaw Lagrangian, the following
conditions for the entries of the Yukawa matrix would hold \cite{Pascoli:2006ci} (no sum over equal indices):
\begin{equation}\label{eq:CP_Y}
Y_{\alpha j} = -i Y_{\alpha j}^* \eta_j^{N\text{CP}}.
\end{equation}

The condition of CP-symmetry applied to the charged current interaction term involving light neutrinos implies the following condition on the entries of the PMNS matrix
\begin{equation}
\label{eq:CP_U}
U_{\alpha a}^* = -iU_{\alpha a}\eta_a^{\nu\text{CP}}.
\end{equation}

The relations on the Yukawa matrix, when written in terms of the CI parameterisation, and using the condition on $U$, imply for the elements of the CI matrix,
\begin{equation} \label{eq:CP_O}
O_{aj} = -O_{aj}^*\eta_j^{N\text{CP}}\eta_a^{\nu\text{CP}}.
\end{equation}
The factors $\eta_j^{N\text{CP}}=\pm i$ and $\eta_a^{\nu\text{CP}}=\pm i$ in the above equations are respectively the CP-parities that the heavy Majorana neutrino $N_j$ and light neutrino $\nu_a$ would carry in the case of unbroken CP-symmetry. The signs of $\eta_j^{N\text{CP}}$ and $\eta_a^{\nu\text{CP}}$ can be determined from the structure of the corresponding RH
and LH flavour neutrino Majorana mass matrices \cite{Bilenky:1987ty}. 
If Eq.~\eqref{eq:CP_Y} does not hold, or, equivalently, either of the two conditions in Eqs.~\eqref{eq:CP_U} and \eqref{eq:CP_O} on 
the entries of $U$ and $O$ is not satisfied, CP is violated by the seesaw Lagrangian.

According to Eq.~\eqref{eq:CP_U}, 
CP-invariance
implies that the columns of the PMNS matrix 
are either real or purely imaginary. Such condition is satisfied
for the following CP-conserving values of the Dirac and Majorana phases $\delta = k_\delta \pi$, $k_\delta = 0, 1$, $\alpha_{21} = k_{21}\pi$, $k_{21} = 0, 1, 2, 3$,
$\alpha_{31} = k_{31}\pi$, $k_{31} = 0, 1, 2, 3$. 

Regarding the CI matrix, the condition on the CI matrix entries imposed by Eq.~\eqref{eq:CP_O} implies that each
element of the matrix $O$ must be individually either real or purely imaginary, thus effectively reducing the number of free parameters by a factor of two. We discuss further how this condition translates in terms of the parameters in \eqref{eq:alt_O}.

\subsection{CP-conserving non-real Casas-Ibarra matrix}\label{sec:LECPV_CI}
Within the parameterisation in \eqref{eq:alt_O}, the condition $y=0$ leads to a real CI matrix, which trivially satisfies the condition in Eq.~\eqref{eq:CP_O}. However, here we are interested in keeping $y\neq 0$ to allow for relatively large Yukawa couplings. 
By direct analysis we find that a general parameterisation for a  CP-conserving non-real $O$, i.e.~that satisfies Eq.~\eqref{eq:CP_O} with $y\neq 0$, is satisfied if 
\begin{equation}\label{eq:CPconsCI}
O = P_\text{rows}\begin{pmatrix}
    i c_N \,\sh_y&i s_N \,\sh_y  & \,\ch_y   \\
        -s_N c_\nu - c_N s_\nu \,\ch_y  & c_N c_\nu - s_N s_\nu \,\ch_y & i s_\nu \,\sh_y \\
    s_N s_\nu - c_N c_\nu \,\ch_y & -c_N s_\nu - s_N c_\nu \,\ch_y  & i c_\nu \,\sh_y
\end{pmatrix}P_\text{columns},
\end{equation}
where $c_\nu\equiv \cos x_\nu$, $s_\nu\equiv \sin x_\nu$, $c_N\equiv \cos x_N$ and $s_N \equiv \sin x_N$, $0\leq x_\nu < 2\pi$ and $0\leq x_N < 2\pi$; $\sh_y = \sinh y$ and $\ch_y \equiv \cosh y$, $y \neq 0$ a free real parameter, while $P_\text{rows}$ and $P_\text{columns}$ are generic $3\times 3$ signed permutation matrices. More details on the derivation of the above form are given in Appendix \ref{app:CPonCI_explicit}. 

 We note that sign changes of the columns in Eq.~\eqref{eq:CPconsCI}, leading to ~$(Y_{e j},\,Y_{\mu j},\, Y_{\tau j}) \to (-Y_{e j},\,-Y_{\mu j},\, -Y_{\tau j})$ for certain $j = 1,\,2,\,3$, would not affect the physics, as one can always transform the heavy Majorana neutrino fields $\tilde{N}_j(x) \to - \tilde{N}_j(x)$ leaving the Lagrangian in Eq.~\eqref{eq:Lseesaw2} invariant. Also, any permutations of columns can be understood as a renumbering of the heavy neutrinos, which is irrelevant in our subsequent analysis since we allow for any of the following possibilities: $\tilde{M}_1\leq \tilde{M}_2 \leq \tilde{M}_3$, $\tilde{M}_2\leq \tilde{M}_3 \leq \tilde{M}_1$, $\tilde{M}_3\leq \tilde{M}_1 \leq \tilde{M}_2$, $\tilde{M}_1\leq \tilde{M}_3 \leq \tilde{M}_2$, $\tilde{M}_3\leq \tilde{M}_2 \leq \tilde{M}_1$ or $\tilde{M}_2\leq \tilde{M}_1 \leq \tilde{M}_3$. Moreover, the second and third rows in Eq.~\eqref{eq:CPconsCI} can be obtained from one another, up to an overall sign shift, by redefining $x_{\nu}\to x_{\nu} + 3\pi/2$, so that $\cos x_{\nu}\to \sin x_{\nu}$ and $\sin x_{\nu} \to -\cos x_{\nu}$. Lastly, since the Yukawa matrix can be defined up to an overall sign, only the relative signs between the rows are physically meaningful. For definiteness, we fix the sign of the first row and control the signs of the second and third rows by extending the range of the Majorana phases $\alpha_{21(31)}$ from  $[0, 2\pi]$ to $[-2\pi, 2\pi]$ \cite{Molinaro:2009lud}.

 We are finally left with three disconnected forms of CP-conserving CI matrices, which we further refer to as cases CP1, CP2, and CP3:
\begin{eqnarray}\label{eq:CPconsCI_1}
\bullet~~\textbf{CP1)}\qquad~~ O^{(1)} &=& \begin{pmatrix}
    i c_N \,\sh_y&i s_N \,\sh_y  & \ch_y   \\
        -s_N c_\nu - c_N s_\nu \,\ch_y  & c_N c_\nu - s_N s_\nu \,\ch_y & i s_\nu \,\sh_y \\
    s_N s_\nu - c_N c_\nu \,\ch_y & -c_N s_\nu - s_N c_\nu \,\ch_y  & i c_\nu \,\sh_y
\end{pmatrix},\\
\label{eq:CPconsCI_2}
\bullet~~\textbf{CP2)}\qquad~~ O^{(2)} &=& \begin{pmatrix}
     -s_N c_\nu - c_N s_\nu \,\ch_y  & c_N c_\nu - s_N s_\nu \,\ch_y & i s_\nu \,\sh_y \\
      i c_N \,\sh_y&is_N \,\sh_y  & \ch_y \\
    s_N s_\nu - c_N c_\nu \,\ch_y & -c_N s_\nu - s_N c_\nu \,\ch_y  & i c_\nu \,\sh_y
\end{pmatrix},\\
\label{eq:CPconsCI_3}
\bullet~~\textbf{CP3)}\qquad~~ O^{(3)} &=& \begin{pmatrix}
       -s_N c_\nu - c_N s_\nu \,\ch_y  & c_N c_\nu - s_N s_\nu \,\ch_y & i s_\nu \,\sh_y \\
    s_N s_\nu - c_N c_\nu \,\ch_y & -c_N s_\nu - s_N c_\nu \,\ch_y  & i c_\nu \,\sh_y\\
    i c_N \,\sh_y&i  s_N \,\sh_y  & \ch_y 
\end{pmatrix}.
\end{eqnarray}
These three discrete choices also differ in how they are connected to the scenario with only two Majorana neutrinos,
in which one heavy neutrino decouples.
In the IH limit $(m_3 \rightarrow 0)$, this decoupling occurs for CP1 and CP2 when $x_\nu \in \{\pi/2, 3\pi/2\}$,
whereas
in the NH limit $(m_1 \rightarrow 0)$ it occurs for CP2 or CP3 when $x_\nu \in \{0, \pi\}$; thus CP2 connects to both limits.
In all of these limits $x_N \in \{0, \pi/2, \pi, 3 \pi/2\}$, with the specific choice determining which of the heavy neutrinos decouples.

\subsubsection{CP-violation uniquely from the Dirac phase} 
It is worth noting that the conditions in Eqs.~\eqref{eq:CP_U} and \eqref{eq:CP_O} are necessary conditions for the CP-invariance of the type-I seesaw Lagrangian, i.e.~for the condition in Eq.~\eqref{eq:CP_Y} to be satisfied, but not sufficient. Indeed, for certain choices of the CP-parities of the light and heavy neutrinos $\eta^{\nu\text{CP}}_a$ and $\eta^{N\text{CP}}_j$, there exist the possibility that the Eqs.~\eqref{eq:CP_U} and \eqref{eq:CP_O} are satisfied separately while Eq.~\eqref{eq:CP_Y} is not. In such situation, CPV arises from the CPV arises from the \textit{interplay} between the CP-conserving PMNS and CI matrices \cite{Pascoli:2006ci}.
To have CPV uniquely from the Dirac phase $\delta$, any contribution to the CPV from the interplay between CP-conserving PMNS and CI matrices $U$ and $O$ should be avoided. This can be achieved by fixing the Majorana phases to specific CP-conserving values, which depend
on the form of the CP-conserving CI matrix.
To ensure that all the CPV is coming solely from the 
Dirac phase $\delta$,
we have to make the following choices for the 
Majorana phases:
\begin{eqnarray}\label{eq:CP1a21a31}
    \textbf{CP1)} \quad 
    \alpha_{21}& =& \pm \pi,\quad \alpha_{31} = \pm \pi;\\\label{eq:CP2a21a31}
    \textbf{CP2)} \quad 
    \alpha_{21} &=& \pm \pi,\quad \alpha_{31} = 0, \pm 2 \pi;\\\label{eq:CP3a21a31}
    \textbf{CP3)} \quad 
    \alpha_{21} &=& 0, \pm 2 \pi,\quad \alpha_{31} = \pm \pi.
\end{eqnarray}
These are the only choices for the Majorana phases that yield a set of $\eta_a^{\nu\text{CP}}$, $a=1,\,2,\,3$, which is consistent with the conditions on the combinations $\eta_j^{N\text{CP}}\eta_a^{\nu\text{CP}}$, $a, j=1,\,2,\,3$, imposed by Eq.~\eqref{eq:CP_O} when $O$ is of the discussed form, and thus ensure CP-conservation in the Yukawa couplings $Y$ if, additionally, $\delta = 0, \pi,\,2\pi$.\footnote{In the case of a real CI matrix, the combinations for the Majorana phases ensuring Dirac-phase CPV would be $\alpha_{21} =0,\,\pm 2\pi$ and $\alpha_{31} = 0, \pm 2\pi$.}

If $m_{1(3)}=0$ as in the NH (IH) case, any overall phase in the third (first) column of the PMNS matrix can be eliminated through field redefinition. Thus, only the combination of phases $\alpha_{23}\equiv \alpha_{21}-\alpha_{31}$ (the phase $\alpha_{21}$) is physical.
This yields the following conditions that guarantee no CPV from the interplay between CP-conserving PMNS and CI matrices in the hierarchical case:
\begin{eqnarray}\label{eq:CP1a23}
\textbf{CP1)} \quad\alpha_{23} &=& 0,\pm2\pi \,(\alpha_{21}=\pm\pi);\\\label{eq:CP2a23}
    \textbf{CP2)}\quad \alpha_{23} &=& \pm \pi ,\pm 3\pi\, (\alpha_{21} = \pm \pi);\\\label{eq:CP3a23}
    \textbf{CP3)} \quad\alpha_{23} &=& \pm \pi,\pm3\pi \,(\alpha_{21} = 0,\pm2\pi).
\end{eqnarray}

\subsection{The observable parameter space assuming Dirac CP-violation}\label{sec:LECPV_parameter_spaces}
The constraints on the CI matrix discussed above, which enforce low-energy CP violation solely from the Dirac phase $\delta$ while allowing for large Yukawa couplings through a complex structure of the matrix, correspond to specific subregions of the allowed parameter space relevant to low-energy experiments, as we shall see in this section.

\subsubsection{Parameter space for heavy neutral lepton searches}\label{sec:LECPV_HNLs}
The mixing $\Theta_{\alpha j}$
sets the strength of the charged current (CC) and neutral current (NC) weak interaction couplings of the heavy mass eigenstates $N_j$ to the $W^\pm$ bosons and
the charged lepton $\alpha$, and to the $Z$ boson and
the LH flavour neutrino $\nu_{\alpha L}$,
$\alpha=e,\mu,\tau$, in the weak interaction Lagrangian:
%%%%%%%%%%%%%%%%%%%%%%%%%%%%%%%%%%%%%%%%%%%
\begin{eqnarray}
 \mathcal{L}_\text{CC}^N(x)&=& -\frac{g_w}{\sqrt{2}}
\overline{\alpha_L}(x)\,\slashed{W}(x)\,\Theta_{\alpha j}\,N_{jL}(x)
+{\rm h.c.},
\label{eq:NCC}\\
\mathcal{L}_\text{NC}^N(x) &=& -\frac{g_w}{2 c_{w}}
\overline{\nu_{\alpha L}}(x)\,\slashed{Z}(x)\,\Theta_{\alpha j}\,N_{jL}(x)
+{\rm h.c.},\;\;\;
\label{eq:NNC}
\end{eqnarray}
%%%%%%%%%%%%%%%%%%%%%%%%%%%%%%%%%%%%%%
%
where $g_w$ is the $SU(2)_L$ gauge coupling, $c_w \equiv \cos\theta_w$, $\theta_w$ being the weak mixing angle; $\alpha_L(x)$ is the LH component of the charged lepton field $\alpha(x)$ with  mass $m_\alpha$, $\alpha = e,\,\mu,\,\tau$; $\slashed{W}(x)$ and $\slashed{Z}(x)$ are respectively the fields describing the $W^{\pm}$ and $Z$ bosons, contracted with the Dirac gamma matrices. \footnote{The same coupling $\Theta_{\alpha j}$ regulates the interaction of $N_j$  with the active neutrino of flavour $\alpha$ and the Higgs boson according to $\mathcal{L}_\text{Higgs}^N(x) = - (M_j/v)\,
    \overline{\nu_{\alpha L}}(x)\Theta_{\alpha j}N_{jR}(x) H(x) + {\rm h.c.}$, 
$H(x)$ being the field of the Higgs boson $H$.}
 Neutral fermions with masses much above the eV scale that interact weakly through CC and NC interactions are commonly referred to as \textit{heavy neutral leptons} (HNLs). A wealth of
testable phenomenology is related to HNLs \cite{%Leung:1983ix, Gronau:1984ct, 
Han:2006ip, delAguila:2007qnc, Gorbunov:2007ak, delAguila:2008cj, delAguila:2008hw, Atre:2009rg, Ibarra:2010xw,Ibarra:2011xn, Dinh:2012bp, Cely:2012bz, Penedo:2017knr, Bondarenko:2018ptm, Bolton:2019pcu, Urquia-Calderon:2022ufc} 
(see, %also
e.g., \cite{Abdullahi:2022jlv, Antel:2023hkf} and ample literature referenced therein). Notably, if HNLs' masses are in the range $\mathcal{O}(0.1 - 100)\,\text{GeV}$, and their couplings are sufficiently large, 
they can be produced with observable rates at colliders and beam-line facilities. 

The magnitude of the couplings $\Theta_{\alpha j}$ is crucial for the possibility of producing and detecting the states $N_j$. Depending on the process, different combinations of the couplings $\Theta_{\alpha j}$ appear in the expressions of the relevant observables, such as production cross section and decay rates, see e.g.~\cite{Antusch:2016ejd, Bondarenko:2018ptm}. In this work, we concentrate on the following phenomenologically relevant quantities:
\begin{equation}\label{eq:Theta2s}
\Theta^2_j \equiv \sum_{a=e,\mu,\tau} |\Theta_{\alpha j}|^2,\quad
\Theta^2_\alpha \equiv \sum_{j=1}^3|\Theta_{\alpha j}|^2\quad\text{and}\quad\Theta^2 \equiv \sum_{\alpha = e,\,\mu,\,\tau}\Theta_\alpha^2 = \sum_{j=1}^3 \Theta_j^2.
\end{equation}
and on the individual $|\Theta_{\alpha j}|^2$. 

Care is needed when computing the mixings of the states $N_j$ starting from
the seesaw parameters at high-energy. For the numerical analysis that
follows,  we always take the parameters of the Lagrangian
in Eq.~\eqref{eq:Lseesaw2} as input, because it is more practical for our
leptogenesis studies. That is, working in the basis in which
$M_R$ is diagonal (i.e.,~$\tilde{V} = \mathbb{1}$),
we first specify the masses of the heavy Majorana
neutrinos $\tilde{N}_j$, namely $\tilde{M}_1$, $\tilde{M}_2$ and
$\tilde{M}_3$, and the Yukawa
couplings $Y_{\alpha j} = \lambda_{\alpha j}$
according to the CP-conserving CI parameterisation discussed
in subsection \ref{sec:LECPV_CI}.
This allows us to construct the matrix $R$ with entries
$R_{\alpha j} = (v/\sqrt{2}) Y_{\alpha j}/\tilde{M}_j$. We then perform
the diagonalisation of the matrix in Eq.~\eqref{eq:tMNMN_2} to get
the masses of the states $N_j$, i.e.~$M_1$, $M_2$ and $M_3$, and the matrix
$V$, and finally we compute the mixing as $\Theta = RV$.

For the  analytical description of the mixing 
it is convenient to consider the quantity
\begin{equation}\label{eq:wTheta}
    \wTheta_{\alpha j} \equiv R_{\alpha j} \tilde{M}_j(\tilde{M}_j^{\rm loop}\wMav)^{-1/2},
\end{equation}
where $\tilde{M}_{\rm av} \equiv \sum_{j=1,2,3} \tilde{M}_j/3$, and the corresponding parameters 
\begin{equation}
    \wTheta_{j}^2 = \sum_{\alpha=e,\mu,\tau} |\wTheta_{\alpha j}|^2, 
    \quad 
    \wTheta_{\alpha}^2 = 
    \sum_{j=1}^3 |\wTheta_{\alpha j}|^2,\quad
    \wTheta^2 = \sum_{\alpha=e,\mu,\tau} \wTheta_\alpha^2.
    \end{equation}
 These mixing parameters do not depend on the mass of the heavy neutrinos
 but only on the average mass $\wMav$.
 As we will show below, to a good approximation we have 
   $\wTheta_\alpha^2 \simeq \Theta_{\alpha}^2$ and
$\wTheta^2 \simeq \Theta^2$.
It proves useful to analyse the dependence
 of  $\wTheta^2_j$, $\wTheta^2_{\alpha}$ and $\wTheta^2$
   on the CI matrix: 
\begin{equation}
    \wMav\wTheta_j^2 = \left[O^\dagger \hat{m}_\nu O\right]_{jj},\quad  
    \wMav\wTheta_{\alpha}^2 =\left|\left[U\sqrt{\hat{m}_\nu} O O^\dagger \sqrt{\hat{m}_\nu}U^\dagger\right]_{\alpha\alpha}\right|\quad\text{and}\quad
    \wMav\wTheta^2 = \text{Tr}\left[OO^\dagger \hat{m}_\nu\right].
\end{equation}
The expanded expressions for $\wTheta^2_j$, $\wTheta_{\alpha}^2$ and $\wTheta^2$ are given in Appendix \ref{app:Theta2_Expr}. As can be checked, within the adopted parameterisation, the matrix product $OO^\dagger$, and thus the quantities $\wTheta_{\alpha}^2$ and $\wTheta^2$, do not depend on the angle $x_N$. In contrast, $\wTheta^2_j$ and $\wTheta^2$ do not depend on the PMNS angles and phases. 

The radiative corrections in $\tilde{M}_j^{\rm loop}$
in Eq.~\eqref{eq:wTheta}, which,  according to
Eq.~\eqref{eq:Mloop} and in the mass range of interest
$0.1\leq \wMav/\text{GeV} \leq 100$, give at most corrections of the order of
$\mathcal{O}(1\%)$ of $\wMav$,
can  be neglected. The effects of the difference of masses of 
the heavy Majorana neutrinos on $\wTheta^2_j$, $\wTheta^2_\alpha$ and
$\wTheta^2$, even if they are $\mathcal{O}(10\%)$ of $\wMav$, can
to a good approximation also be neglected.\footnote{If, e.g., they are $\mathcal{O}(10\%)$ of $\wMav$,
they would give a $\mathcal{O}(5\%)$ correction in  Eq.~\eqref{eq:wTheta}.
} With these rather precise approximations we have
$M^{\rm loop}_{1,2,3} \simeq\tilde{M}_{1,2,3} \simeq \wMav$.
Thus, for all the practical purposes, we can approximate $\wTheta_\alpha^2 \simeq \sum_j |R_{\alpha j}|^2$, $\wTheta_j^2 \simeq \sum_{\alpha} |R_{\alpha j}|^2$ and $\wTheta^2 \simeq \sum_{\alpha, j} |R_{\alpha j}|^2$. 
Furthermore, due to unitarity of the matrix $V$, we also have
$\Theta_{\alpha}^2 = \sum_{j} |R_{\alpha j}|^2 $ so that
$\wTheta_\alpha^2 \simeq \Theta_{\alpha}^2$ and
$\wTheta^2 \simeq \Theta^2$, while, in general,
$\Theta_{\alpha j} \neq R_{\alpha j}$ and
thus $\wTheta_{\alpha j} \neq \Theta_{\alpha j}$ and
$\Theta^2_j \neq \wTheta^2_j$, unless the mass splittings
of the heavy neutrinos are sufficiently larger than
the second order seesaw corrections, so that
$V = \mathbb{1} + \mathcal{O}(R^2)$. For the sake
of simplicity, we concentrate the discussion in this section on the quantities $\wTheta_\alpha^2 \simeq \Theta^2_\alpha$ and $\wTheta^2 \simeq \Theta^2$, for which no such ambiguity arises. Focusing on
these quantities is sufficient for illustrating 
analytically our
main point: future measurements can potentially
test and falsify the hypothesis of
Dirac phase low-energy CPV.

\paragraph{Total mixing squared.} The total mixing squared $\wTheta^2$ controls the overall production rate of heavy Majorana neutrinos at experiments. It is therefore insightful to determine its minimal value, and the corresponding lower bound on $|y|$, required for the mixings to lie within reach of future facilities. The quantity $\wMav \wTheta^2$ is minimised for given $y$ at different values of the angle $x_\nu$, depending on the ordering and the considered forms of the CP-conserving CI matrix. 
\begin{itemize}
\item{\bf CP1)} for $x_\nu = (2n+1)\pi/2\, (x_\nu = n\pi)$ in the case of NO (IO), so that
\begin{equation}
\wMav\wTheta^2 \geq (\ch_y^2 + \sh_y^2) \sum_a m_a - 2m_{3(2)} \,\sh_y^2 = \frac{e^{2|y|}}{2}\left(1+e^{-4|y|}\right)(m_1+m_{2(3)}) + m_{3(2)};
\end{equation}
\item {\bf CP2)} for $x_\nu = (2n+1)\pi/2\, (x_\nu = n\pi)$ in the case of NO (IO), so that
\begin{equation}
\wMav\wTheta^2 \geq (\ch_y^2 + \sh_y^2) \sum_a m_a - 2m_{3(1)} \,\sh_y^2 = \frac{e^{2|y|}}{2}\left(1+e^{-4|y|}\right)(m_{1(3)}+m_2) + m_{3(1)};
\end{equation}
\item{\bf CP3)} for $x_\nu = n\pi$ in either scenario with NO or IO, so that
\begin{equation}
\wMav\wTheta^2 \geq (\ch_y^2 + \sh_y^2) \sum_a m_a - 2m_{2} \,\sh_y^2 = \frac{e^{2|y|}}{2}\left(1+e^{-4|y|}\right)(m_1+m_3) + m_2.
\end{equation}
\end{itemize}
The minimum is found for $y=0$, and we can write the following inequality
\begin{equation}\label{eq:naiveSeesawLimit}
\wTheta^2 \geq \frac{\sum_a m_a}{\wMav} = 10^{-10} \left(\frac{\sum_a m_a}{0.1 \,\text{eV}}\right)\left(\frac{1 \,\text{GeV}}{\wMav}\right),
\end{equation}
for each of the cases. This is the ordinary \textit{seesaw limit} on the total mixing squared that arises from the requirement of reproducing the correct neutrino masses via the seesaw mechanism. The case of a real CI matrix, i.e.~$y=0$, inevitably leads to mixings lying along the seesaw limit, while the CI matrix in Eq.~\eqref{eq:OCP_general} with $y\neq 0$ that we are considering is a general CP-conserving form that allows for mixings above the seesaw bound.

Similar bounds can be derived for each individual $\wTheta_\alpha^2$, while no such bound applies to individual $|\Theta_{\alpha j}|^2$. Values of the mixing deviating substantially from the seesaw limit imply an appreciable number of events at colliders.  Within the adopted parameterisation, sufficiently large couplings can be obtained when $e^{2|y|}\gg1$.
In the case of a hierarchical light  neutrino mass spectrum
 imposing $|y|\geq y_{\text{min}}$, with $y_{\text{min}} = 2$ for CP1 and CP2 or $y_{\text{min}} = 1.15$ for CP3 ($y_{\text{min}} = 1.15$ for each of the cases with IO), we get $\wTheta^2 \gtrsim \Theta^2_{\text{min}} \equiv 3\times 10^{-10}\,\text{GeV}/\wMav$, which is the minimum squared mixing within the parameter reach of, e.g., the planned SHiP \cite{SHiP:2018xqw} at $\mathcal{O}(1\,\text{GeV})$ and future proposed circular colliders FCC-ee/CEPC \cite{Blondel:2022qqo, CEPCStudyGroup:2018ghi} at $\mathcal{O}(10\,\text{GeV})$. For the QD case with $\mnulight \simeq 0.1\,\text{eV}$, it is enough to have $y\gtrsim y_{\text{min}} = 0.6$ so that $\wTheta^2 \gtrsim \Theta^2_{\text{min}}$.

\begin{figure}[t!]
    \centering
        \includegraphics[width = 0.47\textwidth]
    {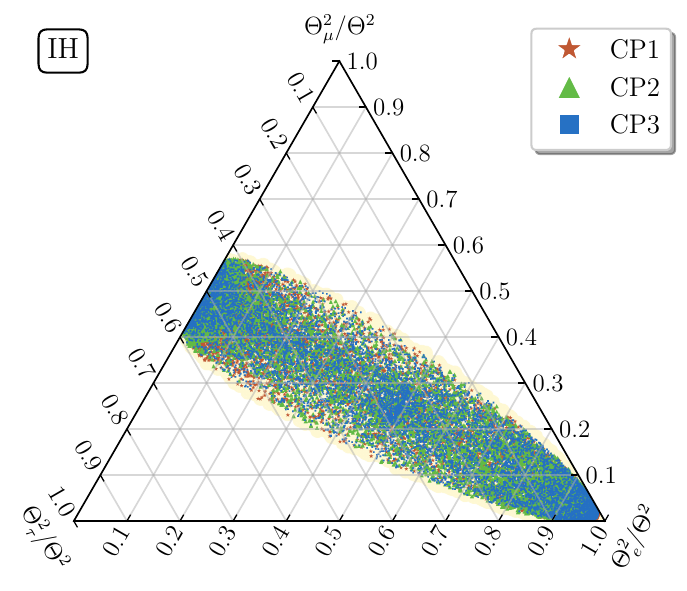}\\
\includegraphics[width=0.47\textwidth]{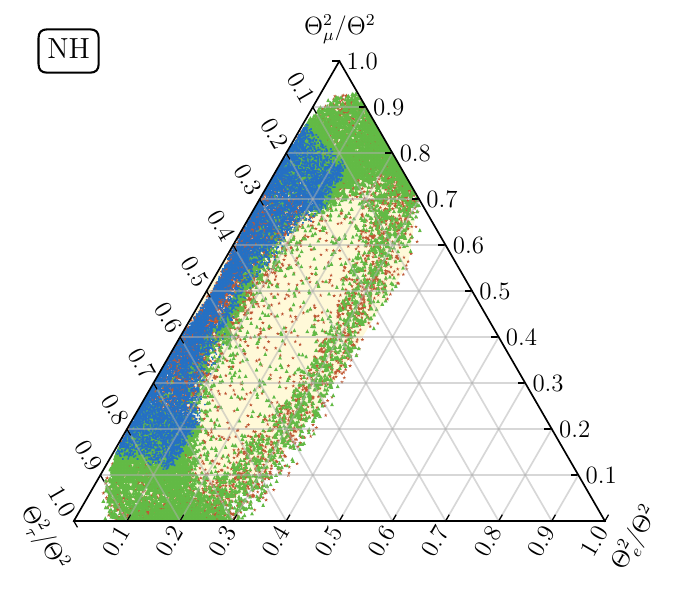}
\includegraphics[width=0.47\textwidth]{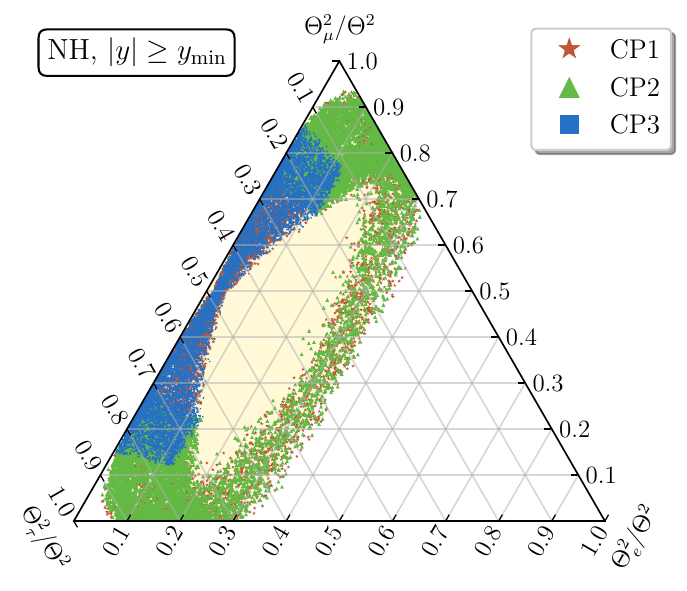}
    \caption{The parameter spaces in the $\Theta^2_e/\Theta^2$-$\Theta^2_\mu/\Theta^2$-$\Theta^2_\tau/\Theta^2$ ternary plane associated to the three different cases of CP-conserving CI matrix; the top (bottom) panel(s) is (are) for IH (NH), $m_{3(1)}=0$. The red stars, green triangles and blue circles correspond to the discussed CP1, CP2 and CP3 cases, respectively. The angle $x_\nu$ is varied within $[0,2\pi]$, while $\delta$ within $0 < \delta <\pi$ and $\pi < \delta < 2\pi$. The parameter $|y|$ is varied within $(0,10]$ in the top and bottom-left panels, while in the lower-right one we set $|y|\geq 2$ for CP1 and CP2, $|y|\geq = 1.15$ for CP3, so that the values of the total mixing squared is reachable by future experiments. Imposing the analogous restriction in the IH case does not lead to any appreciable difference in the plot and thus we do not show it. The Majorana phases are chosen to avoid CPV from the interplay between CP-conserving CI and PMNS matrices. The yellow regions correspond to the full parameter space of the type-I seesaw with three quasi-degenerate heavy Majorana neutrinos. The PMNS angles are varied within the $3\sigma$ ranges as obtained with the \texttt{NuFit 6.0} global analysis \cite{Esteban:2024eli} (including Super-Kamiokande data).}
\label{fig:ternaryplots_H}
\end{figure}

\begin{figure}[t!]
\centering
\includegraphics[width=0.45\textwidth]{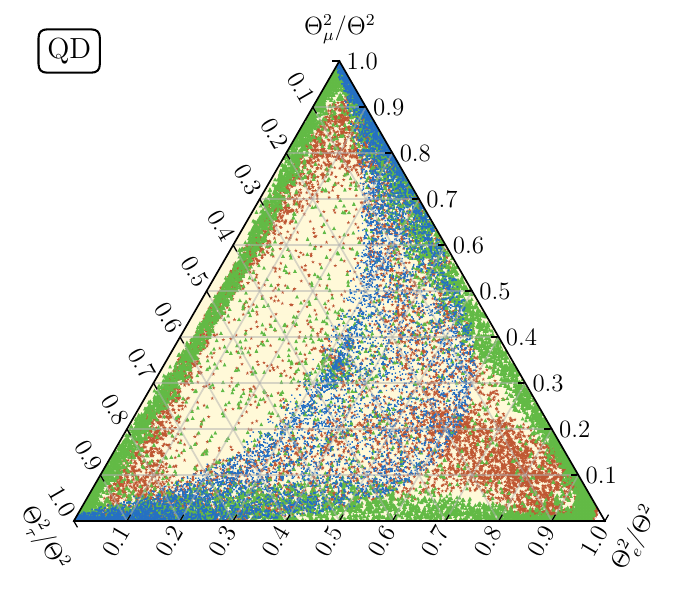}
    \includegraphics[width = 0.45\textwidth]
    {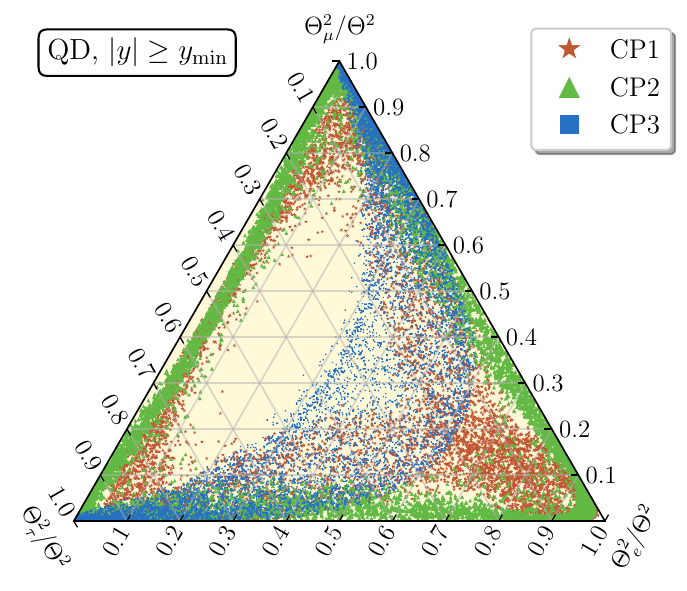}
    \caption{The parameter spaces in the $\Theta^2_e/\Theta^2$-$\Theta^2_\mu/\Theta^2$-$\Theta^2_\tau/\Theta^2$ ternary plane associated to the three different cases of CP-conserving CI matrix in the case   $\mnulight = 0.1\,\text{eV}$, corresponding to a QD spectrum. The right plot is obtained after applying the constraint $|y| \geq y_{\text{min}} = 0.6$. All other details are as in Fig.~\ref{fig:ternaryplots_H}.}
\label{fig:ternaryplots_mnu}
\end{figure}

\paragraph{Flavour ratios.}  We now examine the corresponding flavour ratios in the scenarios under consideration. We show in Fig.~\ref{fig:ternaryplots_H} and \ref{fig:ternaryplots_mnu} the parameter spaces in the $\Theta^2_e/\Theta^2$-$\Theta^2_\mu/\Theta^2$-$\Theta^2_\tau/\Theta^2$ ($\wTheta_\alpha^2 \sim \Theta_\alpha^2$, $\alpha = e,\mu,\tau$) ternary plane associated to the discussed forms of CP-conserving CI matrix.  While the total mixing controls the overall production rate of heavy Majorana neutrinos, their flavour composition encodes information on the structure of the PMNS and CI matrices. The dependence of the parameter $\Theta^2_\alpha/\Theta^2$ on the CI parameters is peculiar to the scenario with 3 RHNs, while in the scenario with 2 RHNs, as it is well known, they do not depend on the CI matrix at all. In Fig.~\ref{fig:ternaryplots_H} we set $\mnulight = 0$, while in Fig.~\ref{fig:ternaryplots_mnu} $\mnulight = 0.1\,\text{eV}$. The PMNS angles are varied within the 3$\sigma$ range of the global analysis \cite{Esteban:2024eli}, while $\delta$ is treated as a free parameter and is varied within $0<\delta < \pi$ and $\pi < \delta < 2\pi$. The red stars, green triangles and blue squares correspond to low-energy CPV solely from $\delta$ in the CP1, CP2 and CP3 scenarios, respectively. 
The yellow region, instead, is obtained for generic forms of the CI matrix without the condition of CP-conservation, with the Dirac phase varied within $[0,2\pi]$ and the Majorana phases within $[-2\pi,2\pi]$. The relevant real angles of the CI parameterisation ($x_\nu$ for CP1, CP2 and CP3; $x$ and $x_{1,2}^\nu$ for the generic case) are varied within $[0, 2\pi]$. The left and right panels of both figures are obtained for NO and IO, respectively. The bottom-right (right) panel of the Fig.~\ref{fig:ternaryplots_H} (Fig.~\ref{fig:ternaryplots_mnu})
is obtained for $|y|\geq y_{\text{min}}$. We note that, for sufficiently large values of $|y|$ so that $e^{2|y|}\gg 1$, the dependence on $y$ of $\wTheta_{\alpha}^2/\wTheta^2$ drops out (see Appendix \ref{app:Theta2_Expr} for the explicit analytical expressions).

It is clear that, depending on the ordering, $\mnulight$ and $y$, the parameter spaces in the ternary planes associated to the CP1, CP2 and CP3 cases can form distinct subsets of the full allowed region. Focusing on the hierarchical case ($\mnulight=0$), for NH, the CP3 case is confined to a region where $\Theta_e^2/\Theta^2 \lesssim 0.1$,  $0.15\lesssim\Theta_\mu^2/\Theta^2 \lesssim 0.85$ and  $0.15\lesssim \Theta_\tau^2/\Theta^2 \lesssim 0.85$, while CP1 and CP2 points scatter over almost the entire allowed region. When no restriction on $y$ is imposed, there is significant overlapping between the various cases, as well as between CP1 and CP2 and the full region that extends in the following ranges: $0 \lesssim \Theta_e^2/\Theta^2 \lesssim 0.35$,  $0\lesssim\Theta_\mu^2/\Theta^2 \lesssim 0.92$ and  $0\lesssim \Theta_\tau^2/\Theta^2 \lesssim 0.92$.

In the IH case, CP1, CP2, and CP3 are completely degenerate with each other and also with the full region that extends in the following ranges: $0 \lesssim \Theta_e^2/\Theta^2 \lesssim 1$,  $0\lesssim\Theta_\mu^2/\Theta^2 \lesssim 0.6$ and  $0\lesssim \Theta_\tau^2/\Theta^2 \lesssim 0.6$.

If the restriction $|y| \geq y_{\text{min}}$ is imposed, i.e., in the region of the parameter space that can be
tested experimentally, most of the degeneracies in the NH case vanish. The CP1 and CP2 cases start forming distinct subsets of the full region without fully overlapping with it (the region where there is no overlap extends within the ranges $0.05 \lesssim \Theta_e^2/\Theta^2 \lesssim 0.25$,  $0.1\lesssim\Theta_\mu^2/\Theta^2 \lesssim 0.7$ and  $0.1\lesssim \Theta_\tau^2/\Theta^2 \lesssim 0.7$), while the CP3 region remains roughly the same and degenerate with both CP1 and CP2. 

Even with the restriction $|y| \geq y_{\text{min}}$, the IH cases remain fully degenerate and no appreciable difference is found -- for this reason we do not show the corresponding plot. 

A similar discussion applies when  $\mnulight$
differs significantly from zero.
According to Fig.~\ref{fig:ternaryplots_mnu}, if $\mnulight= 0.1\,\text{eV}$,
the entire region nearly covers the entire triangle, while the CP1, CP2 and
CP3 regions exhibit degeneracies similar to those in the hierarchical case,
which largely disappear when the condition $y\geq y_{\text{min}}$
is fulfilled, i.e., in the experimentally testable region.
We note that for this choice of $\mnulight$,
the results for the NO and IO cases are practically identical as the light neutrino mass spectrum is QD.\footnote{For definiteness, in this case we have considered the results of the global analysis on neutrino oscillation data for a light neutrino mass spectrum with NO.}

The above discussion suggests that, based on the flavour structure $\Theta_e^2/\Theta^2:\Theta_\mu^2/\Theta^2:\Theta_\tau^2/\Theta^2$, the scenario where
the CPV arises solely from the Dirac phase $\delta$ through a non-real CI matrix is,
in principle, testable and falsifiable 
from scenarios with additional CP-violating sources, such as the Majorana phases and/or high-energy phases in the CI matrix.
Additional data on $\mnulight$ and the light neutrino mass ordering
would facilitate the tests of the considered scenarios. 

\subsubsection{Parameter space for neutrinoless double-beta decay searches}\label{sec:nubb}

With the neutrinos having a Majorana nature, processes like the neutrinoless double-beta ($\nubb$-)decay mediated by the exchange of a virtual Majorana neutrino can happen (see, e.g., \cite{Agostini:2022zub} for a review). Given the mixing of the light neutrinos with the heavy ones, also the eigenstates $N_j$ can contribute to the same process. The $\nubb$-decay rate $\Gamma^{\nubb}_{A}$ for a given nucleus with mass number $A$, in the type-I seesaw scenario of interest, is given by \cite{Blennow:2010th, Faessler:2014kka, Barea:2015zfa, Dekens:2023iyc, Dekens:2024hlz, deVries:2024rfh}
\footnote{ We do not include here the contribution from short-range operators \cite{Cirigliano:2018hja}, the sign and size of which is still under investigation.}:
%%%%%%%%%%%%%%%%%
\begin{equation}
\Gamma_{\nubb} \simeq G_{0\nu} g_A^4 |V_{ud}^2|^2\left|\sum_{a=1,2,3} U_{e a}^2 \frac{m_a}{m_e} \mathcal{M}_{\nubb}(A) + \sum_{j=1,2,3} \Theta_{e j}^2 \frac{M_j}{m_e} \mathcal{M}^{(N)}_{\nubb}(M_j, A)\right|^2
\end{equation}
%%%%%%%%%%%%%%%%%%%%%%
%
where $G_{0\nu}\approx 10^{-14}\,\text{yr}^{-1}$ is a well-known kinematical factor, whose precise value depends on the nucleus, see e.g.~\cite{Kotila:2012zza, Stoica:2013lka};  $g_A \simeq 1.27$ is the nucleon axial coupling; $V_{ud} \simeq 0.97$ is the up-down quark mixing matrix element; $m_e$ 
is the electron mass;
$\mathcal{M}_{\nubb}(A)$ and
$\mathcal{M}^{(N)}_{\nubb}(M_j, A)$ are respectively 
the light and heavy Majorana neutrino exchange  
nuclear matrix elements, which are functions of the nucleus mass
number and contain most of the
theoretical uncertainty. While $\mathcal{M}_{\nubb}(A)$ practically does not depend on the light neutrino mass $m_a$, $\mathcal{M}^{(N)}_{\nubb}(M_j, A)$
is a non-trivial function of $M_j$. For $M_j \gg 200~{\rm MeV}$,
$\mathcal{M}^{(N)}_{\nubb}(M_j, A) \propto M^{-2}_j$. 
In what follows, for definiteness, we focus on ${}^{136}\text{Xe}$ using $A  =136$ and further drop the dependence on $A$. 
The matrix element of the light neutrino exchange can be factored out and the $\nubb$-decay rate can then be re-written as:
%%%%%%%%%%%%%
\begin{equation}
  \Gamma_{\nubb} = G_{0\nu}g_A^4 |V_{ud}^2|^2 |\mathcal{M}_{\nubb}(A)|^2
  (m^{\rm eff}_{\beta\beta})^2/m_e^2\,,
  \label{eq:Gamma2}
\end{equation}
%%%%%%%%%%%%%%%%%%%%%%%%
%
where $m^{\rm eff}_{\beta\beta}$ is the effective Majorana mass,
%%%%%%%%%%%%%%%%%%
\begin{equation}
  \label{eq:mbbeff_full}
m^{\rm eff}_{\beta\beta} = \left|(m_\nu)_{ee} + \sum_{j=1,2,3} \Theta_{e j}^2 M_j \mathcal{F}(M_j)\right|\,,
\end{equation}
%%%%%%%%%%%%%%%%%%%
%
and $\mathcal{F}(M_j) \equiv \mathcal{M}^{(N)}_{\nubb}(M_j)/\mathcal{M}_{\nubb}$.

\paragraph{The case of negligible heavy neutrino contributions.} 
Because of the smallness of the nuclear matrix element
$\mathcal{M}^{(N)}_{\nubb}(M_j)$ at relatively large values of
the heavy neutrino masses $M_j$,
the heavy neutrino contributions for, e.g., $M_j \gtrsim 10$ GeV,
would be strongly suppressed.\footnote{At $M_j \gtrsim 10$ GeV, for example, we have $\mathcal{F}(M_j) \simeq 4.4\times 10^{-4} (10\,\text{GeV}/M_j)^2$ \cite{deVries:2024rfh} (see also \cite{Blennow:2010th}).
While a larger contribution may be possible by increasing the right-handed neutrino masss splittings,
this inevitably leads to large radiative corrections to the light neutrino masses.
}
In this case the $\nubb$-decay rate
would be determined by the standard contribution,
i.e.~$m_{\beta\beta}^{\rm eff} \simeq |(m_\nu)_{ee}|$.
The measurement of $|(m_\nu)_{ee}|$ can provide information 
about the neutrino mass ordering \cite{Pascoli:2002xq},
as well as about the value of the lightest neutrino mass 
\cite{Pascoli:2001by}. 

As is well known, the uncertainties in the predictions for  $|(m_\nu)_{ee}|$ are largely due to the lack of any experimental information on the values of the Majorana CP-violating phases, the existing poor constraints on the Dirac phase, and the still undetermined value of the lightest neutrino mass and of the neutrino mass ordering. In the scenario considered by us, the values of the Majorana phases are fixed, and thus the ranges of possible values of $|(m_\nu)_{ee}|$ are reduced and there are certain correlations between the predictions in the three cases - CP1, CP2 and CP3.
  
More specifically, in the scenarios with low-energy Dirac CPV and non-real CI matrix discussed in this work, the element $|(m_\nu)_{ee}|$ is restricted to take the following forms:
 \begin{eqnarray}\label{eq:mnubbCP1}
\bullet~~\textbf{CP1)}\quad~~
|(m_\nu)_{ee}| &=& \left|m_1 |U_{e1}|^2 - m_2 |U_{e2}|^2 - m_3 |U_{e3}|^2e^{-2i\delta}\right|,\\
\label{eq:mnubbCP2}
\bullet~~\textbf{CP2)}\quad~~
|(m_\nu)_{ee}| &=& \left|m_1 |U_{e1}|^2 - m_2 |U_{e2}|^2 + m_3 |U_{e3}|^2e^{-2i\delta}\right|,\\
\label{eq:mnubbCP3}
\bullet~~\textbf{CP3)}\quad~~
|(m_\nu)_{ee}| &=& \left|m_1 |U_{e1}|^2 + m_2 |U_{e2}|^2 - m_3 |U_{e3}|^2e^{-2i\delta}\right|.
\end{eqnarray}
In the parametrisation of the PMNS matrix we are using $|U_{e1}|^2 = \cos^2\theta_{12}\cos^2_{13}$, $|U_{e2}|^2 = \sin^2\theta_{12} \cos^2\theta_{13}$ and $|U_{e3}|^2 = \sin^2\theta_{13}$. We recall that for the NO (IO) light neutrino mass spectrum expressions for $m_{2,3}$ ($m_{1,2}$) in terms of the lightest mass $m_1$ ($m_3$) and the measured $\Delta m^2_{21}$ and $\Delta m^2_{31}$ ($\Delta m^2_{32}$) are given in Subsec.~\ref{sec:nu_osc_data}. Using the \texttt{NuFit 6.0} values of the neutrino mixing angles and of the neutrino mass squared differences, quoted in Subsec.~\ref{sec:nu_osc_data}, it is not difficult to show that $|(m_\nu)_{ee}|$ in the CP3 case, $|(m_\nu)_{ee}|_{\rm CP3}$, i) in the NO (IO) case satisfies $|(m_\nu)_{ee}|_{\rm CP3} \gtrsim 10^{-3}$ eV (0.05 eV) for any $m_1$ ($m_3$),\footnote{The general conditions leading to 
$|(m_\nu)_{ee}| \gtrsim 10^{-3}$ eV in the NO case have been  investigated in \cite{Pascoli:2007qh, Penedo:2018kpc,Penedo:2026fpv}.} and ii) for non-negligible values of $m_1$ ($m_3$), $|(m_\nu)_{ee}|_{\rm CP3} > |(m_\nu)_{ee}|_{\rm CP1},|(m_\nu)_{ee}|_{\rm CP2}$. If $m_1 \cong 0$ ($m_3 \cong 0$), we have $|(m_\nu)_{ee}|_{\rm CP3} = |(m_\nu)_{ee}|_{\rm CP2} < |(m_\nu)_{ee}|_{\rm CP1}$ ($|(m_\nu)_{ee}|_{\rm CP1} = |(m_\nu)_{ee}|_{\rm CP2} < |(m_\nu)_{ee}|_{\rm CP3}$). For $m_3 \lesssim 10^{-2}$ eV in the IO case, for example,  $|(m_\nu)_{ee}|_{\rm CP3} \cong \sqrt{\Delta m^2_{23}}\cos^2\theta_{13}$,  $|(m_\nu)_{ee}|_{\rm CP1} = |(m_\nu)_{ee}|_{\rm CP2} \cong \sqrt{\Delta m^2_{23}}\cos^2\theta_{13}\cos2\theta_{12}$, and thus  $|(m_\nu)_{ee}|_{\rm CP1(2)}/|(m_\nu)_{ee}|_{\rm CP3} \cong \cos2\theta_{12} \cong 0.38$, where we have used the \texttt{NuFit 6.0} best fit value of $\theta_{12}$.

\begin{figure}[t!]
\centering
\includegraphics[width=0.45\textwidth]{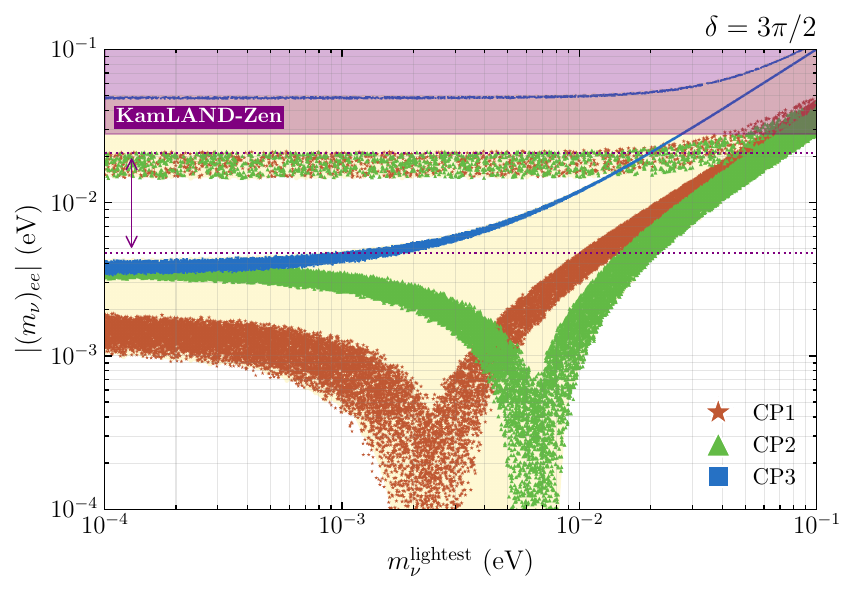}
\includegraphics[width=0.45\textwidth]{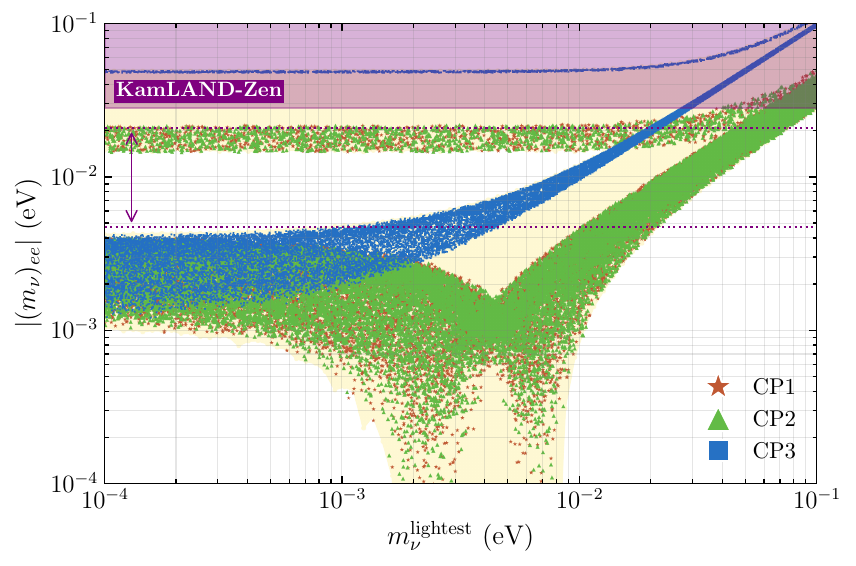}
    \caption{
    The values of $(m_\nu)_{ee}$ versus $\mnulight$ in the three different cases of CP-conserving CI matrix -- red stars for CP1, green triangles for CP2, blue squares for CP3 -- with the Majorana phases fixed accordingly as in Eqs.~(\ref{eq:CP1a21a31}-\ref{eq:CP3a21a31}). 
      The Dirac phase is fixed at $3\pi/2$ on the left, and varied within $[0, 2\pi]$ on the right. The PMNS angles are varied within the $3\sigma$ ranges of the \texttt{NuFit 6.0} global analysis \cite{Esteban:2024eli}; the yellow region is the full one associated to generic CPV; the purple region corresponds to the KamLAND-Zen limit on $m_{\beta \beta}^{\rm eff}$ \cite{KamLAND-Zen:2024eml}, and the dotted horizontal lines mark the expected sensitivity reach of LEGEND-1000 \cite{LEGEND:2021bnm} and CUPID \cite{CUPID:2022jlk}.
      For the ranges of masses ($\Mav \geq 0.1 \,\text{GeV}$), mass splittings ($\Delta \tilde{M}_{21}/\wMav, \Delta \tilde{M}_{31}/\wMav \leq 10^{-4}$) and squared mixings ($\Theta^2\lesssim 10^{-5}$) of interest to this work, we find that $m_{\beta\beta}^{\rm eff}$ is reduced compared to $|(m_\nu)_{ee}|$ due to heavy neutrino contributions by less than $3\%$ for $\Mav\geq 
    1\,\text{GeV}$ (the reduction decreases with $\Mav$), between $3\%-15\%$ for $0.5\leq \Mav/\text{GeV}\leq 1$, and $15\%-70\%$ for $0.1\leq \Mav/\text{GeV}\leq.5$. See the text for further details.}
\label{fig:0nubb}
\end{figure}

We illustrate the preceding discussion in Fig.~\ref{fig:0nubb}, where we depict the regions in the $(m_\nu)_{ee}$-$\mnulight$ plane corresponding to $|(m_\nu)_{ee}|_{\rm CP1}$, $|(m_\nu)_{ee}|_{\rm CP2}$  and
  $|(m_\nu)_{ee}|_{\rm CP3}$ together with the full region of the standard
  contribution to the effective Majorana mass  $(m_\nu)_{ee}$.
We show the regions  for $\delta = 3\pi/2$ in the left panel,
 so that the hierarchies  between  $|(m_\nu)_{ee}|_{\rm CP1}$, $|(m_\nu)_{ee}|_{\rm CP2}$  and $|(m_\nu)_{ee}|_{\rm CP3}$ are manifest, and after scanning over $\delta$ within the range $0< \delta < \pi$ and $\pi < \delta < 2\pi$ on the right panel. In the figure, we also depict in purple the most stringent constraint on the $\nubb$-decay rate coming from KamLAND-Zen, setting $m_{\beta\beta}^{\rm eff}<(0.028-0.122)\,{\rm eV}$ \cite{KamLAND-Zen:2024eml} -- the interval reflects the uncertainty in the nuclear matrix element -- which translates into a bound directly on $(m_\nu)_{ee}$ when the only the standard contribution dominates. We also display the expected sensitivity reach of the upcoming experiments 
LEGEND-1000\cite{LEGEND:2021bnm} and CUPID\cite{CUPID:2022jlk}, $m_{\beta\beta}^{\rm eff}=(0.0047-0.021)\,{\rm eV}$.
     
In the discussed case of negligible contributions from the
  heavy neutrinos, future $\nubb$-decay experiments could potentially
 test the considered CP1, CP2 and CP3 scenarios by measuring $(m_\nu)_{ee}$.
The test would be particularly effective if the light neutrino mass ordering is known, the value of the lightest neutrino mass is determined or constrained to lie in a certain interval, and the value of the Dirac CPV phase is measured.

\paragraph{The case of non-negligible heavy neutrino contributions.} The contribution of the heavy Majorana neutrinos with masses at the GeV
  scale to the  $\nubb$-decay rate can be non-negligible and thus
$m_{\beta\beta}^{\rm eff}$ can differ significantly from $(m_\nu)_{ee}$.
Using the relation in Eq.~\eqref{eq:seesaw6} with the inclusion of the
radiative one-loop correction to $m_\nu$, and keeping terms up to
$\mathcal{O}(R^2)$ order, we rewrite $m_{\beta\beta}^{\rm eff}$ as
%%%%%%%%%%%%%%%%%%%%%%
\begin{equation}
m_{\beta\beta}^{\rm eff} \simeq |(m_\nu)_{ee}|\, \Bigg|1 - \frac{(m^{\rm tree}_\nu)_{ee}}{(m_\nu)_{ee}}\mathcal{F}(\Mav) + \,\sum_{j=1,2,3} \frac{\Theta^2_{e j} M_j}{(m_\nu)_{ee}} (\Mav-M_j)\mathcal{F}'(\Mav)\Bigg|
\end{equation}
%%%%%%%%%%%%%%%%%%
%
where $(m^{\rm tree}_\nu)_{ee} = \sum_{a=1,2,3}U^2_{ea}\,m_a$, and we have Taylor-expanded $\mathcal{F}(M_j)$ around $\Mav$, namely,
  $\mathcal{F}(M_j) \simeq \mathcal{F}(\Mav) + \mathcal{F}'(\Mav) (\Mav-M_j)$,
  $\mathcal{F}'$ being the first derivative with respect to the mass.
  
The first correction proportional to $(m_\nu^{\rm tree})_{ee}/(m_\nu)_{ee}$ usually suppresses the standard contribution, except in fine-tuned configurations for which $|(m_\nu^{\rm tree})_{ee}|$, $|(m_\nu^{\rm loop})_{ee}| \gg |(m_\nu)_{ee}|$, where one can have either a suppression or an enhancement from the first correction, depending on the phase of $(m_\nu^{\rm tree})_{ee} / (m_\nu)_{ee}$.
The second correction can also either suppress or enhance the standard contribution and its size is governed by the combination 
$\Theta^2_{e j}(M_j-\Mav)/\Mav$, with $j = 1,2,3$   \cite{Drewes:2016lqo, deVries:2024rfh} 
(see also \cite{Abada:2018oly}).

When this second correction can be neglected and also $(m_\nu^{\rm tree})_{ee} \simeq (m_\nu)_{ee}$, the effective Majorana mass parameter reduces to  
\begin{equation}\label{eq:mbbeff_reduced}
    m_{\beta \beta}^{\rm eff} \simeq |(m_\nu)_{ee}[1-\mathcal{F}(\Mav)]|,
\end{equation} 
receiving a suppression from the contribution of the heavy neutrinos that only depends on the mass scale associated to them.
We perform an extensive scan ($10^6$ points) of the parameter space,
varying, in particular, $\wMav$ in the range $(0.1 - 100)\,\text{GeV}$, the splittings within $ 10^{-11} \leq |\Delta \tilde{M}_{21}|/\wMav, |\Delta \tilde{M}_{31}|/\wMav \leq 10^{-1}$, where $\Delta \tilde{M}_{jk} \equiv \tilde{M}_j - \tilde{M}_k$, $j,k =1,\,2,\,3$
and $|y|\leq 0.5\,\text{ln}\left(\wTheta_\text{max}^2 \wMav/\sum_a m_a\right)$, $\wTheta^2_{\rm max} = 10^{-5}$, so that $\wTheta^2 \leq \wTheta^2_{\rm max}$.

As long as $|\Delta \tilde{M}_{21}|/\wMav, |\Delta \tilde{M}_{31}|/\wMav \leq 10^{-4}$, we find that both the one-loop corrections and the second contributions from non-vanishing mass splittings can be neglected,
and that the effective Majorana mass parameter is accurately given by Eq.~\eqref{eq:mbbeff_reduced}, except for a small fraction of points at low masses and large splittings. In particular, we find that the Majorana mass parameter agrees with Eq.~\eqref{eq:mbbeff_reduced} up to corrections of at most $\mathcal{O}(2)$ for $0.1 \leq \Mav/\text{GeV} \leq 10$, and at most $\mathcal{O}(10\%)$ for $10 \leq \Mav/\text{GeV} \leq 100$. However, the vast majority of points -- $99.72\%$ in the former mass range and $99.95\%$ in the latter -- exhibit much smaller deviations, below $\mathcal{O}(10\%)$ and $\mathcal{O}(1\%)$, respectively. For larger mass splittings, $\Delta \tilde{M}_{21}/\wMav, \Delta \tilde{M}_{31}/\wMav \gtrsim 10^{-4}$, we instead find that a larger set of points deviate more significantly from the approximation given in Eq.~\eqref{eq:mbbeff_reduced}. In the mass range $0.1 \leq \Mav/\text{GeV} \leq 10$, $14.6\%$ ($0.7\%$) of the scan points exhibit deviations larger than $\mathcal{O}(10\%)$ ($\mathcal{O}(2)$), respectively. Correspondingly, in the range $10 \leq \Mav/\text{GeV} \leq 100$, the fraction decrease to $3.2\%$ ($0.2\%$).

Overall, our scan shows that sizeable corrections to Eq.~\eqref{eq:mbbeff_reduced} are confined to a relatively small fraction of the sampled points, and arise predominantly for masses $0.1 \lesssim \Mav/\text{GeV} \lesssim 10$ and mass splittings $\Delta \tilde{M}_{21}/\wMav, \Delta \tilde{M}_{31}/\wMav \gtrsim 10^{-4}$.

When $\Delta \tilde{M}_{21}/\wMav, \Delta \tilde{M}_{31}/\wMav \gtrsim 10^{-4}$, the large contributions (although these are relatively rare in our scan) allow the scan to populate the entire phenomenologically allowed range of $m_{\beta\beta}^{\rm eff}$, from values close to zero up to current bounds. In particular, $m_{\beta\beta}^{\rm eff}$ can span the full range covered by the $|(m_\nu)_{ee}|_{\rm CP1}$, $|(m_\nu)_{ee}|_{\rm CP2}$, and $|(m_\nu)_{ee}|_{\rm CP3}$ regions, as well as the standard contribution region. This results in substantial overlap, making an experimental falsification of the considered scenarios very challenging.  Most importantly, heavy neutrino contributions can shift points that would normally lie in the NO region, in the standard contribution, into the IO region, and vice versa.

To keep these corrections under control and maintain the possibility of testing the considered scenarios at $\nubb$-decay searches, in the following leptogenesis analysis we focus on splittings $\Delta \tilde{M}_{21}/\wMav$, $\Delta \tilde{M}_{31}/\wMav \leq 10^{-4}$, while $\Theta_{ej}^2 \lesssim 10^{-5}$ will be basically required by current experimental bounds, so that the effective mass is to a good approximation given by Eq.~\eqref{eq:mbbeff_reduced}.\footnote{It is not strictly necessary for both splittings to be small to prevent such corrections; one of the three neutrinos can have a larger mass provided that the corresponding electron mixing is also sufficiently small. However, to facilitate the numerical scan, we choose to work with both splittings small.} We extrapolate the values of $\mathcal{F}(M)$ as obtained in \cite{deVries:2024rfh}, finding $\mathcal{F}(M) \approx 0.70, 0.15$, $0.03$  for $M/\text{GeV} = 0.1, 0.5, 1$, respectively. Consequently, in the mass range $0.1\lesssim \Mav/\text{GeV} \lesssim0.5$ the reduction of the effective Majorana mass parameter compared to $(m_\nu)_{ee}$ vary between $70\%$ and $15\%$, while for $0.5\lesssim \Mav/\text{GeV} \lesssim 1$ we can approximate $m_{\beta \beta}^{\rm eff} \simeq |(m_\nu)_{ee}|$ up to corrections that vary from $15\%$ to $3\%$. For larger masses the reduction is much smaller since $\mathcal{F}(M) \approx 4.4\times 10^{-4}(10\,\text{GeV}/M)^2$, and thus the approximation $m_{\beta \beta}^{\rm eff} \simeq |(m_\nu)_{ee}|$ holds even more precisely. The results shown in Fig.~\ref{fig:0nubb} can therefore be reinterpreted by a simple $\Mav$-dependent rescaling based on the corrections mentioned above.

%%%%%%%%%%%%%%%%%%

\section{Low-scale leptogenesis with Dirac-phase CP-violation}\label{sec:LG}
We study here the viability of low-scale leptogenesis in reproducing the observed BAU in the heavy neutrino mass range of interest, under the assumption of Dirac-phase low-energy CPV. Solving numerically the state-of-the-art evolution equations, we perform a scan of the parameter space and analyse the dependence of successful leptogenesis on the Dirac phase $\delta$.

\subsection{The density matrix equations for low-scale leptogenesis}
\label{sec:DMEs}
%%%%%%%%%%%%%%%%%%%%%%%%%%%%
%

The momentum-averaged \textit{density matrix equations} (DMEs) relevant to the scenario of low-scale leptogenesis are given by \cite{Abada:2018oly, Granelli:2026goh}: %\cite{Drewes:2016gmt,Hernandez:2016kel,Ghiglieri:2017gjz, Eijima:2017anv, Antusch:2017pkq} for derivations; 
%(we write them in a fashion similar to that of \cite{Hernandez:2022ivz}, but neglecting non-linear terms):
%
\allowdisplaybreaks
\begin{subequations}
\begin{align}
\label{eq:DME_N}
%%%%%    rN    %%%%%
\frac{dr_N}{dx} = &-\frac{i}{Hx}\left[\langle \mathcal{H}\rangle,r_N\right]-\frac{r_N}{N_{N_1}^{\rm eq}}\frac{dN_{N}^{\rm eq}}{dx}- \frac{\langle\gamma_{\LNC}^{(0)}\rangle}{2Hx}\left\{Y^\dagger Y,r_N-\mathds{1}\right\}
+ \frac{\langle \gamma_{\LNC}^{(1)}\rangle}{Hx} Y^\dagger \mu Y +\nonumber \\&
- \frac{\langle S_{\LNV}^{(0)}\rangle}{2T^2Hx}\left\{ MY^{\rm T}Y^*M,r_N-\mathds{1}\right\}
- \frac{\langle S_{\LNV}^{(1)}\rangle}{T^2Hx}\,M Y^{\rm T}\mu Y^* M,\\
%%%%%    rNbar   %%%%%
\frac{dr_{\overline{N}}}{dx} = &-\frac{i}{Hx}\left[\langle \mathcal{H^*}\rangle,r_{\overline{N}}\right]-\frac{r_{\overline{N}}}{N_{N}^{\rm eq}}\frac{dN_{N}^{\rm eq}}{dx}- \frac{\langle\gamma_{\LNC}^{(0)}\rangle}{2Hx}\left\{Y^{\rm T} Y^*,r_{\overline{N}}-\mathds{1}\right\} 
- \frac{\langle \gamma_{\LNC}^{(1)}\rangle}{Hx} Y^{\rm T} \mu Y^* +\nonumber \\&
- \frac{\langle S_{\LNV}^{(0)}\rangle}{2T^2Hx}\left\{ MY^\dagger Y M,r_N-\mathds{1}\right\}
+ \frac{\langle S_{\LNV}^{(1)}\rangle}{T^2Hx}\,M Y^\dagger \mu Y M,\\
%%%%CHEMICAL POTENTIALS%%%%%
\kappa \frac{d\mu_{\Delta_\alpha}}{dx} = &-\frac{\langle\gamma_{\LNC}^{(0)}\rangle}{2Hx}(Yr_NY^\dagger-Y^* r_{\overline{N}}Y^{\rm T})_{\alpha\alpha} + \frac{\langle \gamma_{\LNC}^{(1)}\rangle}{Hx}(YY^\dagger)_{\alpha\alpha}\mu_\alpha
+\nonumber\\
&{}+ \frac{\langle S_{\LNV}^{(0)}\rangle}{2T^2Hx}(Y^*Mr_NMY^{\rm T}-YMr_{\overline{N}}MY^\dagger)_{\alpha\alpha}+ \frac{\langle S_{\LNV}^{(1)}\rangle}{T^2Hx}(YM^2Y^\dagger)_{\alpha\alpha}\mu_\alpha\,,
\label{eq:DME_mu}
 \end{align}
 \end{subequations}
 where $\alpha = e,\,\mu,\,\tau$. The DMEs written above govern the evolution of the chemical potential $\mu_{\Delta_\alpha}$ associated to the charge $\Delta_\alpha \equiv B/3-L_\alpha$, and of the comoving number density matrix of the heavy Majorana neutrinos with positive (negative) helicity $\rho_N$ ($\rho_{\overline{N}}$), having defined the ratio $r_N=\rho_N/N_{N}^\text{eq}$ ($r_{\overline{N}}=\rho_{\overline{N}}/N_{N}^\text{eq}$), with $N_{N}^\text{eq}$ the number of heavy Majorana neutrinos, per generation, in a comoving volume when in thermal equilibrium. The variable $x$ is defined as $x\equiv T_{\sph}/T$, with $T_{\sph}\simeq 131.7\,\text{GeV}$ the temperature at which sphalerons depart from equilibrium \cite{DOnofrio:2014rug}, while $H$ is the Hubble rate. The matrices $M=\text{diag}(\tilde{M}_1, \tilde{M}_2, \tilde{M}_3)$ and $\mu=\text{diag}(\mu_e,\, \mu_\mu,\,\mu_\tau)$ are the mass matrix of the heavy Majorana neutrinos in the mass basis and that of the lepton doublet chemical potentials $\mu_\alpha$, respectively. The latter is given by $\mu_\alpha = -2\sum_{\beta =e,\,\mu,\,\tau}\chi_{\alpha\beta}\mu_{\Delta_\beta}$, where the susceptibility matrix $\chi$ accounts for the effects of spectator processes. The susceptibility matrix, neglecting the dependence on the temperature, is given by \cite{Ghiglieri:2017gjz} (see also \cite{Eijima:2017cxr})
 \begin{equation}
    \chi= \frac{1}{711}\begin{pmatrix} 257&20&20\\
    20&257&20\\
    20&20&257
    \end{pmatrix}\,.
\end{equation}
The thermally averaged Hamiltonian $\left\langle\mathcal{H}\right\rangle$ is given by \cite{Ghiglieri:2017gjz, Antusch:2017pkq} (see also \cite{Hernandez:2016kel, Abada:2018oly})
\begin{equation}
\frac{\langle \mathcal{H}\rangle}{T} \simeq \left\langle\frac{1}{y_0}\right\rangle \frac{\Delta M_\text{diag}^2}{2T^2}\,
+ (Y^\dagger Y)\langle h^{\LNC}\rangle 
+ (Y^T Y^*)\langle h^{\LNV}\rangle
\label{eq:thermH}
\end{equation}
where we have included both lepton number conserving (LNC) and lepton number violating (LNV) thermal contributions and $\langle \cdot \rangle$ denotes thermal average \footnote{The thermal averages are defined as  $\langle X \rangle \equiv (1/n_N^{\text{eq}}) \int dk \,k^2 f_F X/(2\pi^2)$, with $k \equiv (E_N^2 + \wMav^2)^{1/2}$, $n_N^{\text{eq}}\equiv \int dk \,k^2 f_F/(2\pi^2)$, $f_F = 1/(e^{E_N/T} + 1)$ is the Fermi-Dirac distribution function in momentum space and $X = 1/y_0, \,h^{\LNC},\, h^{\LNV},\, \gamma_{\LNC}^{(0)},\, \gamma_{\LNC}^{(1)}, S_{\LNV}^{(0)}$ or $S_{\LNV}^{(1)}$.}; $y_0=E_N/T$, $E_N$ being the heavy neutrino average energy; $\Delta M_\text{diag}^2 \equiv \text{diag}(0,\tilde{M}_2^2-\tilde{M}_1^2, \tilde{M}_3^2-\tilde{M}_1^2)$. We have subtracted an overall $\tilde{M}_1^2$ factor from the zero-temperature contribution without affecting the dynamics. 
In the relativistic regime of $z \ll 1$, with $z\equiv \wMav x/T_{\sph}$, $\wMav = \sum_{j=1}^3 \tilde{M}_j/3$, the expressions for $\langle 1/y_0\rangle$ and $\langle h^{\LNC}\rangle$ are given, to a very good approximation, by
\begin{eqnarray}
    \left\langle \frac{1}{y_0}\right\rangle &\simeq& \frac{\pi^2}{18\zeta(3)}, \\
    \left\langle h^{\LNC}\right\rangle &\simeq& \frac{1}{8} \left\langle \frac{1}{y_0}\right\rangle,
\end{eqnarray}
while $\langle h^{\LNV}\rangle \simeq 2.5\times 10^{-3} z^2 [3.50 - 0.47\log{(z^2)} + 3.47\log^2{(z^2)}]$ \cite{Antusch:2017pkq}, and is thus suppressed compared to the LNC contribution. Instead, in the non-relativistic regime of temperatures, when $z\gg 1$, $\langle 1/y_0\rangle = K_1(z)/[zK_2(z)]\simeq 8/(15+8z)$ \cite{Buchmuller2005305}, with $K_n(z)$, $n = 1\,,2\,,\,...$, being the modified $n^\text{th}$ Bessel function of the second kind, and the LNC and LNV contributions approach the same value of $\langle h^{\LNV}\rangle\simeq \langle h^{\LNV}\rangle \simeq (1/16)\langle 1/y_0\rangle$. The numerical code that we use to solve the DMEs includes the full mass-temperature dependence of these quantities and corrections from indirect contributions in the broken phase, increasing the overall accuracy of our numerical results.

The quantities $\gamma_{\LNC}^{(0)}$ and $S_{\LNV}^{(0)}$ are the heavy Majorana neutrino production rates respectively for the LNC and LNV processes involving the heavy Majorana neutrinos. The wash-out rates  $\gamma_{\LNC}^{(1)}$ and $S_{\LNV}^{(1)}$ are given in terms of the production ones according to $\gamma_{\LNC}^{(1)} = (1-f_F)\gamma_{\LNC}^{(0)} $ and $ S_{\LNV}^{(1)}= (1-f_F) S_{\LNV}^{(0)}$. The dependence on the momentum and temperature of the LNC and LNV production rates was computed numerically in the relativistic limit \cite{Ghiglieri:2017csp} (see also \cite{Ghiglieri:2017gjz}) -- and tabulated in \href{http://www.laine.itp.unibe.ch/leptogenesis/}{\ttfamily{this site}}. There, the LNC and LNV production rates are denoted with $Q_+$ and $Q_-$, respectively, and can be mapped into our rates according to $\gamma_{\LNC}^{(0)} = (1 + y/y_0)Q_+/2$ and $z^2 S^{(0)}_{\LNV} = 2y^2(1 - y/y_0) Q_-$, after including non-relativistic correction to the on-shell relation and with $y = \sqrt{y_0^2-z^2}$ \cite{Hernandez:2022ivz}. To further increase the accuracy of our numerical results in the non-relativistic regime and allow to extend the study beyond the $100\,\text{GeV}$ heavy neutrino mass scale, we also add the contributions from the heavy neutrino decay in the non-relativistic limit \cite{Klaric:2021cpi}, included the indirect contributions in the broken phase. 

Finally, we adopt a normalisation of the comoving volume such that it contains a single photon when $z\ll 1$ and $N^\text{eq}_{N}(z\ll 1) = 3/8$. We also consider the analytical approximation $N^\text{eq}_{N}(z) \simeq (3/16) z^2 K_2(z)$.
Consequently, the overall constant factor $\kappa$ appearing in Eq.~\eqref{eq:DME_mu} reads \begin{equation}
    \kappa \simeq \frac{4\pi^2}{9\zeta(3)z^2 K_2(z)},
    \end{equation}
 while 
 \begin{equation}
 \frac{1}{N_N^\text{eq}} \frac{dN_{N}^\text{eq}}{dx} \simeq -\frac{\wMav}{T_{\sph}}\frac{K_1(z)}{K_2(z)}.
 \end{equation}

We make use of the Python package \texttt{ULYSSES} \cite{Granelli:2020pim, Granelli:2023vcm, Granelli:2026goh} to solve numerically the DMEs described in this section (see \cite{Granelli:2026goh} for more details on the rates and Hamiltonian corrections). The code, in particular, after solving numerically the DMEs, returns the baryon-to-photon ratio as
%%%%%%%%%%%%%%%%%%%%%%%%%%%%%%%%%
\begin{equation}
    \eta_B = \frac{c_s}{f}\frac{\pi^2}{6 \zeta(3)}
\mu_{B-L}\simeq 0.018\,(\mu_{\Delta_e} + \mu_{\Delta_\mu} +  \mu_{\Delta_\tau})\,,
\label{eq:etaBl}
\end{equation}
%%%%%%%%%%%%%%%%%%%%%%%%%%%%%%%%%%%
%
where $c_s = 28/79$ is the sphaleron conversion coefficient and 
the factor $f = 1/27$ is the common dilution factor (see, e.g., \cite{Buchmuller2005305}). The final results for the baryon-to-photon ratio are compared against the best-fit value $\eta_B^{\rm obs} \simeq 6.1 \times 10^{-10}$ as it stems from measurements of the anisotropies of the cosmic microwave background \cite{Planck2018} and the abundances of light primordial elements \cite{Cooke_2018}. For our numerical analysis we set the initial time of leptogenesis at $x_{\rm in} = 10^{-6}$  and consider the situation for which the initial abundances of heavy neutrinos and the initial flavour asymmetries are vanishing, i.e.~$(\rho_N)_{jk}(x_{\rm in}) = (\rho_{\overline{N}})_{jk}(x_{\rm in})= \mu_{\Delta_\alpha}(x_{\rm in}) = 0$, $j,k=1,\,2,\,3$ and $\alpha=e,\,\mu,\,\tau$.

\subsection{General features of CP-violation}
While the CP-violating combinations used in low-scale leptogenesis do not always transparently relate to the final lepton asymmetries,
they remain useful in illuminating the sources of CPV and the general differences compared to the scenario with two RH neutrinos.
As an illustrative example we will consider CP1, and the corresponding lepton number violating (LNV) and lepton flavour violating (LFV) CPV combinations (see e.g.~\cite{Drewes:2022kap}) in the limit of $y \gg 1$,
which is most directly relevant for direct searches.
In the limit of $\delta$-only CPV, we find that the LFV combination gives:
\begin{align}
\label{eq:cpv_lfv}
    C_\mathrm{LFV}^\alpha =&\,
    i \left(Y [\tilde{M}_N^{d\,2} 
    , Y^\dagger Y] Y^\dagger \right)_{\alpha \alpha}\,,\\\notag
    \approx \,& e^{3 |y|} \frac{(\tilde{M}_2-\tilde{M}_1) \wMav^3}{2 v^4} s_N c_N (m_1 + m_2 s^2_\nu + m_3 c^2_\nu)\times\\\notag
    & \times[ \sqrt{m_1 m_2} \,c_\nu \mathrm{Im}\, (U_{\alpha 1}U_{\alpha 2}^*)
    -\sqrt{m_2 m_3} \mathrm{Im}\, (U_{\alpha 2}U_{\alpha 3}^*)
    - \sqrt{m_1 m_3} \,s_\nu \mathrm{Im}\, (U_{\alpha 1}U_{\alpha 3}^*)]
    \\\notag
    \propto \, & e^{3 |y|} (\tilde{M}_2-\tilde{M}_1) s_N c_N \sin \delta,
\end{align}
whereas the LNV combination is given by
\begin{align}
\label{eq:cpv_lnv}
    C_\mathrm{LNV}^\alpha =&\,
    i \left(Y^* [\tilde{M}_N^{d\,2} 
    , Y^\dagger Y] Y^T \right)_{\alpha \alpha}\,,\\\notag
    \approx \,& e^{3 |y|} \frac{(\tilde{M}_2-\tilde{M}_1)(\tilde{M}_3-\tilde{M}_1)(\tilde{M}_3-\tilde{M}_2) 3 \wMav}{8 v^4} s_N c_N (m_1 + m_2 s^2_\nu + m_3 c^2_\nu)\times\\\notag
    & \times[ \sqrt{m_1 m_2} \, c_\nu \mathrm{Im}\, (U_{\alpha 1}U_{\alpha 2}^*)
    -\sqrt{m_2 m_3} \mathrm{Im}\, (U_{\alpha 2}U_{\alpha 3}^*)
    - \sqrt{m_1 m_3} \,s_\nu \mathrm{Im}\, (U_{\alpha 1}U_{\alpha 3}^*)]
    \\\notag
    \propto\,& e^{3 |y|} (\tilde{M}_2-\tilde{M}_1)(\tilde{M}_3-\tilde{M}_1)(\tilde{M}_3-\tilde{M}_2) s_N c_N \sin \delta\,,
\end{align}
where we expanded to leading order in the mass differences and in $e^{|y|}$.
These combinations show a number of interesting features that are absent in the Dirac-phase CPV two RH neutrino scenario, and the generic three RH neutrino case:
\begin{itemize}
    \item The leading contribution to the CP-asymmetry already appears at $\mathcal{O}(e^{3|y|})$,
    whereas it only appears at $\mathcal{O}(e^{2|y|})$ in the Dirac phase scenario with two RH neutrinos.
    This is directly reflected in the overall $\sin (2x_N)$ dependence of the leading order term, which vanishes in the two RH neutrino limit.
    This explicit $\sin (2x_N)$ dependence is absent from the higher order terms.
    \item The leading LFV combination vanishes if $\tilde{M}_2 = \tilde{M}_1$ while it does not depend on the other splittings, unlike the three RH neutrino scenario without Dirac phase CPV, where having at least one non-zero splitting between any pair of heavy neutrinos is sufficient to produce a $\mathcal{O}(e^{3|y|})$ CPV combination.
    \item The leading order LNV term vanishes unless all three RH neutrino masses are different.
    \item The LFV and LNV terms do not cancel,
    thereby avoiding the suppression of the total CP-asymmetry in the non-relativistic limit of the scenario with two RH neutrinos~\cite{Granelli:2023tcj}.
\end{itemize}

While these general features illustrate some of the differences compared to Dirac-phase leptogenesis with only two RH neutrinos, they cannot substitute for a full numerical exploration of the parameter space, which we present in the following sections.

\subsection{Scans of the leptogenesis parameter space}\label{sec:LG_scans}
We perform a numerical scan of the parameter space to find the regions in which leptogenesis with Dirac-phase CPV is viable. For illustrative purposes, we restrict ourselves to specific ranges of parameters and benchmark configurations. We focus in particular on the CP1 scenario with either NH, IH or a QD light neutrino mass spectrum with $\mnulight = 0.1\,\text{eV}$. We show the results obtained for the other CP-conserving scenarios in Appendix \ref{app:otherscans}.

For the scans, we calculate the BAU for 
points randomly distributed over the ranges
\begin{equation}
\begin{split}
0.1 \leq \tilde{M}_1 / \text{GeV} \leq 100,\quad
10^{-11} \leq |\Delta \tilde{M}_{21}|/\tilde{M}_1 \leq 10^{-4},\quad
10^{-11} \leq |\Delta \tilde{M}_{32}|/\tilde{M}_1\leq 10^{-4},\\
0 < |y| \leq 12,\quad
0 \leq \delta \leq 2\pi,\quad
0 \leq x_\nu\leq 2\pi,\quad
0 \leq  x_N \leq 2\pi.
\end{split}
\end{equation}
The mass splittings and $y$ are allowed to take either sign. The PMNS mixing angles and light neutrino squared mass differences are fixed to their best-fit values from the \texttt{NuFit~6.0} global analysis \cite{Esteban:2024eli}. The considered range of $y$ values implies that $5.6\times 10^{-11} \lesssim \wTheta^2\wMav/\text{GeV} \lesssim 0.1$.\footnote{Note that, while one small mass splitting between two heavy neutrinos forming a pseduo-Dirac pair can be associated with an approximate $B-L$ symmetry, the smallness of the remaining splitting is not necessarily symmetry-protected. Nevertheless, we restrict both independent mass splittings to be below $10^{-4}$. This phenomenological choice allows us to explore a broader range of $x_\nu$ and $x_N$ while keeping radiative corrections under control and sub-leading heavy-neutrino contributions to $\nubb$-decay rate, as discussed in Sec.~\ref{sec:nubb}. Moreover, since the pseudo-Dirac pair can be either two of the heavy neutrinos, considering both independent mass splittings to be small avoids the need to identify the pseudo-Dirac pair point by point and impose the small-splitting condition only on that pair, while allowing the remaining splitting to be large.}

A variety of experimental and cosmological constraints have to be taken into account when exploring the parameter space of type-I seesaw models in the mass range $(0.1-100)\,\text{GeV}$. These include: constraints from experiments on HNL production via meson decays (PS191 \cite{Bernardi:1985ny, Bernardi:1987ek}, BEBC \cite{Barouki:2022bkt}, PIENU \cite{PIENU:2017wbj}, E949 \cite{E949:2014gsn}, NA62  \cite{NA62:2020mcv, NA62:2021bji}, T2K \cite{T2K:2019jwa}, NuTeV \cite{NuTeV:1999kej}, MicroBooNE \cite{MicroBooNE:2022ctm, MicroBooNE:2023eef} CHARM \cite{CHARM:1985nku, Boiarska:2021yho}, searches at KEK \cite{Hayano:1982wu}), $\tau$-decays (BELLE \cite{Belle:2013ytx, Belle:2022tfo, Belle:2024wyk}), and at colliders (DELPHI \cite{DELPHI:1996qcc}, CMS \cite{CMS:2022fut, CMS:2023jqi, CMS:2024xdq}, ATLAS \cite{ATLAS:2019kpx, ATLAS:2022atq, Tastet:2021vwp, ATLAS:2025uah}).
Such current constraints (also the future sensitivities mentioned later) are often calculated assuming either $\Theta^2 \equiv \Theta^2_e$, i.e.~$\Theta^2_e : \Theta^2_\mu : \Theta^2_\tau = 1:0:0$ ($e$-channel), $\Theta^2 \equiv \Theta^2_\mu$, i.e.~$\Theta^2_e : \Theta^2_\mu : \Theta^2_\tau = 0:1:0$ ($\mu$-channel), or $\Theta^2 \equiv \Theta^2_\tau$, i.e.~$\Theta^2_e : \Theta^2_\mu : \Theta^2_\tau = 0:0:1$ ($\tau$-channel), and in a simplified framework with a single HNL. In more realistic contexts, as the one considered in this manuscript, these limits and sensitivities should be properly recast (see e.g., \cite{Abada:2022wvh}). We also include bounds from flavour and electroweak precision observables at 95$\%$ C.L.~as obtained in the global fit analysis of \cite{Blennow:2023mqx}. The analysis in \cite{Blennow:2023mqx} was carried out in a type-I seesaw with three heavy neutrinos, as in our scenario, but the lightest neutrino mass was free to vary and the Dirac phase was not the unique source of CPV. The assumption of Dirac-phase CPV and the choices of $\mnulight$ we are considering may imply different bounds, because the corresponding regions of the flavour-mixing triangle are restricted with respect to the generic one. We do not attempt a detailed recasting in our framework of all these bounds (neither of the future sensitivities), but rather consider the constraints as they are computed in the referenced analyses and apply them to the individual squared mixings $|\Theta_{\alpha j}|^2$, $\alpha = e,\,\mu,\,\tau$ and $j = 1,\,2,\,3$. Consequently, these bounds are only meant to be indicative in our context.

\begin{figure}[t!]
    \centering
\includegraphics[width=0.45\textwidth]{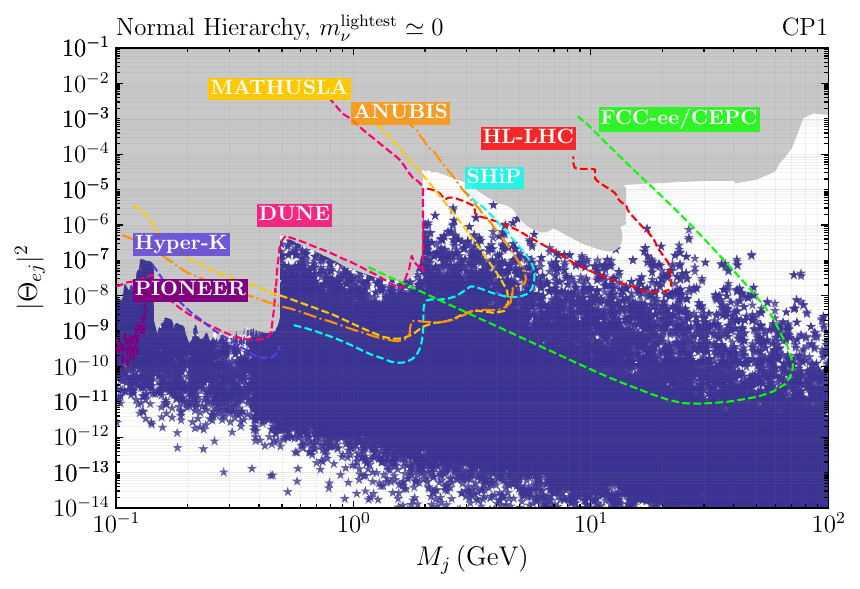}
\includegraphics[width=0.45\textwidth]{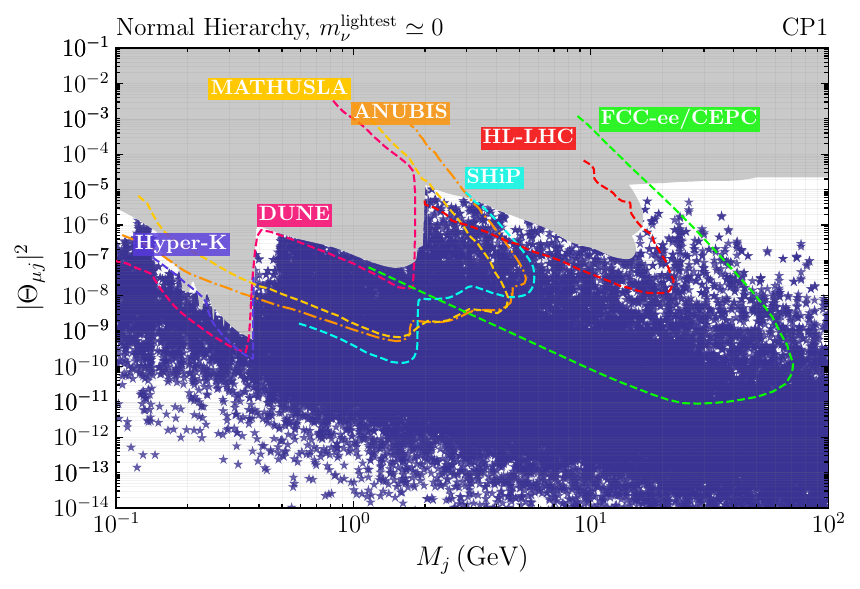}
\includegraphics[width=0.45\textwidth]{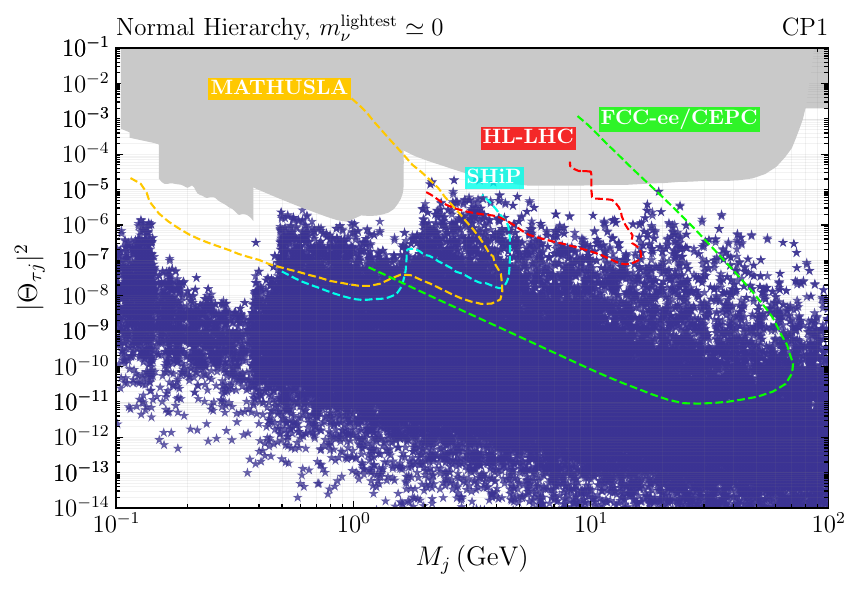}
\hspace{1.2em}
\includegraphics[width=0.4\textwidth]{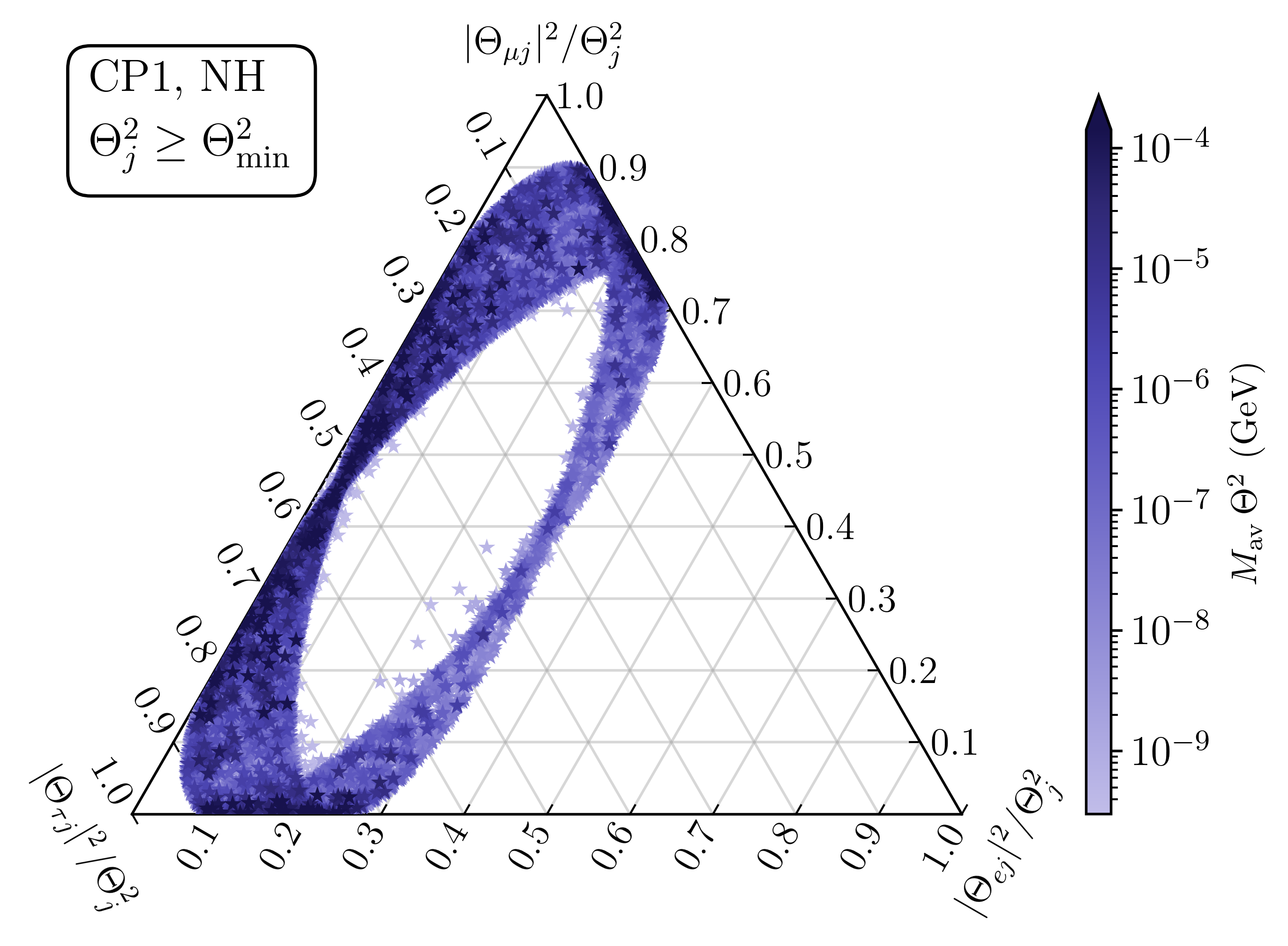}
    \caption{Scan of the parameter space where leptogenesis is successful in the case of low-energy CPV solely from the Dirac phase $\delta$, for the benchmark case CP1 with NH, i.e.~NO and $\mnulight \simeq 0$ (see Figs.~\ref{fig:CP1s_scan_IH_QD} for the scans in the IH and QD cases). We show the scan in the $|\Theta_{\alpha j}|^2$ versus $M_j$ plane, for the flavour $\alpha = e,\,\mu,\,\tau$ in the panel on the top-left, top-right and bottom-left, respectively. In the bottom-right panel the scan is shown in the ternary plane $|\Theta_{ej}|^2/\Theta^2_j$-$|\Theta_{\mu j}|^2/\Theta^2_j$-$|\Theta_{\tau j}|^2/\Theta^2_j$ with the brightness of the colour indicating $\Mav \Theta^2$. Only points yielding a BAU larger (in modulus) than the observed value and within the existing constraints are shown. Each scan point is represented by three stars in the figure, corresponding either to $N_1$, $N_2$ and $N_3$. The parameters are randomly varied in the following ranges: $0.1\leq \tilde{M}_1/\text{GeV} \leq 100$, $10^{-11} \leq|\Delta \tilde{M}_{21}|/\tilde{M}_1,\, |\Delta \tilde{M}_{32}|/\tilde{M}_1 \leq 10^{-4}$, $0 < |y| \leq 12$, $0 \leq \delta \leq 2\pi$ and $0 \leq x_\nu,\, x_N \leq 2\pi$, while the remaining oscillation parameters are fixed to their best-fit values from the \texttt{NuFit~6.0} global analysis \cite{Esteban:2024eli}. 
    Details and references for the displayed constraints and sensitivities are given in the main text.}
\label{fig:CP1s_scan_NH}
\end{figure}

\begin{figure}
    \centering
\includegraphics[width=0.45\textwidth]{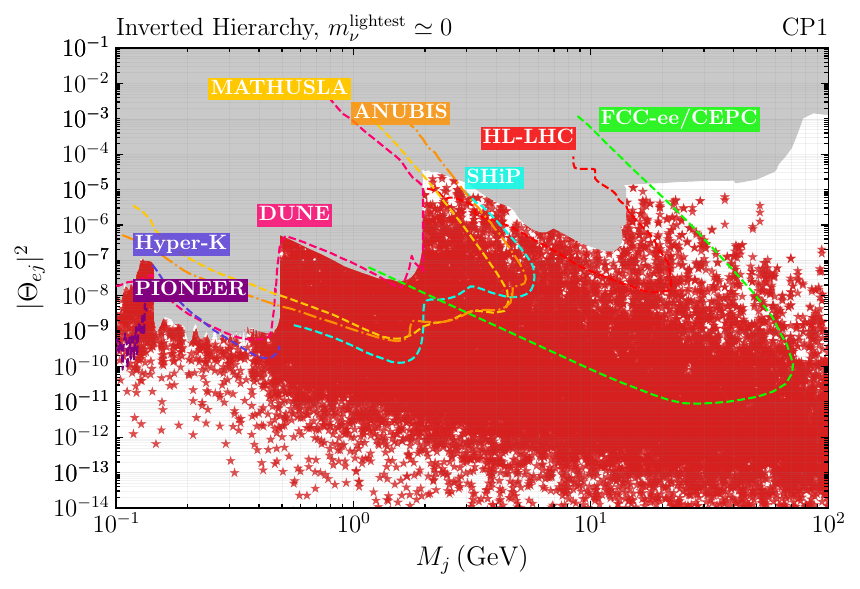}
\includegraphics[width=0.45\textwidth]{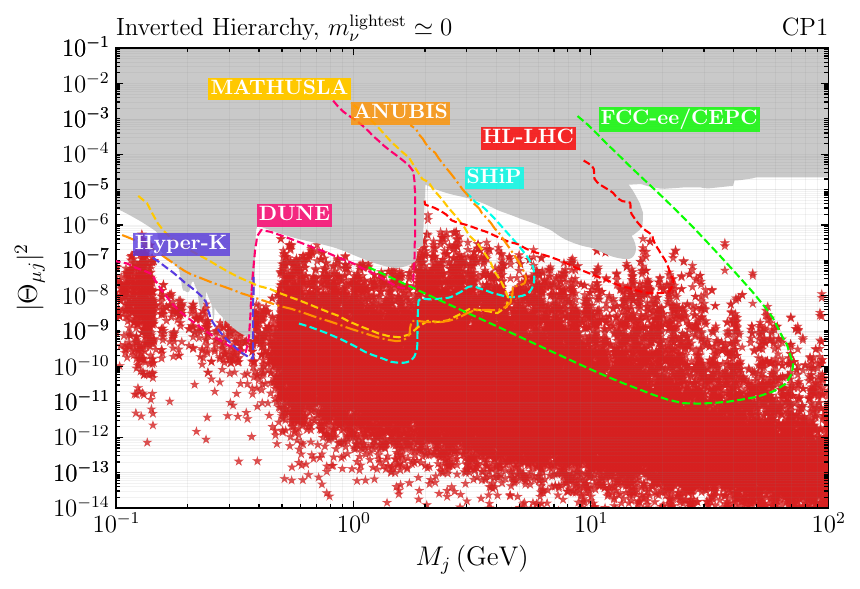}
\includegraphics[width=0.45\textwidth]{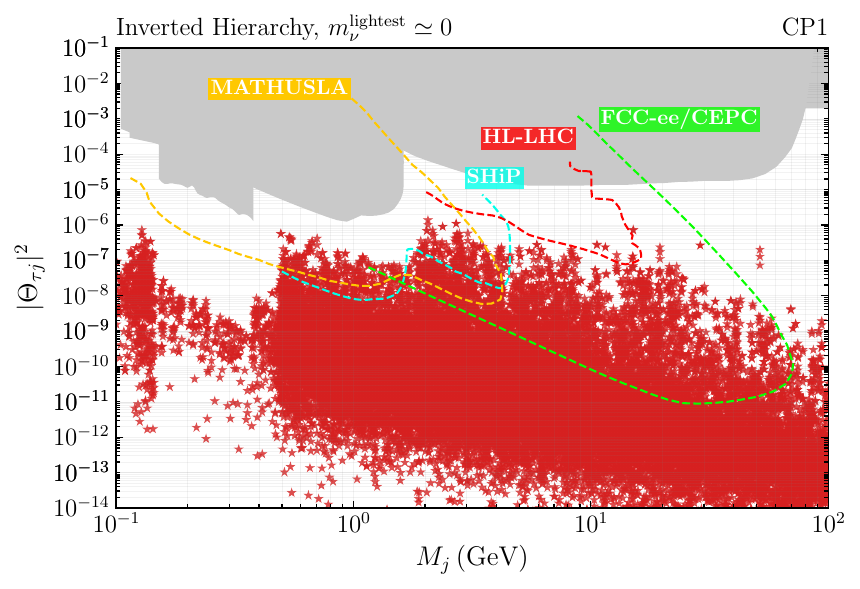}
\hspace{3.5em}
\includegraphics[width=0.35\textwidth]{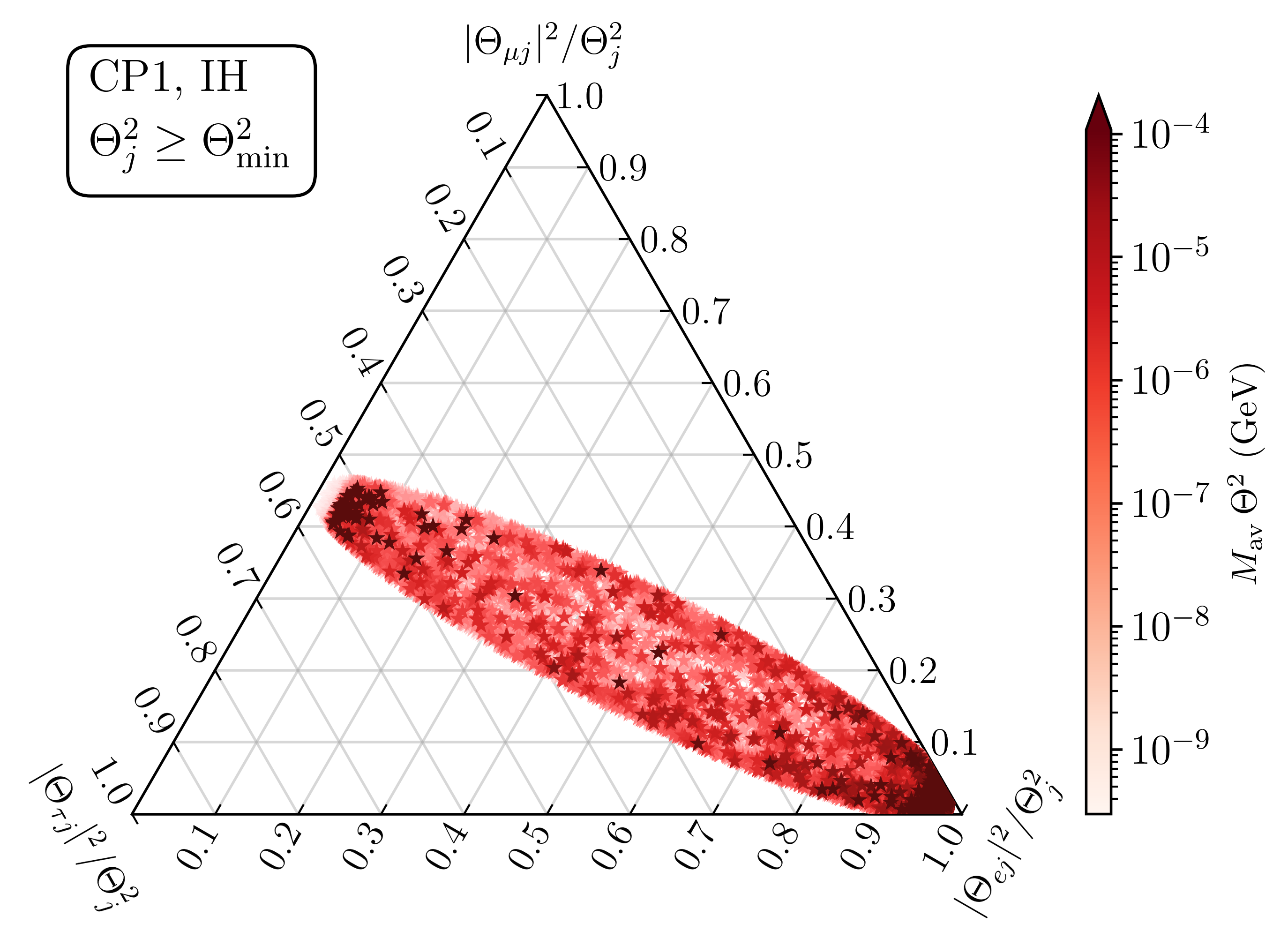}
\includegraphics[width=0.45\textwidth]{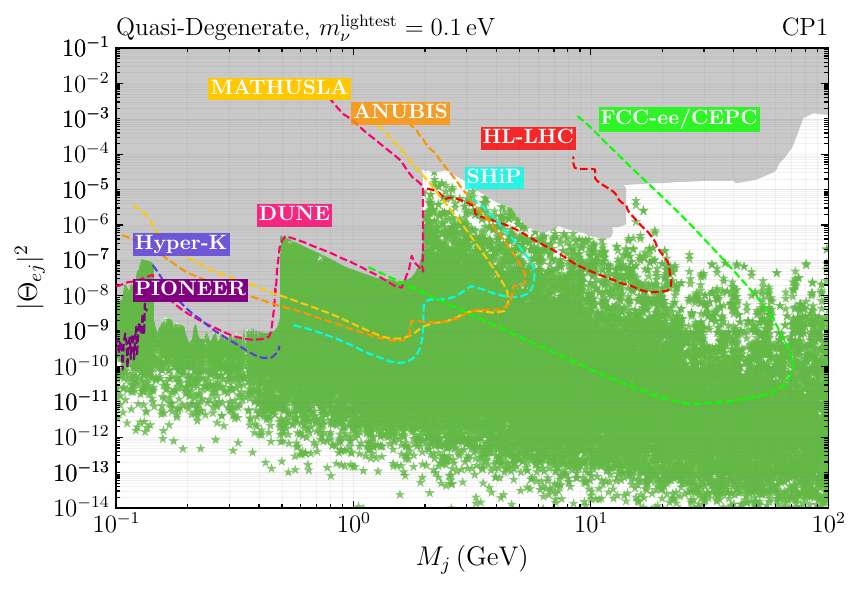}
\includegraphics[width=0.45\textwidth]{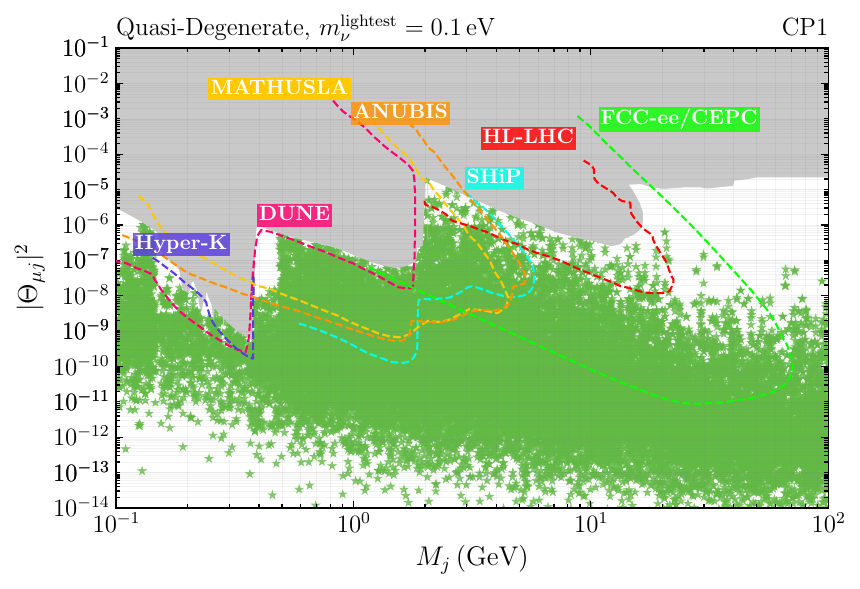}
\includegraphics[width=0.45\textwidth]{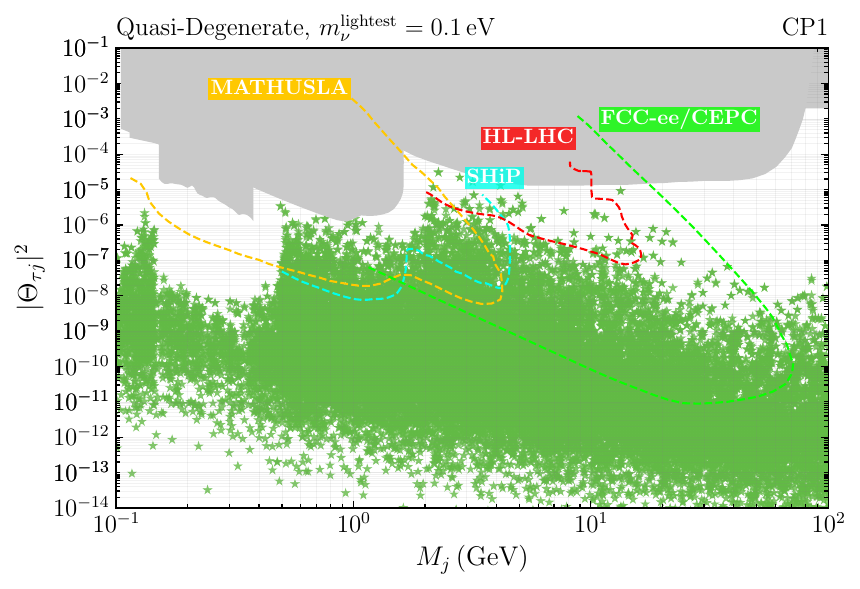}
\hspace{3.5em}
\includegraphics[width=0.35\textwidth]{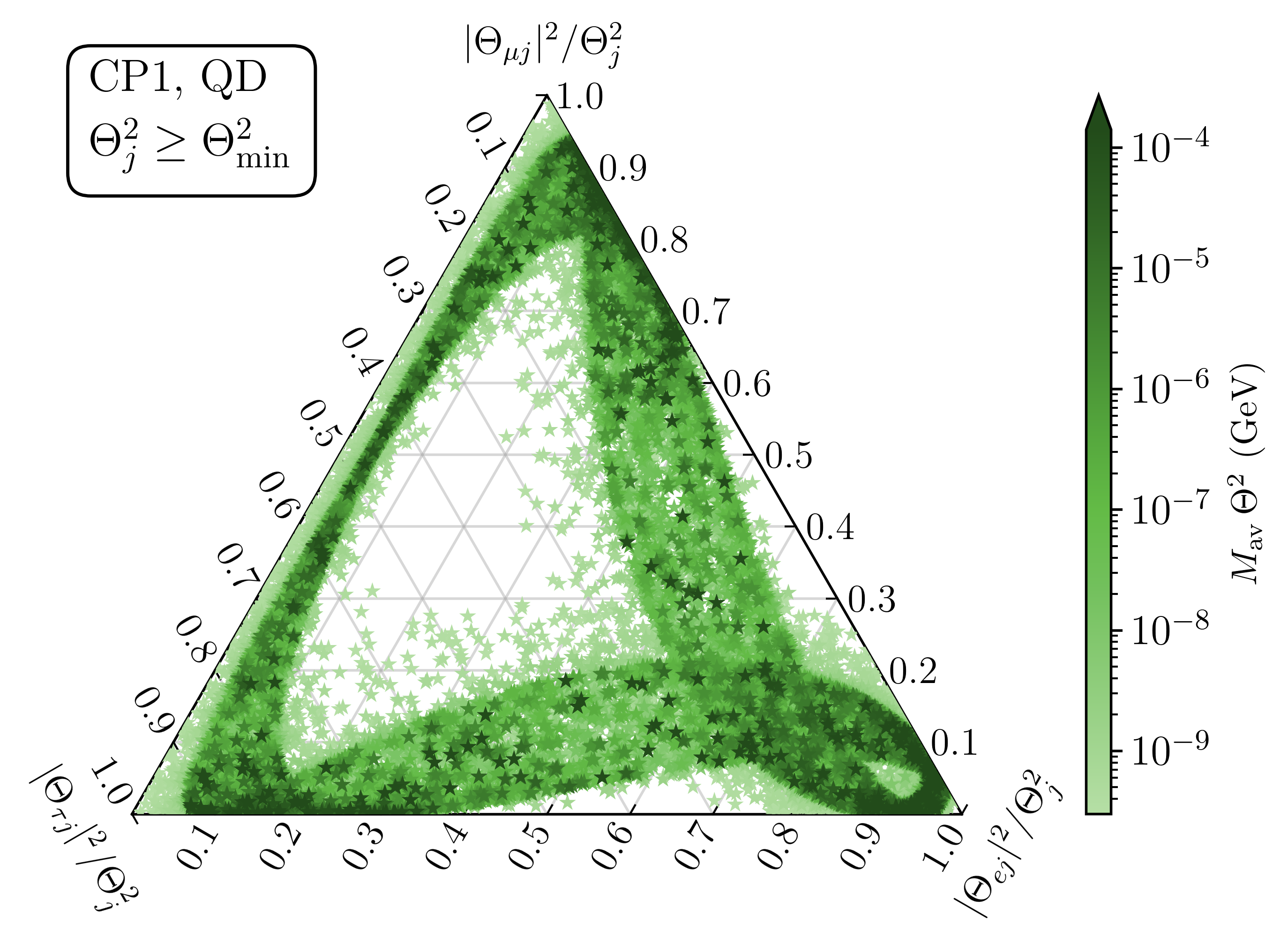}
    \caption{The same scan as in Fig.~\ref{fig:CP1s_scan_NH} but for IH (first four plots, red colour) and QD with $\mnulight=0.1\,\text{eV}$ (last four plots, green colour). All other details are as in Fig.~\ref{fig:CP1s_scan_NH}.}
\label{fig:CP1s_scan_IH_QD}
\end{figure}
Parameter points that violate any of the aforementioned constraints are discarded from the scan. The remaining points are then filtered depending on the value of the final BAU, namely we discard points if $|\eta_B| < \eta_B^{\rm obs}$, keep them if $|\eta_B| \geq \eta_B^{\rm obs}$. The final points thus correspond to configurations in which the generated BAU exceeds the observed value. These are classified as points of successful leptogenesis because there is enough degeneracy in the parameters to eventually yield $\eta_B = \eta_B^{\rm obs}$ (e.g., by lowering the deviation of $\delta$ from CP-conserving values 0, $\pi$, $2\pi$).

The result of the scan is depicted in Fig.~\ref{fig:CP1s_scan_NH} for NH, Fig.~\ref{fig:CP1s_scan_IH_QD} for IH (first four panels) and QD (last four panels). In the top-left, top-right and bottom-left panels of each case we show the scan in the $|\Theta_{ej}|^2 - M_j$, $|\Theta_{\mu j}|^2 - M_j$, $|\Theta_{\tau j}|^2 - M_j$ planes, respectively. In the bottom-right of the figures we show the scan in the ternary plane $|\Theta_{ej}|^2/\Theta^2_j$-$|\Theta_{\mu j}|^2/\Theta^2_j$-$|\Theta_{\tau j}|^2/\Theta^2_j$ with a colour mapping based on the value $\Theta^2 \Mav$. Each scan entry gives rise to a triplet of points in the plots corresponding to the three heavy neutrinos $N_{1,\,2,\,3}$. We further display the current bounds from HNL searches in gray and the projected sensitivities of the upcoming, planned and proposed experiments PIONEER \cite{PIONEER:2022yag} (purple), Hyper-K \cite{T2K:2019jwa} (blue), DUNE \cite{Breitbach:2021gvv} (pink), MATHUSLA \cite{MATHUSLA:2020uve} (orange), ANUBIS \cite{Hirsch:2020klk} (darker orange, dot-dashed), SHiP \cite{SHiP:2018xqw} (cyan), searches at HL-LHC \cite{Drewes:2019fou} (red), and future circular colliders FCC-ee/CEPC \cite{Blondel:2022qqo, CEPCStudyGroup:2018ghi} (green). 

Here we do not include the limits from Big Bang Nucleosynthesis (BBN) \cite{Sabti:2020yrt, Boyarsky:2020dzc}. The BBN limits as computed in \cite{Sabti:2020yrt, Boyarsky:2020dzc} do not apply straightforwardly to our scenarios, because they were computed considering two HNLs and the range of mixing angles that are achievable in that context, while here we are considering three HNLs.  Moreover, BBN bounds do not apply to the individual mixings $|\Theta_{\alpha j}^2|$, since the requirement of decaying sufficiently earlier than BBN cannot set a lower bound on a single flavour component for each heavy neutrino \cite{Drewes:2019mhg} (see also \cite{Domcke:2020ety}). Indeed, even if the squared mixing of $N_j$ with a flavour $\alpha$, $|\Theta_{\alpha j}^2|$, is BBN-disfavoured, the total decay rate can still be sufficiently large provided the other flavour mixings $|\Theta_{\beta j}^2|$, $\beta \neq \alpha$, are sizeable enough. This may not hold punctually in our parameter scan because the flavour structure differs for each point. A proper treatment would require recomputing the BBN constraint for each parameter choice, but such an analysis lies beyond the scope of the present work. For these reasons, we do not display BBN limits in the plots. 

The results of the scans show that successful leptogenesis driven only by the CP-violating Dirac phase $\delta$ occurs in regions well within reach of future experiments and where the Dirac-phase CPV hypothesis can be potentially falsified, as discussed in Sec.~\ref{sec:LECPV_parameter_spaces}, through the measurement of flavour ratios.

\subsection{Benchmark points of viable Dirac-phase driven leptogenesis}
We identify a set of benchmark points in the CP1 case that are within reach of future experiments and for which we obtain the observed BAU for $\delta = 212^\circ$ (NH), $274^\circ$ (IH) and $315^\circ$ (QD). We further analyse for these benchmark the dependence on the Dirac phase $\delta$. Here we show the results for just one of these benchmarks, the one for NH, while the results for the others can be found in Appendix \ref{app:BMC}. The choices of $\delta$ made for these points are exemplary, as we can obtain the correct BAU for a wide range of the Dirac phase $\delta$ while keeping the mass and the total mixing squared fixed at the benchmark values.

\paragraph{A benchmark in the NH case for CP1.} 

\begin{figure}[t!]
\centering
\includegraphics[width=0.65\linewidth]{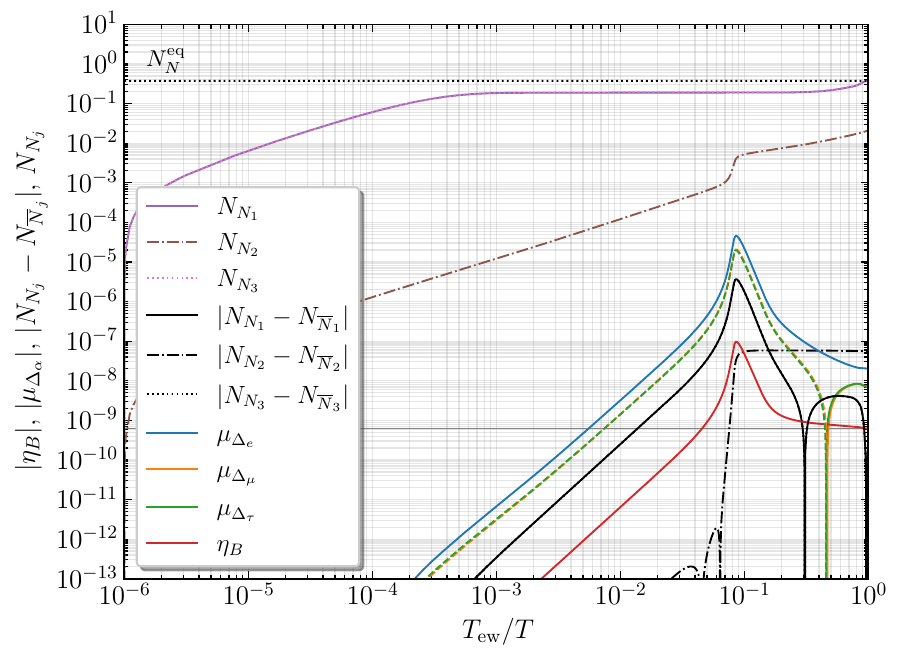}
    \caption{Evolution of the heavy neutrino abundances
$N_{N_j} \equiv (\rho_N)_{jj}$ (purple solid for $j=1$, brown dot--dashed for $j=2$, pink dotted for $j=3$),
their CP-asymmetries $|N_{N_j}-{N}_{\overline{N}_j}|$, $N_{\overline{N}_j} \equiv (\rho_{\overline{N}})_{jj}$ (black curves with the same line styles),
the flavour asymmetries $\mu_{\Delta_\alpha}$ with $\alpha=e,\mu,\tau$
(blue, orange and green curves, respectively; solid/dashed lines denote positive/negative values),
and the baryon-to-photon ratio $\eta_B$ (red curve),
as functions of $T_{\rm ew}/T$, for the benchmark point in Eq.~\eqref{eq:params_BMC2_NH}. The solid horizontal gray line marks the observed BAU $\eta_B^{\rm obs} = 6.10\times 10^{-10}$, while the dotted black one is for $N_N^{\rm eq} = 3/8$.}
    \label{fig:LG_Plot_BMC2_NH}
\end{figure}
We identify the following benchmark set of parameters:

\begin{equation}\label{eq:params_BMC2_NH}
\begin{split}
    \tilde{M}_1 = 2\,\text{GeV}, \quad \Delta \tilde{M}_{21} = 1.36\times 10^{-5}\,\text{GeV},\quad \Delta \tilde{M}_{31} = 0, \quad \mnulight = 0,\\
    x_\nu = \pi/12,\quad x_N = 0,\quad y = 5.44,\\
    \delta = 212^\circ\,\text{(best-fit of } \texttt{NuFit  6.0}\text{)},\quad
    \alpha_{23} = -2\pi.
\end{split}
\end{equation}
Note that this peculiar configuration corresponds to $\tilde{N}_1$ and $\tilde{N}_3$ being exactly degenerate in mass (but not $N_1$ and $N_3$). The PMNS angles and light neutrino squared mass differences are fixed according to their best-fit values of the \texttt{NuFit 6.0} global analysis in the NO case (with Super-Kamiokande data). This set of parameters fixes the Yukawa $Y$ and the $R$-matrix to be
\begin{equation}
    Y = \begin{pmatrix}
-1.20\times10^{-6} + 5.05\times10^{-7}i & 1.08\times10^{-8} + 1.17\times10^{-9}i & 5.05\times10^{-7} + 1.20\times10^{-6}i \\
3.88\times10^{-6} - 2.14\times10^{-8}i & 2.52\times10^{-8} + 6.93\times10^{-10}i & -2.14\times10^{-8} - 3.88\times10^{-6}i \\
4.98\times10^{-6} - 2.27\times10^{-8}i & -1.29\times10^{-9} + 7.35\times10^{-10}i & -2.27\times10^{-8} - 4.98\times10^{-6}i
\end{pmatrix}
\end{equation}
\begin{equation}
    R = \begin{pmatrix}
-1.04\times10^{-4} + 4.40\times10^{-5}i & 9.38\times10^{-7} + 1.02\times10^{-7}i & 4.39\times10^{-5} + 1.04\times10^{-4}i \\
3.37\times10^{-4} - 1.86\times10^{-6}i & 2.19\times10^{-6} + 6.03\times10^{-8}i & -1.86\times10^{-6} - 3.37\times10^{-4}i \\
4.33\times10^{-4} - 1.97\times10^{-6}i & -1.12\times10^{-7} + 6.40\times10^{-8}i & -1.97\times10^{-6} - 4.33\times10^{-4}i
\end{pmatrix}
\end{equation}
Then, after the procedure outlined around Eq.~\eqref{eq:tMNMN_2}, we obtain $M_1 = (2 + 3.17898\times 10^{-7})\,\text{GeV}$, $\Delta M_{21} = 1.35 \times 10^{-5}\,\text{GeV}$ and $\Delta M_{31} = 3.87 \times 10^{-7} \,\text{GeV}$,
\begin{equation}
V \simeq
\begin{pmatrix}
0.9716 & -5.333 \times 10^{-3} & -0.237 \\
-9.163 \times 10^{-5} & 0.9997 & -0.02290 \\
0.237 & 0.02227 & 0.9713
\end{pmatrix},
\end{equation}
up to sign changes and permutations of columns, and
\begin{equation}
\Theta \simeq \begin{pmatrix}
-9.10 \times 10^{-5} + 6.74 \times 10^{-5}i & 2.47 \times 10^{-6} + 2.19 \times 10^{-6}i & 6.74 \times 10^{-5} + 9.09 \times 10^{-5}i \\
3.27 \times 10^{-4} - 8.17 \times 10^{-5}i & 3.51 \times 10^{-7} - 7.44 \times 10^{-6}i & -8.17 \times 10^{-5} - 3.27 \times 10^{-4}i \\
4.20 \times 10^{-4} - 1.05 \times 10^{-4}i & -2.47 \times 10^{-6} - 9.57 \times 10^{-6}i & -1.04 \times 10^{-4} - 4.20 \times 10^{-4}i
\end{pmatrix}
\end{equation}
with 
\[
|\Theta_{e1}|^2 = 1.28 \times 10^{-8},\quad 
|\Theta_{e2}|^2 = 1.09 \times 10^{-11},\quad
|\Theta_{e3}|^2 = 1.28 \times 10^{-8}, 
\]
\[
|\Theta_{\mu 1}|^2 = 1.14 \times 10^{-7},\quad 
|\Theta_{\mu 2}|^2 = 5.55 \times 10^{-11},\quad
|\Theta_{\mu 3}|^2 = 1.14 \times 10^{-7},
\]
\[
|\Theta_{\tau 1}|^2 = 1.88 \times 10^{-7},\quad
|\Theta_{\tau 2}|^2 = 9.77 \times 10^{-11},\quad
|\Theta_{\tau 3}|^2 = 1.88 \times 10^{-7},
\]
\[
\Theta_e^2 = 2.564 \times 10^{-8},\quad
\Theta_\mu^2 = 2.276 \times 10^{-7}, \quad
\Theta_\tau^2 = 3.754 \times 10^{-7},
\]
\[
\Theta_1^2 = 3.143 \times 10^{-7},\quad
\Theta_2^2 = 1.641 \times 10^{-10},\quad
\Theta_3^2 = 3.141 \times 10^{-7},
\]
and $\Theta^2 = 6.29 \times 10^{-7}$. The BAU for this benchmark point roughly equals the observed value $\eta_B^{\rm obs}$, i.e.~$\eta_B \simeq 6.1 \times 10^{-10}$. We note that for this particular benchmark $N_2$ has much smaller individual couplings than $N_1$ and $N_3$. The behaviour of the quantities $N_{N_j}\equiv (\rho_N)_{jj}$ and $|N_{N_j} - \overline{N}_{N_j}|$, $j=1,\,2,\,3$,  $\mu_{\Delta_\alpha}$, $\alpha = e,\,\mu,\,\tau$ and $\eta_B$ against $x = T_{\rm ew}/T$ for this benchmark is shown in the top-right panel of Fig.~\ref{fig:LG_Plot_BMC2_NH}. Note that, for this benchmark point, we get a level crossing effect discussed in \cite{Abada:2018oly} that induces a resonance in the asymmetries around $x \simeq 8.5 \times 10^{-2}$. The effective Majorana mass parameter for this benchmark reads $m_{\beta\beta}^{\rm eff} \simeq 3.23\,\text{meV}$.

\begin{figure}[t]
    \centering
    \includegraphics[width=0.5\linewidth]{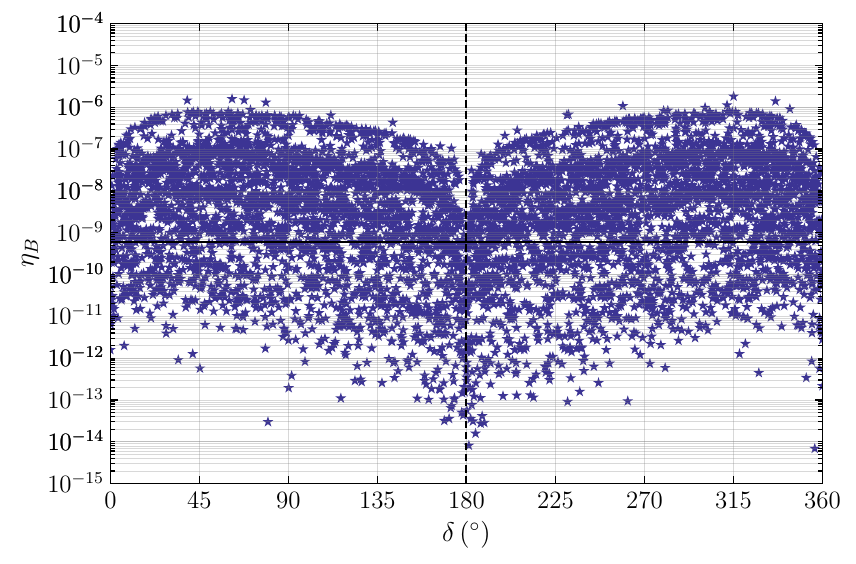}
    \hspace{.5em}
        \includegraphics[width=0.4\linewidth]{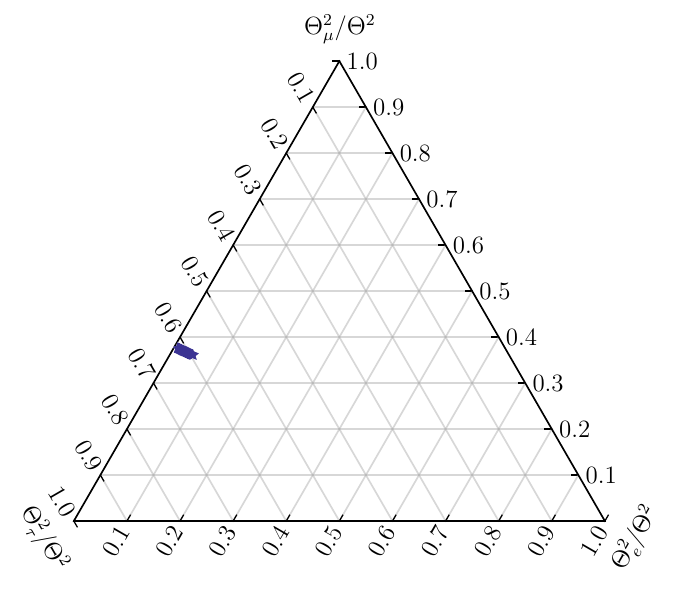}
    \caption{On the left, the numerical scan around the benchmark point of Eq.~\eqref{eq:params_BMC2_NH} in the $\delta$-$\eta_B$ plane, with $\tilde{M}_1 = 2\,\text{GeV}$, $\Theta^2 \simeq 6.29\times 10^{-7}$, $\alpha_{23} = -2\pi$ and $x_\nu = \pi/12$ while $10^{-11} \leq |\Delta \tilde{M}_{21}|/\tilde{M}_1,\,
|\Delta \tilde{M}_{32}|/\tilde{M}_1 \leq 10^{-4}$, 
$0 \leq x_N \leq 2\pi$ and $0 \leq \delta \leq 2\pi$. On the right, the same points of the scan but in the ternary plane $\Theta^2_e/\Theta^2$-$\Theta^2_\mu/\Theta^2$-$\Theta^2_\tau/\Theta^2$.}
    \label{fig:BAUvsdelta}
\end{figure}

\paragraph{Scan around the benchmark.} We perform a numerical scan around the discussed benchmark point, in order to further investigate the role of the Dirac phase $\delta$ in generating the BAU.
In this scan, we keep fixed $\tilde{M}_1 = 2\,\text{GeV}$, $y = 5.44$, $\alpha_{23} = -2\pi$ and $x_\nu = \pi/12$,
while the remaining parameters are randomly varied within the ranges
$10^{-11} \leq |\Delta \tilde{M}_{21}|/\tilde{M}_1,\,
|\Delta \tilde{M}_{32}|/\tilde{M}_1 \leq 10^{-4}$, 
$0 \leq x_N \leq 2\pi$ and $0 \leq \delta \leq 2\pi$. In this way we set the total squared mixing $\Theta^2 \simeq 6.29\times 10^{-7}$ and identify a specific region in the $\Theta^2_e/\Theta^2$-$\Theta^2_\mu/\Theta^2$-$\Theta^2_\tau/\Theta^2$ plane. The results of the scan are shown in Fig.~\ref{fig:BAUvsdelta}, on the left in the $\delta$-$\eta_B$ plane, on the right in the $\Theta^2_e/\Theta^2$-$\Theta^2_\mu/\Theta^2$-$\Theta^2_\tau/\Theta^2$ triangle. In the figure, we keep only the points that are not ruled out by current experiments and for which $\eta_B \geq 0$.

We do not observe any correlation between the sign of the BAU and the Dirac phase $\delta$,
in contrast to what is found in other regimes
(see e.g.~\cite{Granelli:2021fyc, Granelli:2023tcj}).
This behaviour is a consequence of the degeneracy with the parameters
$\Delta \tilde{M}_{21}$, $\Delta \tilde{M}_{32}$ and $x_N$.
Such degeneracies could in principle be lifted if the heavy neutrinos were kinematically
resolved at experiments, allowing for a direct determination of the mass splittings and
of $x_N$ through measurements of the individual squared couplings
$\Theta_1^2,\, \Theta_2^2,\, \Theta_3^2$.
In the absence of any experimental information on these parameters,
no correlation between the sign of the BAU and $\delta$ can be established.

Interestingly, we can get the correct BAU for an extended range of values of the Dirac phase. The BAU of course vanishes when the Dirac phase takes the CP-conserving values
$\delta = 0,\,\pi,\,2\pi$, as expected, but
it is quite remarkable to us
that successful leptogenesis remains possible  within the testable region for a broad range of values of $\delta$, and in particular for values that are significantly close to CP-conserving points.
In what follows, we further investigate how much $\delta$ can deviate from CP-conserving values while still leading to successful leptogenesis in the experimentally accessible  region of the parameter space.

\subsection{Viable and testable leptogenesis with suppressed Dirac-phase CP-violation}

By looking at the density plots in the triangles of Figs.~\ref{fig:CP1s_scan_NH} and \ref{fig:CP1s_scan_IH_QD} (see darker regions there), we note that the points of maximal squared mixings are concentrated around regions of the parameter space where one of the flavoured mixings is suppressed with respect to the other two. Intriguingly, in our scenario with low-energy CPV from $\delta$ and in the limit of large total squared mixing $\Theta^2$ -- i.e.~large $y$, neglecting terms of order $\mathcal{O}(e^{-2|y|})$ -- it is possible to set one of the individual flavoured mixings $\Theta_e^2$, $\Theta^2_\mu$ or $\Theta^2_\tau$ (approximately) to zero only if $\delta$ is (close to) $0,\,\pi,\,2\pi$ and for certain conditions on $x_\nu$.  These observations suggest that the points of successful leptogenesis that maximise the heavy neutrino mixings correspond to $\delta$ being close to CP-conserving values.
This fact is quite remarkable since one might have naively expected the opposite behaviour, namely that the parameter space in the $\Mav$-$\Theta^2$ plane compatible with successful leptogenesis shrinks as the Dirac phase $\delta$ approaches CP-conserving values.
This can be understood by noting that a small value of the Dirac phase $\delta$ suppresses the CP-asymmetries only linearly (cf.~Eqs.~\eqref{eq:cpv_lfv} and~\eqref{eq:cpv_lnv}),
whereas a large washout damps them exponentially. Hence, a large flavour hierarchy enabled by an almost CP-conserving value of the Dirac phase can lead to an enhancement of the overall asymmetry.

We illustrate here the combination of $\delta$, $\alpha_{23}$ and $x_\nu$ that give vanishing $\Theta_\tau^2\ll \Theta_\mu^2, \Theta_e^2$ for the NH and CP1 case. Such combination, as well as those that give $\Theta_e^2\simeq 0$ or $\Theta_\mu^2\simeq 0$ in the same case, and the corresponding ones in the other CP-conserving scenarios and choices of $\mnulight$, can be derived starting from the analytical expressions in Appendix \ref{app:Theta2_Expr}. The condition $\Theta_\tau^2 \simeq 0$ is satisfied at large $y$ for $\delta = 0, \,\pi,\,2\pi$ and $x_\nu=\bar{x}_\nu$, where
\begin{equation}
    \bar{x}_\nu \equiv \arctan\left[+ (-) \sqrt{\frac{m_3}{m_2}}\, \frac{c_{23}c_{13}}{c_{12}s_{23}\pm s_{12}c_{23}s_{13}}\right],
\end{equation} 
for $\alpha_{23} = 0$ ($\pm 2\pi$) and with the $+,-$ sign in the denominator corresponding to $\delta = 0,\,2\pi$ and $\delta = \pi$, respectively.

\begin{figure}[t!]
    \centering
    \includegraphics[width=0.52\linewidth]{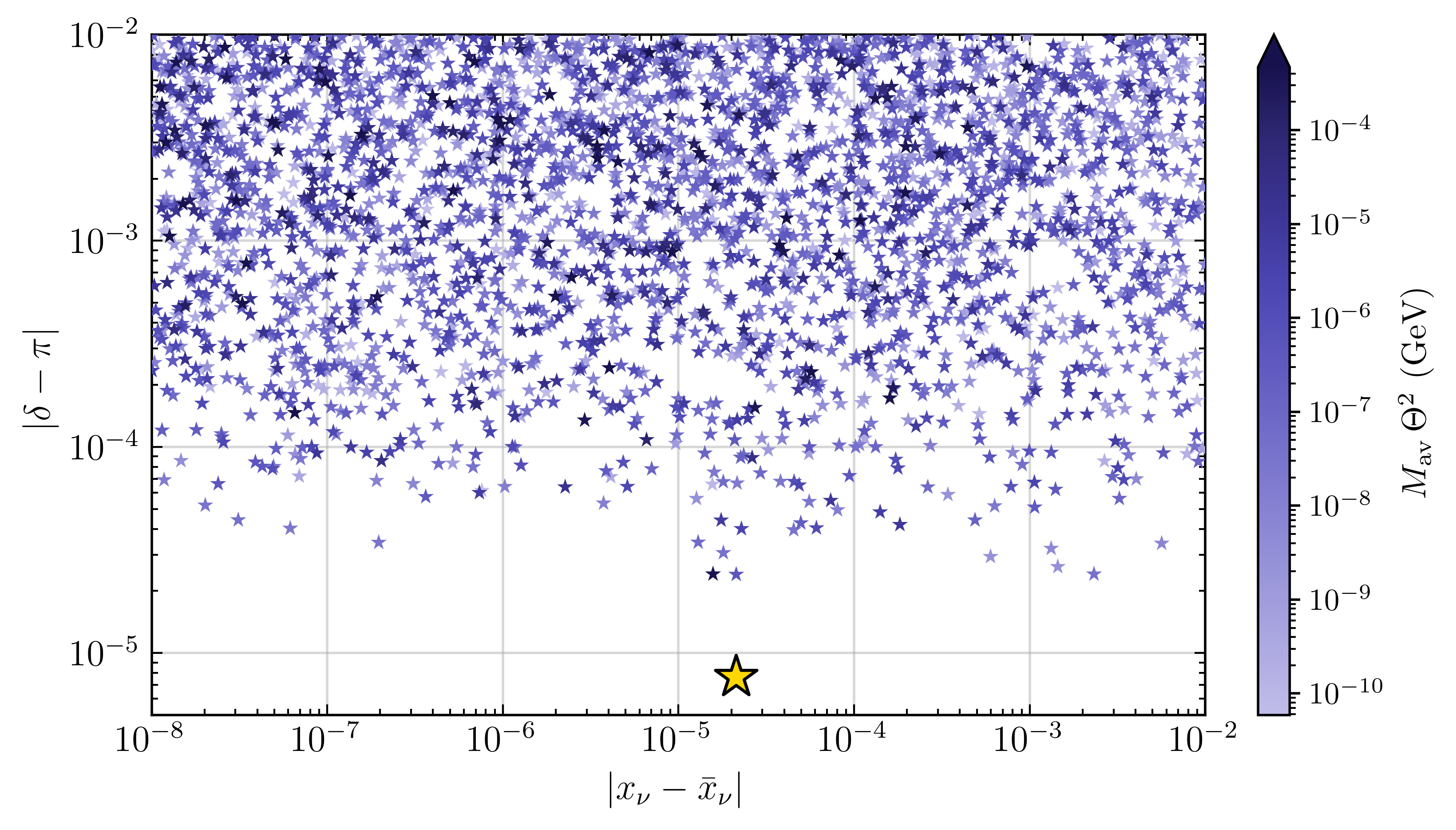}
\includegraphics[width=0.45\linewidth]{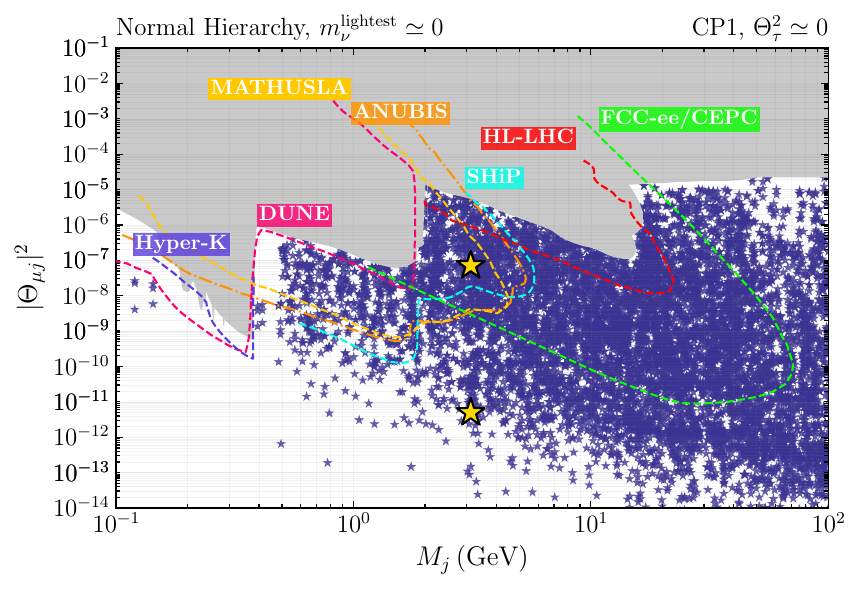}
    \caption{Parameter scan for NH and CP1 around the region where $\Theta^2_\tau \ll \Theta^2_\mu, \Theta^2_e$. The scan is obtained for $\alpha_{23}=0$, $\delta = \pi \pm \epsilon_\delta$ and $x_\nu = \bar{x}_\nu \pm \epsilon_\nu$, where $\bar{x}_\nu$ is the angle that allows for $\Theta^2_\tau = 0$ when $\delta = \pi$ and $\epsilon_\delta, \epsilon_\nu$ small parameters. The parameters are randomly varied in the following ranges: $0.1\leq \tilde{M}_1/\text{GeV} \leq 100$, $10^{-11} \leq|\Delta \tilde{M}_{21}|/\tilde{M}_1,\, |\Delta \tilde{M}_{32}|/\tilde{M}_1 \leq 10^{-4}$, $2 \leq |y| \leq 12$, $10^{-8} \leq \epsilon_\delta, \epsilon_\nu \leq 10^{-2}$. The remaining oscillation parameters are fixed to their best-fit values from the \texttt{NuFit~6.0} global analysis. The brightness of the points indicates $\Mav \Theta^2$, as illustrated in the bar legend. The points shown in this plot satisfy $|\eta_B|\geq \eta_B$ and the current experimental constraints on HNL individual mixings. On the right, we show the same scan point in the $\Theta^2_\mu$-$\Mav$ plane (same details as in Fig.~\ref{fig:CP1s_scan_NH}). The yellow stars mark the benchmark in Eq.~\eqref{eq:params_BMC_flav}. In the right panel, each heavy neutrino is represented by a star. For the benchmark, two stars overlap almost completely and are related to $N_1$ and $N_3$ with similar mixings, while the third one at lower mixing corresponds to the neutrino $N_2$ that has suppressed mixing.}
    \label{fig:tauzero_scan}
\end{figure}

To find how much $\delta$ can deviate from CP-conserving values while still being compatible with successful and testable leptogenesis, we perform a dedicated scan of the parameter space around the region for which $\Theta_\tau^2 \ll \Theta_e^2, \Theta_\mu^2$ in CP1 and NH case. The scan is performed by perturbing the Dirac phase as $\delta = \pi \pm \epsilon_\delta$ and $x_\nu = \bar{x}_\nu \pm \epsilon_\nu$, 
for $\alpha_{23}=0$. The parameters $\epsilon_\delta$ and $\epsilon_\nu$ are both varied within the range $[10^{-8}, 10^{-2}]$. For this particular scan we also concentrate on relatively large mixings, setting $y\geq 2$. The results of the scan are shown in Fig.~\ref{fig:tauzero_scan}, where we display only the points of the scan for which $|\eta_B| \geq \eta_B^{\rm obs}$ after imposing the experimental constraints. In the left panel we show the scan in the $|\delta - \pi|$ - $|x_\nu-\bar{x}_\nu|$ plane. The colour brightness in the plot indicates the value of $\Mav\Theta^2$. The results show that successful leptogenesis can be achieved for deviations of $\delta$ from the CP-conserving value $\pi$ smaller than $\sim 10^{-5}$ while the heavy neutrino squared mixings remain well within reach of future experiments. In the plot, the yellow star marks an exemplary benchmark point for which $\eta_B \simeq \eta_B^{\rm obs}$ for $\delta = \pi + 7.68\times10^{-6}$. We give the parameters of such benchmark further.

In the right panel of the same figure, we show the scan points in the $\Theta^2_\mu$-$\Mav$ plane. Interestingly, the condition of $\Theta^2_\tau \simeq 0$ makes the scan more efficient in the $\Mav\sim \mathcal{O}(20-100) \,\text{GeV}$ and $\Theta^2 \gtrsim \mathcal{O}(10^{-7})$ regime, resulting in a higher density of points in that region. We emphasize that the tendency towards values of $\delta$ close to CP-conservation is not a requirement
for successful leptogenesis in this region. Rather, our scan indicates that suppressing one flavour direction (which in our case is possible for $\delta$ sufficiently close to CP-conserving values) can help avoid \virg{unnatural} fine-tuned cancellations in the generation of the BAU, thus making our numerical scan more efficient in this region of the parameter space. A detailed explanation of this is feature, however, would require a deeper analytical understanding of three heavy neutrino leptogenesis in this regime first, which is beyond the scope of the present work. This observation is in any case crucial for the possibility of experiments that search for charged lepton flavour violating processes involving muons to probe this part of the parameter space of Dirac-phase driven leptogenesis (see next subsection). 

\paragraph{A benchmark with suppressed CPV.} We identify the following benchmark of viable and testable leptogenesis with suppressed Dirac-phase CPV: \begin{equation}\label{eq:params_BMC_flav}
\begin{split}
    \tilde{M}_1 = 3.12\,\text{GeV}, \quad \Delta \tilde{M}_{12} \simeq \Delta \tilde{M}_{13} \simeq 3.12\times 10^{-4}\,\text{GeV}, \quad \mnulight = 0,\\
    x_\nu = 73.575^\circ,\quad x_N = 0,\quad y = 5.67,\\
    \delta = \pi + 7.68 \times 10^{-6},\quad
    \alpha_{23} = 0.
\end{split}
\end{equation}
The PMNS angles and light neutrino squared mass differences are fixed according to their best-fit values of the \texttt{NuFit 6.0} global analysis in the NO case (with Super-Kamiokande data). This set of parameters fixes the Yukawa $Y$ and the $R$-matrix to be
\begin{equation}
    Y \simeq \begin{pmatrix}
 - 1.85\times10^{-6}i &  1.49\times10^{-8}i & -1.85\times10^{-6} \\
 - 4.77\times10^{-6}i & - 4.12\times10^{-8}i & -4.77\times10^{-6} \\
1.67\times10^{-10}i &  - 5.39\times10^{-8}i & 1.67\times10^{-10}
\end{pmatrix},
\end{equation}
\begin{equation}
    R \simeq \begin{pmatrix}
    - 1.03\times10^{-4}i &   8.31\times10^{-7}i & -1.03\times10^{-4} \\
 - 2.66\times10^{-4}i &  - 2.30\times10^{-6}i & -2.66\times10^{-4} \\
 + 9.34\times10^{-9}i &- 3.01\times10^{-6}i & 9.34\times10^{-9} 
\end{pmatrix}
\end{equation}
Note that $Y_{\alpha 1}\simeq i Y_{\alpha 3}$, i.e.~$\tilde{N}_1$ and $\tilde{N}_3$ form a pseudo-Dirac pair. In this case $V \simeq \mathbb{1}$, so that $\Delta M_{12} \simeq \Delta \tilde{M}_{12}$, $  \Delta M_{13} \simeq \Delta \tilde{M}_{13}$  and $\Theta \simeq R$. The squared mixings are
\[
|\Theta_{e1}|^2 = 1.07 \times 10^{-8},\quad |\Theta_{e2}|^2 = 6.9 \times 10^{-13},\quad |\Theta_{e3}|^2 = 1.07 \times 10^{-8},
\]

\begin{figure}[t!]
    \centering
    \includegraphics[width=0.65\linewidth]{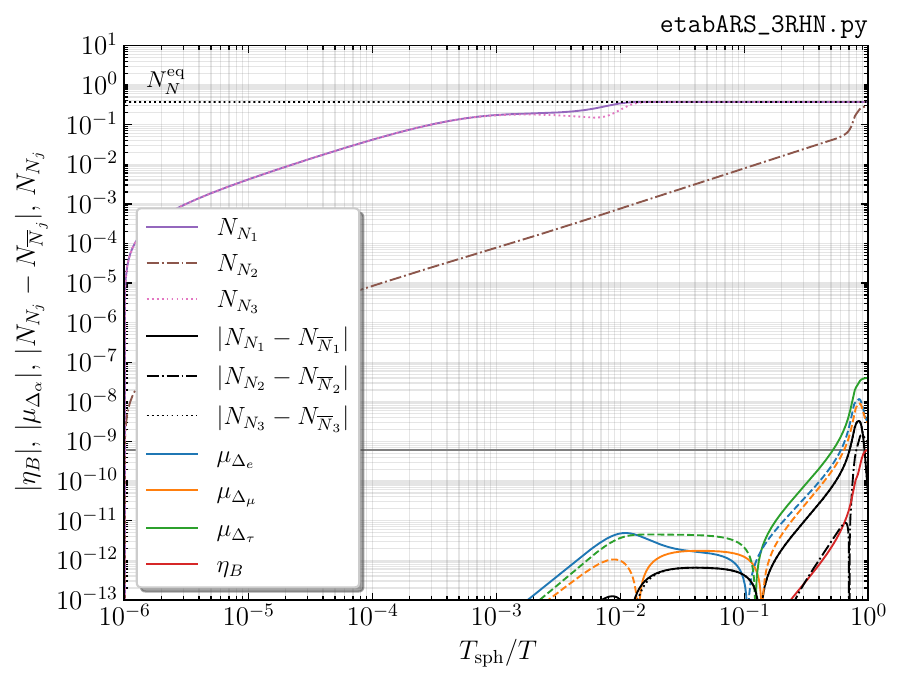}
    \caption{Evolution of the heavy neutrino abundances $N_{N_j} \equiv (\rho_N)_{jj}$,
their CP-asymmetries $|N_{N_j}-{N}_{\overline{N}_j}|$, $N_{\overline{N}_j}$, the flavour asymmetries $\mu_{\Delta_\alpha}$ with $\alpha=e,\mu,\tau$,
and the baryon-to-photon ratio $\eta_B$,
as functions of $T_{\rm ew}/T$, for the benchmarks point in Eq.~\eqref{eq:params_BMC_flav}. All other details are as in Fig.~\ref{fig:LG_Plot_BMC2_NH}.}
    \label{fig:tauzero_scan_BMC_ev}
\end{figure}
\begin{figure}[h!]
    \centering
    \includegraphics[width=0.95\linewidth]{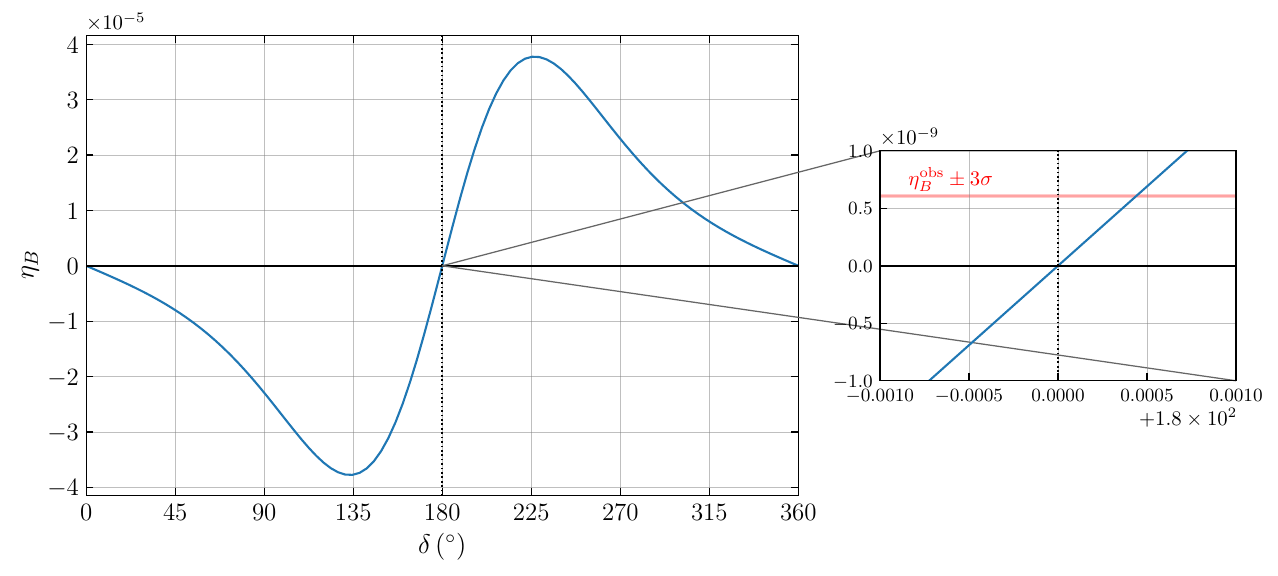}
    \caption{BAU versus $\delta$ for a specific benchmark point for which we get successful leptogenesis for $\delta = \pi + 1.05\times 10^{-5}$. The parameters are  as in Eq.~\eqref{eq:params_BMC_flav}. The red band corresponds to the $3\sigma$ range of the observed BAU, $\eta_B^{\rm obs} = 6.10^{+0.08}_{-0.08}\times 10^{-10}$.
    }
    \label{fig:tauzero_scan_BAUdelta}
\end{figure}
\[
|\Theta_{\mu 1}|^2 = 7.09 \times 10^{-8},\quad |\Theta_{\mu 2}|^2 = 5.28 \times 10^{-12},\quad |\Theta_{\mu 3}|^2 = 7.09 \times 10^{-8},
\]
\[
|\Theta_{\tau 1}|^2 \simeq 0,\quad |\Theta_{\tau 2}|^2 = 9.06 \times 10^{-12},\quad |\Theta_{\tau 3}|^2 \simeq 0,
\]
\[
\Theta_e^2 = 2.125 \times 10^{-8},\quad \Theta_\mu^2 = 1.417 \times 10^{-7},\quad \Theta_\tau^2 = 9.05 \times 10^{-12},
\]
\[
\Theta_1^2 = 8.17 \times 10^{-8},\quad \Theta_2^2 = 1.50 \times 10^{-11},\quad \Theta_3^2 = 8.17 \times 10^{-8},
\]
$\Theta^2 = 1.63 \times 10^{-7}$. We show the evolution of the BAU, lepton chemical potentials, heavy neutrino abundances and asymmetries during leptogenesis for this benchmark in Fig.~\ref{fig:tauzero_scan_BMC_ev}. The final BAU for this point equals the observed value, $\eta_B \simeq \eta_B^{\rm obs}$. We also depict the behaviour of the asymmetry versus the Dirac phase for this benchmark in Fig.~\ref{fig:tauzero_scan_BAUdelta}. The effective Majorana mass parameter for this benchmark reads $m_{\beta\beta}^{\rm eff} \simeq 3.68\,\text{meV}$.

\subsection{Parameter space of charged lepton flavour violation}
The interactions in
Eqs.~\eqref{eq:NCC} and \eqref{eq:NNC} also induce charged lepton flavour violating (cLFV)
processes at one loop, such as $\mu\to e\gamma$, $\mu\to eee$ and $\mu$--$e$ conversion in
nuclei, as well as analogous channels involving the $\tau$ lepton
\cite{Petcov:1976ff,Bilenky:1977du}.
Their phenomenology depends on flavour-off-diagonal combinations of mixings,
$\Theta_{\alpha \beta}^2 \equiv \big|\sum_{j=1}^3 \Theta_{\alpha j}^* \Theta_{\beta j}\big|$, $\alpha,\,\beta = e,\,\mu,\,\tau$. These processes provide indirect probes of heavy neutral leptons and leptogenesis. Future experiments on cLFV processes involving muons,  namely MEG II~\cite{MEGII:2018kmf} on $\mu\to e\gamma$ decay, Mu3e Project~\cite{Arndt:2009} on $\mu \to eee$ decay, Mu2e~\cite{Bartoszek:2015} and COMET~\cite{Abramishvili:2020} (PRISM/PRIME~\cite{Barlow:2011zza}) on $\mu$\,--\,$e$ conversion in aluminium (titanium), are expected to have enough sensitivity to probe the parameter space of viable leptogenesis for heavy neutrino above the GeV scale and beyond in the TeV-PeV realm \cite{Urquia-Calderon:2022ufc, Granelli:2022eru}, while experiments such as BELLE II \cite{BELLEII, BELLEIIbook} looking at cLFV in processes involving $\tau$ are less competitive in this regard.

\begin{figure}[t!]
    \centering
    \includegraphics[width=0.65\linewidth]{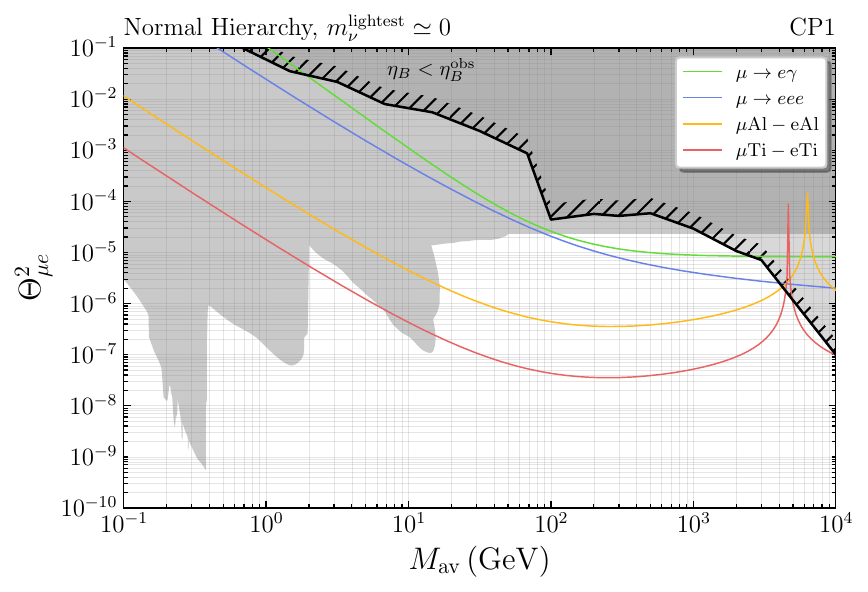}
    \caption{The scan in the $\Theta_{\mu e}^2$-$\Mav$ plane extending to $\Mav = 10 \,\text{TeV}$. For illustrative purposes, we depict only the case of NH (NO with $\mnulight \simeq 0$), CP1, i.e.~CI matrix as in \eqref{eq:CPconsCI_1}, and fix the Majorana phases as in Eq.~\eqref{eq:CP1a21a31} to ensure the CPV only comes from the Dirac phase $\delta$. The black curve indicate the maximal allowed combination $\Theta_{\mu e}^2$ that is compatible with successful leptogenesis with CPV uniquely from the Dirac phase $\delta$. In the region above the black curve, the BAU $\eta_B$ is too small compared to the observed value $\eta_B^{\rm obs}$. The gray region up to $\Mav = 50\,\text{GeV}$ corresponds to the current constraints on $\Theta^2_\mu$, while for $\Mav > 50\,\text{GeV}$ to the bound on $\Theta^2_{\mu e}$ at $95\%$ C.L.~as derived in \cite{Blennow:2023mqx}. The green, blue and yellow (red) curves are, from top to bottom, the sensitivities of MEG II~\cite{MEGII:2018kmf} on $\mu\to e\gamma$ decay, Mu3e Project~\cite{Arndt:2009} on $\mu \to eee$ decay, Mu2e~\cite{Bartoszek:2015} and COMET~\cite{Abramishvili:2020} (PRISM/PRIME~\cite{Barlow:2011zza}) on $\mu$\,--\,$e$ conversion in aluminium (titanium), respectively.}
    \label{fig:cLFV_scan}
\end{figure}

We show in Fig.~\ref{fig:cLFV_scan} the results of our leptogenesis scan in the $\Theta_{\mu e}^2$-$\Mav$ plane.  For illustrative purposes, we depict only the case of NH and CP1. We project the sensitivities of upcoming experiments on cLFV processes involving muons as derived in \cite{Granelli:2022eru}. To showcase the potential of these experiments beyond the $100\,\text{GeV}$ limit, we extend our scan to the $10\,\text{TeV}$ mass scale. Moreover, to maximise the combination $\Theta_{\mu e}^2$ in the scan, we focus on points that satisfy the condition $\Theta_\tau^2\ll \Theta_\mu^2, \Theta_2$. The gray region in Fig.~\ref{fig:cLFV_scan} up to $\Mav = 50\,\text{GeV}$ corresponds to the current constraints on $\Theta^2_\mu$ which are only meant to be indicative as they do not directly apply to the combination $\Theta_{\mu e}^2$, while for $\Mav > 50\,\text{GeV}$ we show the bound on $\Theta^2_{\mu e}$ at $95\%$ C.L.~as derived in \cite{Blennow:2023mqx}. \footnote{The scan below 100 GeV is the same as the one shown in Fig.~\ref{fig:tauzero_scan}. Here we extended that scan to higher masses. The difference between Fig.~\ref{fig:cLFV_scan} and the right panel Fig.~\ref{fig:tauzero_scan}, apart from the extended mass range in the latter, is the presentation of the points: in Fig.~\ref{fig:tauzero_scan} we removed the points already excluded by current experimental constraints, while in Fig.~\ref{fig:cLFV_scan} we kept them and displayed the maximal values of $\Theta_{\mu e}^2$ for the considered flavour structure.}

The results show that, in the CP1 case, if CPV comes solely from the Dirac phase $\delta$, the experiment MEG II~\cite{MEGII:2018kmf} on $\mu\to e\gamma$ (Mu3e~\cite{Arndt:2009} on $\mu \to eee$) has the potentiality to probe the leptogenesis parameter space in the $100 \,\text{GeV}\lesssim \Mav \lesssim 2\,(4)\,\text{TeV}$ mass window, while COMET~\cite{Abramishvili:2020} and  PRISM/PRIME~\cite{Barlow:2011zza} on $\mu$\,--\,$e$ conversion in aluminium and titanium can do it within $15 \,\text{GeV}\lesssim \Mav \lesssim 4\,\text{TeV}$ and $2\,\text{GeV}\lesssim \Mav \lesssim 10\,\text{TeV}$ mass windows, respectively. We also find that the parameter space of Dirac-phase driven viable leptogenesis is smaller compared to that one associated with additional CPV from the CI matrix and/or the Majorana phases. In particular, the experiment MEG II~\cite{MEGII:2018kmf} on $\mu\to e\gamma$ (Mu3e~\cite{Arndt:2009} on $\mu \to eee$) has enough sensitivity to probe the latter for $2\,(4)\lesssim \Mav/\text{TeV} \lesssim 20\,(70)$ and \cite{Granelli:2023tcj} (see also \cite{Drewes:2021nqr}), but not the former. This allows for the possibility to falsify experimentally the Dirac-phase CPV hypothesis for successful leptogenesis in upcoming future experiments on cLFV processes involving muons.

We expect the results to be similar for the cases of CP1 with IH or QD, CP2 with NH or IH or QD, and CP3 with IH or QD, while the results might be quite different in the CP3 case in NH because of its flavour structure that renders impossible to set $\Theta_\tau^2\ll \Theta_\mu^2,\,\Theta_e^2$. The potential of testing these cases using data from the upcoming cLFV experiments with muons requires an independent investigation, which is beyond the scope of this study.

\subsection{Testability and falsifiability of Dirac-phase driven leptogenesis}

The discussed experiments offer a clear path towards testing this leptogenesis scenario,
fundamentally relying on a combination of direct searches and neutrino experiments including  $\nubb$-decay.
While full reconstruction of parameters is in principle possible~\cite{Drewes:2024bla},
in practice it remains a significant challenge due to the large number of free parameters in the general seesaw model.
The Dirac-phase leptogenesis scenario significantly simplifies this, reducing the number of free parameters from 13, down to merely 8:
the phase $\delta$, the lightest neutrino mass, the CI-angles $x_N$, $x_\nu$ and $y$, and the three heavy neutrino masses.

As shown in detail over the previous sections,
as well as in Appendix~\ref{app:Theta2_Expr},
the information about these parameters is contained in the heavy neutrino mixing angles.
Specifically, the value of the total mixing angle $\wTheta^2$ provides information about the combination of the angle $y$,  $x_\nu$ and the mass of the lightest neutrino.
On the other hand, the phase $\delta$, the angle $x_\nu$ and the mass of the lightest neutrino could directly be inferred from a measurement of the heavy neutrino branching ratios
$\wTheta^2_\alpha/\wTheta^2$ at direct search experiments.
If the latter two are measured from the branching ratios, this would also determine the size of the angle $y$.
Since the HNL branching ratios only provide two independent observables,
there remains a degeneracy between the phase $\delta$, the angle $x_\nu$ and $\mnulight$.
Furthermore, there is a discrete degeneracy for the phase $\delta$, as the mixings $\wTheta^2_{\alpha j}$ are insensitive to $\delta \rightarrow - \delta$ in the scenario where the Dirac phase is the sole origin of CP violation, as can be verified by appropriate substitutions in Appendix~\ref{app:Theta2_Expr}.
The remaining degeneracy between the parameters can be broken either by the direct measurement of the phase $\delta$ at neutrino experiments such as T2HK~\cite{Hyper-Kamiokande:2025fci} and DUNE \cite{DUNE:2021mtg}, or by an indirect measurement of the mass of the lightest neutrino.
The latter can however pose a significant challenge as neither the KATRIN experiment, nor cosmological observation can probe arbitrarily small neutrino masses.
For the benchmark point in Table~\ref{tab:testable_benchmark} (see furhter), the two dominant branching ratios could be measured to sub-percent precision at future Tera-Z factories (see e.g.~\cite{Caputo:2016ojx,Antusch:2017pkq,Drewes:2025ocf}), and the strongly suppressed $\tau$ fraction to about $10\%$.

The remaining heavy neutrino parameters pose the greatest challenge: the angle $x_N$, and the three Majorana masses.  
In an ideal scenario the three heavy neutrino states would be measured directly, which would simultaneously determine their masses, and the angle $x_N$,
by measuring the ratios between the mixing angles of the three states.
For all three scenarios this also imposes an important consistency check, as in the limit of large mass splittings and large $y$,
the mixing of one of the heavy neutrinos is exactly equal to the sum of the remaining two, whereas the ratio of their mixings determines $x_N$.

To explicitly demonstrate a scenario where baryogenesis could be tested, we introduce a benchmark with $\mathcal{O}(10\%)$ mass splittings, $\tilde{M}_j = 9,\,10,\,11\,\text{GeV}$, and a sizable overall mixing $\Theta^2 \simeq 2.1\times 10^{-7}$.
At a Tera-Z run with $N_Z = 6\times 10^{12}$ this corresponds to $\simeq 1.5\times 10^5$ displaced-vertex events,
distributed over three kinematically distinguishable heavy neutrinos and three lepton flavours.
In contrast to the rest of the parameter scans, where we conservatively restrict the mass splittings to $\Delta \tilde{M}_{21}/\wMav,\, \Delta \tilde{M}_{31}/\wMav \leq 10^{-4}$,
here we deliberately choose a benchmark whose splittings are large enough to exceed the expected resolution of a few percent~\cite{Aleksan:2024hyq}.\footnote{
We verified that this benchmark remains stable under radiative corrections,
the one-loop contribution of Eq.~\eqref{eq:Mloop} amounting to $\|m_\nu^{\rm loop}\| \simeq 0.05\sum_i m_i$,
and that the heavy Majorana neutrino contribution to the effective Majorana mass parameter stays at the $\lesssim 10\%$ level,
so that $m_{\beta\beta}^{\rm eff} \simeq |(m_\nu)_{ee}|$ remains a good approximation.
}
Since the three heavy neutrinos can appear as separate peaks, the mixings $\wTheta^2_{\alpha j}$ can be measured individually rather than only in the flavour-summed combinations $\wTheta^2_\alpha$.

Using the event number estimates at FCC-ee from~\cite{Drewes:2025ocf}, and the sensitivity estimates from~\cite{Drewes:2024bla},
we perform a simulated parameter reconstruction under the assumption of a Dirac-phase driven leptogenesis scenario.
We find that the chosen CP1 benchmark is not uniquely reconstructed: three further solutions, listed in Table~\ref{tab:testable_benchmark}, reproduce the same nine mixings $\wTheta^2_{\alpha j}$ to within the reconstructed uncertainties.
One of them, denoted CP1$'$, lies within case CP1 itself and is the second root of the mixings in the lightest neutrino mass, which enters $\wTheta^2_{\alpha j}$ quadratically in $\sqrt{\mnulight}$; the other two belong to case CP2.
All four solutions predict a baryon asymmetry of the order of the observed one, so that the magnitude of the BAU is common to all of them, but they differ in the lightest neutrino mass and in the Dirac phase, and it is these differences that allow them to be told apart.

The value of the lightest neutrino mass separates the benchmark from all three alternatives at once: $\mnulight \approx 19.91$ meV for the benchmark, against $\mnulight \lesssim 1$ meV for each of the others.
This directly impacts the sum of the light neutrino masses: $\sum_i m_i \simeq 0.096$ eV for the benchmark,
close to the current cosmological bounds~\cite{DESI:2024mwx, DESI:2025zgx, DES:2026fyc}, against $\simeq 0.059$ eV for each of the others.
Alternatively, this degeneracy can be broken by the value of the effective Majorana mass, with $3.3$ meV for CP1$'$, $2.5$ meV for CP2 and $2.2$ meV for CP2$'$, against $7.3$ meV for the benchmark itself, within the expected sensitivity of the next generation of experiments $m_{\beta\beta}^{\rm eff} \simeq 5$ meV~\cite{LEGEND:2021bnm, CUPID:2022jlk, nEXO:2021ujk}.

\begin{figure}[htbp]
    \centering
    \includegraphics{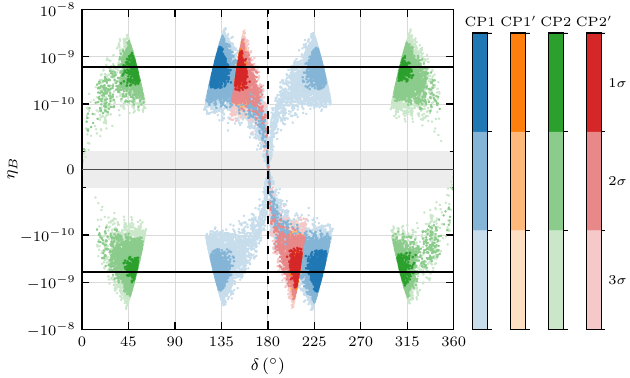}\hfill
    \includegraphics{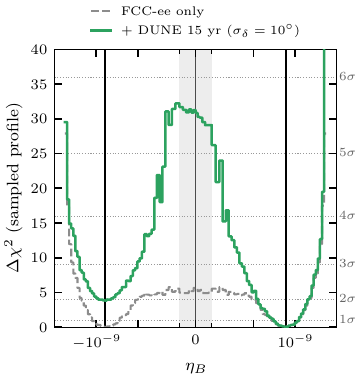}
    \caption{Left: the baryon asymmetry as a function of the Dirac phase, over the parameter
    values compatible with the simulated Tera-Z measurement of the benchmark of
    Tab.~\ref{tab:testable_benchmark}, on a symmetric logarithmic axis that is linear for
    $|\eta_B| < 10^{-11}$. The four degenerate solutions of
    Tab.~\ref{tab:testable_benchmark} are shown separately: the benchmark itself (CP1, blue), the
    second root of the mixings in $\mnulight$ (CP1$'$, orange), and the two solutions of case CP2
    (CP2 in green, CP2$'$ in red), with the shading indicating the $1\sigma$, $2\sigma$ and
    $3\sigma$ regions ($\Delta\chi^2 < 2.30,\,6.18,\,11.83$). Since the squared mixings are even functions of
    $\delta$ while $\eta_B$ is odd, each solution appears twice, at $(\delta,\,\eta_B)$ and at
    $(2\pi - \delta,\,-\eta_B)$. Away from the CP-conserving values the reconstruction confines $|\eta_B|$ to
    within a factor of a few of the observed value; the solutions, however, extend towards
    $\delta = 0$ and $\pi$, where the asymmetry vanishes identically. No
    solutions exist in the gap $0.34 \lesssim |\delta|/\pi \lesssim 0.65$, which the measured
    mixings alone exclude at more than $5\sigma$.
    Right: the profile $\Delta\chi^2$ of $\eta_B$ from the reconstruction alone (dashed), and combined
    with a measurement of $\delta$ with a resolution of $10^\circ$ (solid), as expected after 15
    years of DUNE~\cite{DUNE:2020ypp}. The measurement of $\delta$ removes the path towards
    $\eta_B = 0$ and the mirror solutions at $2\pi-\delta$, leaving only the wrong-sign solution
    at the same value of $\delta$, which the mixings disfavour at just below $2\sigma$.}
    \label{fig:BAUdcpTestable}
\end{figure}

The value of the Dirac phase proves to be a much stronger constraint.
A determination of the Dirac phase removes the CP2 solution, which sits at $|\delta| \simeq 0.26\pi$, some $\sim \pi/2$ away from the benchmark.
In Fig.~\ref{fig:BAUdcpTestable} (left), we explicitly show points consistent with the reconstruction of the HNL parameters based on a simulated measurement at the FCC-ee.
The dependence of the baryon asymmetry on the Dirac phase is weak over most of the allowed range, but $\eta_B$ vanishes at the CP-conserving values $\delta = 0,\,\pm\pi$, and the reconstruction on its own does not exclude them: while the measured mixings already disfavour $|\delta| \simeq \pi/2$ at more than $5\sigma$, the CP-conserving values remain allowed at $\Delta\chi^2 \simeq 4$, where a solution with $\eta_B = 0$ exists.
An independent determination of $\delta$ at experiments such as DUNE~\cite{DUNE:2021mtg} and T2HK~\cite{Hyper-Kamiokande:2025fci} therefore turns the parameter reconstruction into a determination of the magnitude of the baryon asymmetry.
This is shown in Fig.~\ref{fig:BAUdcpTestable} (right), where we include a measurement of $\delta$ with a resolution of $10^\circ$, as expected after 15 years of DUNE~\cite{DUNE:2020ypp}: the reconstruction then confines $|\eta_B|$ to $[0.1,\,5]\,\eta_B^{\rm obs}$ at $3\sigma$.
In contrast to the magnitude, the sign of the BAU is less constrained.
Without information on $\delta$ it is entirely undetermined; once the sign of $\sin\delta$ is known, the best wrong-sign reconstruction lies at $\Delta\chi^2 \approx 3.8$, just below $2\sigma$, and since it sits at the same value of $\delta$ as the benchmark, a more precise measurement of $\delta$ does not improve on it.

\begin{table}[htbp]
\centering
\small
\setlength{\tabcolsep}{4.5pt}
\begin{tabular}{lcc|ccc}
\hline\hline
 & \multicolumn{2}{c|}{\bf CP1 benchmark} & \multicolumn{3}{c}{\bf degenerate solutions}\\
 & {\bf input} & {\bf reconstructed} & {\bf CP1$'$} & {\bf CP2} & {\bf CP2$'$}\\
\hline
$\tilde{M}_{1,2,3}$ [GeV] & \multicolumn{5}{c}{$9,\,10,\,11$}\\
\hline
$y$ & $5.587$ & $[5.574,\,5.610]$ & $[6.3541,\,6.3606]$ & $[6.175,\,6.199]$ & $[6.3042,\,6.3142]$\\
$x_N/\pi$ & $0.2420$ & $[0.2406,\,0.2424]$ & $[0.2401,\,0.2406]$ & $[0.2400,\,0.2431]$ & $[0.2390,\,0.2400]$\\
$x_\nu/\pi$ & $0.2495$ & $[0.2482,\,0.2556]$ & $[0.4025,\,0.4046]$ & $[0.3645,\,0.3711]$ & $[0.3970,\,0.3997]$\\
$|\delta|/\pi$ & $0.720$ & $[0.705,\,0.761]$ & $[0.832,\,0.852]$ & $[0.213,\,0.291]$ & $[0.826,\,0.853]$\\
$\mnulight$ [meV] & $19.91$ & $[18.65,\,20.82]$ & $[0.0285,\,0.0344]$ & $[0.61,\,0.71]$ & $[0.0327,\,0.0420]$\\
\hline
$\sum_i m_i$ [meV] & $95.6$ & $[92.6,\,97.6]$ & $[58.81,\,58.82]$ & $[59.42,\,59.54]$ & $[58.82,\,58.83]$\\
$|(m_\nu)_{ee}|$ [meV] & $7.27$ & $[6.49,\,7.73]$ & $[3.27,\,3.36]$ & $[2.14,\,2.70]$ & $[2.11,\,2.29]$\\
$|\eta_B|/\eta_B^{\rm obs}$ & $1.00$ & $[0.41,\,1.68]$ & $[0.43,\,1.58]$ & $[0.33,\,1.45]$ & $[0.41,\,1.39]$\\
\hline
$\chi^2_{\rm min}$ & $0$ & $0.03$ & $0.31$ & $0.72$ & $0.19$\\
\hline\hline
\end{tabular}
\caption{Parameters of the testable CP1 benchmark, their reconstruction from a simulated
Tera-Z measurement of the nine squared mixings $\wTheta^2_{\alpha j}$, and the three further
solutions that reproduce the same nine mixings. The heavy neutrino masses are taken as measured
from the kinematics of the displaced vertices and are common to all solutions. The quoted
ranges are $68\%$ credible intervals, with the posterior of each solution forming a single
unimodal family in the parameters shown.
Only $|\delta|$ is quoted, since $\delta \to -\delta$ give identical mixings and differ in the sign of the BAU.
CP1$'$ is the second root of the mixings in $\mnulight$ within case CP1, while CP2 and CP2$'$ are the two solutions of case CP2.
All four are indistinguishable from the benchmark at the level of the event rates at FCC-ee,
and all four predict $|\eta_B|$ of the order of the observed value.
They are instead separated by the lightest neutrino mass and by $\delta$.}
\label{tab:testable_benchmark}
\end{table}

Finally, it is worth commenting on testability in the case of highly degenerate HNL masses,
that are often realised in low-scale leptogenesis scenarios.
Even if the mass differences are too small to be directly measured, additional information about $x_N$ can still be contained in the decay distribution of the sterile neutrinos~\cite{Drewes:2024pad}.
Specifically, the decay distribution can differ from a simple exponential law due to the interplay of the three heavy neutrinos with different lifetimes.
This effect is most pronounced in the quasi-degenerate regime, and is therefore complementary to the benchmark discussed above, where the three states are instead resolved kinematically.
The information about the Majorana mass $M_R$ can however be obscured by the large corrections coming from the $R^2$ terms when $\wTheta^2 \gtrsim \Delta \tilde{M}_{jk}/\wMav$.
Another probe worth exploring in this regime is the possibility of directly measuring the RH neutrino oscillations~\cite{Antusch:2017ebe,Cvetic:2018elt,Tastet:2019nqj}.
We leave a detailed study of this limit to future work.

With this, we can conclude that there are direct observables that can lead to the determination of all parameters within this scenario.
A measurement of all nine elements of the mixing matrix $\wTheta^2_{\alpha j}$ could fix all of the continuous parameters, but does not by itself resolve the discrete degeneracies.
Resolving them requires the complementary information of the light neutrino sector --- the sum of the neutrino masses and the rate of $\nubb$-decay --- together with a determination of the Dirac phase, and it is this combination that ultimately makes the scenario testable.

\section{Summary and conclusions}\label{sec:conclusions}
In this work, we have studied the type-I seesaw
with three heavy Majorana neutrinos under the condition that
  the CP-violation in this scenario  originates solely from the
 Dirac phase $\delta$ present in the
  PMNS neutrino mixing matrix.
We have derived a general non-real structure for the orthogonal matrix entering the Casas-Ibarra parameterisation of the neutrino Yukawa couplings -- the Casas-Ibarra matrix --
which respects the CP-symmetry
(Eq.~\eqref{eq:CPconsCI}). The non-real Casas-Ibarra matrix preserving the CP-symmetry can be defined up to permutations of rows and columns, some of which are unphysical, and this ultimately leads to three physically distinct forms of interest of the matrix, CP1, CP2 and CP3
(Eqs.~\eqref{eq:CPconsCI_1}, \eqref{eq:CPconsCI_2}
  and \eqref{eq:CPconsCI_3}). Each of these three forms can be parameterised
uniquely in terms of two real angles, controlling
the mixing between the light and heavy Majorana neutrinos,
and one single purely imaginary parameter
that controls the overall magnitude of the
neutrino Yukawa couplings.

We have then concentrated on the case for which the unique source of CP-violation is the Dirac phase $\delta$. We achieved this by further restricting the Majorana phases to specific CP-conserving values and avoiding the peculiar situation for which one can have CP-violation even in presence of CP-conserving Casas-Ibarra and PMNS matrices \cite{Pascoli:2006ie,Pascoli:2006ci}.
The values of the two Majorana phases, $\alpha_{21}$ and
$\alpha_{31}$, corresponding to the three CP-conserving non-real forms of the CI matrix are the following: $\alpha_{21}=\pm\pi$, $\alpha_{31}=\pm \pi$ for CP1;
$\alpha_{21}= \pm\pi$, $\alpha_{31} = 0,\pm2\pi$ for CP2;
$\alpha_{21}=0,\pm 2\pi$, $\alpha_{31} = \pm \pi$ for CP3. Under these conditions, all the CP-violating phenomena in the type-I seesaw model, including the generation of a baryon asymmetry of the Universe through leptogenesis and CP-violation in flavour neutrino oscillations, are connected through $\delta$, and are not present if $\delta$ takes CP-conserving values, i.e.~$\delta = 0,\,\pi,\,2\pi$. 
   
We have studied the parameter space related to heavy neutrino searches
in the considered scenario,
focusing on the heavy Majorana neutrino mass range
$\mathcal{O}(0.1-100)\,\text{GeV}$. We have demonstrated that, when
the squared mixing $\Theta^2$ of the heavy Majorana
neutrinos are sufficiently large, where
$\Theta^2 = \sum_{\alpha=e,\mu,\tau}\sum_{j=1,2,3} \Theta^2_{\alpha j}$,
$\Theta_{\alpha j}$ being the coupling of the neutrino $N_j$
to the charged lepton $\alpha = e,\,\mu,\,\tau$
(flavour neutrino $\nu_\alpha$) in the weak
charged (neutral) current, the hypothesis of low-energy Dirac-phase CP-violation can be
tested and potentially falsified at future experiments
(e.g.,~SHiP and FCC-ee/CEPC). This hypothesis, in particualr,
leads to a distinct subregions of the full flavour triangle identified by the ratios $\Theta^2_e/\Theta^2$, $\Theta^2_\mu/\Theta^2$ and $\Theta^2_\tau/\Theta^2$, and can thus be falsified against the scenario where additional CP-violation
originates from the Casas-Ibarra matrix and/or the Majorana phases. The flavour structures in  the studied scenario depend on the neutrino mass ordering and the value of the lightest neutrino mass $\mnulight$ (Fig.~\ref{fig:ternaryplots_H} and Fig.~\ref{fig:ternaryplots_mnu}).
Thus, the tests discussed by us would benefit from
independent information on both the ordering and $\mnulight$.

We have also discussed the possibility to test the CP1, CP2 and CP3 scenarios with Dirac-phase CP-violation and non-real CI matrix at neutrinoless double-beta decay
experiments. If the heavy Majorana neutrino contribution
  to the neutrinoless double-beta decay effective Majorana mass, $m_{\beta \beta}^{\rm eff}$, is negligible, for given light neutrino mass ordering
  and value of the lightest neutrino mass, we get very
  specific predictions for $m_{\beta \beta}^{\rm eff}\simeq |(m_\nu)_{ee}|$, that, in principle, e.g.,
  in the CP3 case, can be tested experimentally (Fig.~\ref{fig:0nubb}). However, with a non-negligible contribution of the heavy Majorana neutrinos to $m_{\beta \beta}^{\rm eff}$ the analysis is more involved.

Lastly, we have discussed the implications of the Dirac-phase CP-violation condition for low-scale leptogenesis. Our numerical analysis shows that leptogenesis is viable when $\delta$ is the only source of CP-violation within the entire region that is testable at direct heavy neutral lepton searches (Figs.~\ref{fig:CP1s_scan_NH} and \ref{fig:CP1s_scan_IH_QD} for CP1, Figs.~\ref{fig:CP2_CP3_scan_NH}, \ref{fig:CP2_CP3_scan_IH} and \ref{fig:CP2_CP3_scan_QD} for CP2 and CP3 in Appendix \ref{app:otherscans}), as well as in future experiments looking for charged lepton flavour violating processes involving muons (Fig.~\ref{fig:cLFV_scan}). Remarkably, we find that one can have viable leptogenesis within the testable region even in the case of deviations of $\delta$ from the CP-conserving values 0, $\pi$, $2\pi$ that are as small as or smaller than $\sim 10^{-5}$ (Fig.~\ref{fig:tauzero_scan}). This has significant implications, in particular, for theories of flavour where the CP-symmetry is realised only approximately.

The testability of the scenario relies on the complementarity of direct HNL searches with neutrino experiments. Measurements of the heavy-neutrino masses, mixings and flavour structure constrain the parameters but leave degeneracies that can be resolved by combining heavy-neutrino searches with independent measurements of $\delta$ and/or $\mnulight$ and/or $m_{\beta\beta}^{\rm eff}$, as we demonstrated for specific benchmarks. This synergy provides a quantitative test of the hypothesis that the Dirac phase $\delta$ is the unique source of leptonic CP violation in the type-I seesaw and the sole CP-violating phase responsible for the baryon asymmetry.  
Finally, our work motivates ultraviolet completions of the type-I seesaw in which low-energy CP violation emerges from first principles.

\section*{Acknowledgements}
We thank Dhruv Pasari for improving considerably the speed of the version of the \texttt{ULYSSES} code used in this work \cite{Granelli:2026goh}, allowing for much more rapid numerical scans. We also thank Jacobo López-Pavón for useful clarifications on the global analysis in \cite{Blennow:2023mqx}. A.G.~acknowledges the use of computational resources from the parallel computing cluster of the \href{https://site.unibo.it/openphysicshub/en}{Open Physics Hub} at the Department of Physics and Astronomy of the University of Bologna, Italy.
A.G.~is supported by the Spanish grant PID2023-148162NB-C21 (MCIN/AEI/10.13039/501100011033), co-funded by the European Union (FEDER).
The work of S.T.P.~was supported 
by the Italian INFN program on Theoretical Astroparticle Physics and by the World Premier
International Research Center Initiative (WPI Initiative, MEXT), Japan.
A.G. and S.T.P. are also supported in part by the 
European Union's Horizon Europe programme under Marie Skłodowska-Curie Actions – Staff Exchanges (SE) grant agreement No.~101086085-ASYMMETRY. 
J.K. acknowledges the use of the Supek supercomputer at the University of Zagreb computing centre SRCE. J.K. is supported by the HRZZ grant UIP-2025-02-3909 and the Ministry of Science, Education and Youth (MZOM) Multilateral Scientific and Technological Cooperation in the Danube Region project ``Dark matter, neutrinos and beyond'' (DA$\nu$BE) KLASA: 910-06/25-01/00041.

\appendix

\section{Equivalence between two parameterisations of complex orthogonal matrices}\label{app:ConnectingParams}
For completeness, we demonstrate here that the ordinary parameterisation for an orthogonal complex matrix as in Eq.~\eqref{eq:Euler_O} can always be mapped into the alternative parameterisation given in Eq.~\eqref{eq:alt_O}.
We start by noting that, for a matrix $O$ written as in Eq.~\eqref{eq:alt_O}, we have the following relations:
\begin{eqnarray}
\label{eq:realRotFinder1}
    \myRe ( O O^\dagger ) &=& R_\nu \myRe[ R^{(12)}(iy)R^{(12)^\dagger}(iy)] R_\nu^T\,,\\
    \label{eq:realRotFinder2}
    \myRe ( O^\dagger O ) &=& R_N^T \myRe[R^{(12)^\dagger}(iy) R^{(12)}(iy)] R_N\,,
\end{eqnarray}
where $R_N = R^{(12)}(x)R^{(23)}(x_1^N)R^{(13)}(x_2^N)$ and $R_\nu = R^{(13)}(x_2^\nu)R^{(23)}(x_1^\nu)$ are real orthogonal matrices, while the middle matrix $\myRe[ R^{(12)}(iy)R^{(12)^\dagger}(iy)]= \myRe[R^{(12)^\dagger}(iy) R^{(12)}(iy))] $
is real and diagonal.
The two relations above suggest a procedure to find the sets of angles that parameterise any complex orthogonal matrix $O$ in the form of Eq.~\eqref{eq:alt_O}, namely diagonalising the products $\myRe (O O^\dagger)$ and $\myRe (O^\dagger O)$ to find the matrices $R_N$ and $R_\nu$ that allow to write $O = R_\nu(x_1^\nu, x_2^\nu,x) R^{(12)}(iy) R_N(x_1^N, x_2^N)$.

We now show that any complex orthogonal matrix $O$ parameterised according to Eq.~\eqref{eq:Euler_O} can be written in the form of Eq.~\eqref{eq:alt_O}. We proceed sequentially from the product 
\begin{align}
  R_1^{(23)} R_2^{(13)} =
  R^{(23)}(x_1) R^{(23)}(iy_1) R^{(13)}(iy_2) R^{(13)}(x_2)\,,
\end{align}
where we recall that $R_l^{(jk)} \equiv R^{(jk)}(x_l+iy_l)$, and we have decomposed each complex rotation into the associated real and purely imaginary rotations.\footnote{As can be easily checked, real and purely imaginary rotations in the same plane commute, i.e.~$[R^{(jk)}(x),R^{(jk)}(iy)] = 0$ for given $j,k = 1,2,3$, $j\neq k$, and $x,y$ real.} The intermediate product $R^{(23)}(iy_1) R^{(13)}(iy_2)$ can further be decomposed as $R^{(23)}(iy_1) R^{(13)}(iy_2) = R_A R^{(23)}(iy'_1)R_B$, with $R_A$ and $R_B$ being two real orthogonal matrices and $y'_1$ a real function of $y_1$ and $y_2$. This can be checked explicitly by finding $R_A$ and $R_B$ as the two real orthogonal matrices that diagonalise respectively $$\myRe[R^{(23)}(iy_1) R^{(13)}(iy_2)R^{(13)^\dagger}(iy_2) R^{(23)^\dagger}(iy_1)]\text{ and }\myRe[R^{(13)^\dagger}(iy_2) R^{(23)^\dagger}(iy_1)R^{(23)}(iy_1) R^{(13)}(iy_2)]$$
and defining $y_1'$ such that $\cosh y'_1 = \cosh y_1\cosh y_2$ and  $\sinh y'_1 = \text{sign}(y_1)\sqrt{\cosh^2 y'_1-1}$.

Then, a matrix $O$ parameterised according to Eq.~\eqref{eq:Euler_O} can be rewritten as:
\begin{align}
   O = R_1^{(23)}R_2^{(13)} R_3^{(12)} &=
   R^{(23)}(x_1) R_A R^{(23)}(iy'_1) R_B R^{(13)}(x_2) R^{(12)}(iy_3) R^{(12)}(x_3)\,,\\\notag
   &=
   R'_A R^{(23)}(iy'_1) R^{(13)}(\tilde{x}_2) R^{(12)}(iy_3)R^{(12)}(\tilde{x}_3)\,,
\end{align}
where $R'_A$ is a generic real orthogonal matrix, while $\tilde{x}_2$ and $\tilde{x}_3$ are two additional real angles. In the last step, we have used the fact that the product of two real orthogonal matrices is again real and orthogonal, 
and that $[R^{(jk)}(x),R^{(jk)}(iy)] = 0$ for given $x,y$ real.  In particular, $R_BR^{(13)}(x_2) = R^{(23)}(x'_1)R^{(13)}(\tilde{x}_2) R^{(12)}(x'_3)$, $R^{(12)}(\tilde{x}_3) = R^{(12)}(x'_3)R^{(12)}(x_3)$ and $R_A' = R^{(23)}(x_1)R_AR^{(23)}(x'_1)$, with $x'_1$ and $x'_3$ real. 
To complete the proof it is then only necessary to find the parameterisation of
$R^{(23)}(iy'_1) R^{(13)}(\tilde{x}_2) R^{(12)}(iy_3)$
in terms of a single complex rotation. Such decomposition can always be performed as the matrices $$A = \text{Re}[R^{(23)}(iy'_1) R^{(13)}(\tilde{x}_2) R^{(12)}(iy_3)R^{(12)^\dagger}(iy_3) R^{(13)^\dagger}(\tilde{x}_2) R^{(23)^\dagger}(iy'_1)]$$
and
$$
B = \text{Re}[R^{(12)^\dagger}(iy_3) R^{(13)^\dagger}(\tilde{x}_2) R^{(23)^\dagger}(iy'_1)R^{(23)}(iy'_1) R^{(13)}(\tilde{x}_2) R^{(12)}(iy_3)]
$$
are real and symmetric. Thus, by virtue of the spectral theorem, they admit the decomposition $A = \tilde{R}_AD_A\tilde{R}_A^T$ and $B =\tilde{R}_B D_B \tilde{R}_B^T$, with $\tilde{R}_A$ and $\tilde{R}_B$ two real orthogonal matrices, and $D_A$ and $D_B$ real and diagonal. By direct computation we find that $A$ and $B$ share the same set of eigenvalues $\{1, a, a\}$, where $a\geq 1$ is a function of the angles $y'_1, \tilde{x}_2, y_3$, which we do not report here for brevity.
Therefore, the matrices $D_A$ and $D_B$ can be put in the form $D \equiv D_A = D_B = \text{diag}(a, a, 1)$. By defining the angle $\tilde{y}_3 = \text{arcosh}\left[(a+1)/2\right]^{1/2}$, we can write $D = \text{Re}[R^{(12)}(i\tilde{y}_3)R^{(12)^\dagger}(i\tilde{y}_3)]$ so that $R^{(23)}(iy'_1) R^{(13)}(\tilde{x}_2) R^{(12)}(iy_3) = \tilde{R}_A R^{(12)}(i\tilde{y}_3) \tilde{R}_B$, concluding the proof.

\section{Derivation of the CP-conserving non-real Casas-Ibarra matrix}\label{app:CPonCI_explicit}
We present here how the CP-conserving form of the CI matrix as given in Eq.~\eqref{eq:CPconsCI} can be derived. We start by noting that, evidently, once a specific CP-conserving form of the CI matrix is derived, other forms that satisfy the condition of CP-conservation can be obtained by permuting rows and/or columns, and/or by changing the signs of entire rows and/or columns (also by transposition). 
That is, if $O_{\rm CP}$ is CP-conserving, then also $O_{\rm CP}^T$
and $P_{\rm rows} O_{\rm CP} P_{\rm columns}$ are CP-conserving, where $P_{\rm rows}$ and $P_{\rm columns}$ are generic signed permutation matrices, having exactly one entry equal to $\pm 1$ in each row and each column and all other entries zero. \footnote{Matrices defined up to signed permutations of rows and columns (and transposition)
are equivalent, and we denote this equivalence with $\sim$.} 
It is thus enough to find the general form of a CP-conserving CI matrix up to transposition and/or 
signed permutations of rows and columns. 

As we want to keep $y\neq0$ to allow for sizeable couplings when $|y|\gg 1$, the condition of CP-conservation imposes constraints on the real angles $x_1^\nu$, $x_2^\nu$, $x_1^N$, $x_2^N$ and $x_3^N$. \textit{A posteriori}, it is necessary to fix some of them to the discrete values $0,\pi,\pi/2$ or $3\pi/2$. We find it instructive to understand how the parameterisation in Eq.~\eqref{eq:alt_O} gets simplified under the assumption that one or more of these angles take such fixed values. \begin{itemize}
    \item If, e.g., $x_2^\nu = 0,\,\pi/2,\,\pi,\,3\pi/2$, then $R^{(13)}(x_2^\nu) \sim P_{\rm rows}$ and \begin{equation}
 O \sim P_{\rm rows}R^{(23)}(x_1^\nu) R^{(12)}(iy) R^{(12)}(x_3^N) R^{(23)}(x_1^N)R^{(13)}(x_2^N).
\end{equation}
Similarly, if $x_2^N = 0,\,\pi/2,\,\pi,\,3\pi/2$, then $R^{(13)}(x_2^N) \sim P_{\rm columns}$ and
\begin{equation}
 O \sim R^{(13)}(x_2^\nu)R^{(23)}(x_1^\nu) R^{(12)}(iy) R^{(12)}(x_3^N) R^{(23)}(x_1^N)P_{\rm columns}.
 \end{equation}
\item If instead $x_1^\nu = 0$ ($\pi$, after redefining $x_2^\nu\to \pi-x_2^\nu$,) we can write $
     R^{(13)}(x_2^\nu) R^{(23)}(x_1^\nu) \sim P_{\rm rows} R^{(13)}(x_2^\nu)$ and thus
\begin{equation}
 O \sim P_{\rm rows}R^{(13)}(x_2^\nu) R^{(12)}(iy) R^{(12)}(x_3^N) R^{(23)}(x_1^N)R^{(13)}(x_2^N).
\end{equation}
Analogously, if $x_1^N = 0$ ($\pi$, after redefining $x_2^N\to \pi-x_2^N$,) $R^{(23)}(x_1^N) R^{(13)}(x_2^N) \sim R^{(13)}(x_2^N)P_{\rm columns}$ so that
\begin{equation}
 O \sim R^{(13)}(x_2^\nu)R^{(23)}(x_1^\nu) R^{(12)}(iy) R^{(12)}(x_3^N) R^{(13)}(x_2^N)P_{\rm columns}.
\end{equation}
\item Fixing $x_1^\nu = 3\pi/2$ ($\pi/2$, after redefining $x_2^\nu\to \pi-x_2^\nu$,) we get
\begin{equation}
    O \sim P_{\rm rows} R^{(12)}(x_2^\nu+x_3^N) R^{(12)}(iy)(x_1^N)R^{(13)}(x_2^N)R^{(23)}(x_1^N),
    \end{equation}
since $R^{(12)}(iy)R^{(12)}(x_3^N) = R^{(12)}(x_3^N)R^{(12)}(iy)$ and $R^{(12)}(x_2^\nu) R^{(12)}(x_3^N) = R^{(12)}(x_2^\nu+x_3^N)$. Analogously, when $x_1^N = \pi/2$ ($3\pi/2$, after redefining $x_2^N\to \pi-x_2^N$,) we get
    \begin{equation}
    O \sim  R^{(13)}(x_2^\nu)R^{(23)}(x_1^\nu) R^{(12)}(iy)R^{(12)}(x_3^N+x_2^N)P_{\rm columns}.
    \end{equation}
\item If $x_3^N = 0$ ($\pi$, after proper redefinition of the angles $x_1^N \to -x_1^N$ and $x_2^N \to -x_2^N$,) we get 
\begin{equation}
    O \sim  R^{(13)}(x_2^\nu)R^{(23)}(x_1^\nu) R^{(12)}(iy)R^{(23)}(x_1^N) R^{(13)}(x_2^N)P_{\rm columns}.
    \end{equation}
\item Finally, if
$x_3^N = \pi/2$ (or $3\pi/2$), after the proper redefinition $x_2^N\to -x_2^N$ (or 
 $x_1^N\to -x_1^N$), we get $R^{(12)}(x_3^N)R^{(23)}(x_2^N)R^{(13)}(x_1^N) \sim R^{(13)}(x_2^N)R^{(23)}(x_1^N)P_{\rm columns}$ and thus
 \begin{equation}
     O \sim  R^{(13)}(x_2^\nu)R^{(23)}(x_1^\nu) R^{(12)}(iy)R^{(13)}(x_1^N) R^{(23)}(x_1^N)P_{\rm columns}.
 \end{equation}
 Alternatively, for $x_3^N = \pi/2$ (or $3\pi/2$), after redefining $x_2^\nu\to \pi-x_2^N$ (or 
 $x_1^\nu\to \pi-x_1^N$), we also can rewrite
  \begin{equation}
     O \sim  P_{\rm rows}R^{(23)}(x_2^\nu)R^{(13)}(x_1^\nu) R^{(12)}(iy)R^{(23)}(x_1^N) R^{(13)}(x_1^N).
 \end{equation}
 \end{itemize}
 The above relations will prove useful in the following analysis.

\subsection{Explicit derivation}
Before performing an explicit investigation, we start with an ansatz and consider the following form of the CI matrix:
\begin{equation}\label{eq:O_ansatz}
    O = P_\text{rows} R^{(jk)}(x_\nu) R^{(12)}(iy) R^{(lm)}(x_N) P_\text{columns},
\end{equation}
with $x_\nu$ and $x_N$ real angles and $(jk),\,(lm) \in \{(12),\,(13),\,(23)\}$. We want to find the combinations of $(jk)$ and $(lm)$ that lead to CP-conservation. If $(jk)=(lm)=(12)$, then the only way to have a CP-conserving $O$ would be $R^{(12)}(x_\nu) = [R^{(12)}(x_N)]^T$, which would eventually lead to $O = P_\text{rows} R^{(12)}(iy) P_\text{columns}$. By direct investigation one can check that if $(jk)\neq (lm)$, CP-conservation cannot hold for any value of $x_\nu$, $x_N$ and $y$. Thus, only the combination $(jk)=(lm)=(13)$ or $(jk)=(lm)=(23)$ allows for a CP-conserving CI matrix. However, the two cases are related by permutations of rows and columns. Hence, a general parameterisation for a CP-conserving CI matrix as in Eq.~\eqref{eq:O_ansatz} is given by
\begin{equation}\label{eq:OCP_general}
    O_{\rm CP} =P_\text{rows} R^{(23)}(x_\nu)R^{(12)}(iy) R^{(23)}(x_N)P_\text{columns},
\end{equation}
which eventually leads to the form given in Eq.~\eqref{eq:CPconsCI}, or its transpose, after proper redefinitions of the angles, and signed permutations of rows and columns. 

We now show that the above form is 
the general CP-conserving non-real CI matrix within the adopted parameterisation.
By fully expanding the CI matrix in Eq.~\eqref{eq:alt_O}, we find that the second element in the diagonal reads
\begin{equation}\label{eq:O22}
            O_{22} = 
        \cnuu \cNt \cNu \ch_y - \snuu \sNu -i\,\cnuu \cNu \sNt \sh_y.
\end{equation}
We subdivide the study into two cases, depending on whether $O_{22}$ is real or purely imaginary.

%%%%%%%%%%%%%%%%%%%%%%%    CASE 1    %%%%%%%%%%%%%%%%%%%%%%%
\begin{itemize} 
\item[{\bf 1)}] \underline{$\Re(O_{22})=0$.}

The condition of $O_{22}$, as given in Eq.~\eqref{eq:O22}, being purely imaginary for any $y\neq 0$ imposes that
$\sNu \snuu = 0$ and $\cNt \cNu \cnuu = 0$ must be valid at the same time. There are four combinations of the angles
that we analyse below.

%%%%%%%%%%%    CASE 1.1    %%%%%%%%%%%
\begin{enumerate}
\item[{\bf 1.1)}] \underline{$\tNu = 0,\,\pi$ and $\tnuu = \pi/2,\,3\pi/2$.}

In this case we have that $O\sim P_{\rm rows} R^{(12)}(x_2^\nu + x_3^N) R^{(12)}(iy)R^{(13)}(x_2^N)P_{\rm columns}$. To get CP-conservation we must have that $x_2^\nu + x_3^N = n\pi/2$, $n=0,1,2,...$, leading to $O \sim P_{\rm rows} R^{(12)}(iy)R^{(13)}(x_2^N)P_{\rm columns}$, which is CP-conserving as a sub-case of the form in Eq.~\eqref{eq:OCP_general}.

%%%%%%%%%%%    CASE 1.2    %%%%%%%%%%%
\item[{\bf 1.2)}] \underline{$\tnuu = 0,\,\pi$ and $\tNu = \pi/2,\,3\pi/2$.}

In this case we have that $O\sim P_{\rm rows} R^{(13)}(x_2^\nu) R^{(12)}(iy)R^{(12)}(x_2^N+x_3^N)P_{\rm columns}$. To get CP-conservation we must have that $x_2^N + x_3^N = n\pi/2$, $n=0,1,2,...$, leading to $O \sim P_{\rm rows} R^{(13)}(x_2^\nu)R^{(12)}(iy)P_{\rm columns}$, which is CP-conserving, again as a sub-case of the form in Eq.~\eqref{eq:OCP_general}.

%%%%%%%%%%%    CASE 1.3    %%%%%%%%%%%
\item[{\bf 1.3)}] \underline{$\tNu = 0,\,\pi$ and $\tNt = \pi/2,\,3\pi/2$.}

In this case we have $O\sim R^{(13)}(x_2^\nu) R^{(23)}(x_1^\nu)R^{(12)}(iy)R^{(23)}(x_2^N) P_{\rm columns}$. Taking a closer look at the first row, up to permutations, one element is
\begin{equation}
O_{1\smallerbullet} \sim \cnud \ch_y + i \snud \snuu \sh_y.
\end{equation}
The CP symmetry can be preserved if $\tnud = 0,\,\pi/2,\,\pi,\,3\pi/2$ while $\tnuu$ and $\tNd$ are free, implying 
\begin{equation}
O\sim P_{\rm rows} R^{(23)}(x_1^\nu)R^{(12)}(iy)R^{(23)}(x_2^N)P_{\rm columns},
\end{equation}
equivalent to that in Eq.~\eqref{eq:OCP_general}. Alternatively, one could have $\tnuu=0,\,\pi$. Then, $O\sim P_{\rm rows}R^{(13)}(x_2^\nu)R^{(12)}(iy)R^{(23)}(x_2^N) P_{\rm columns}$ and the CI matrix is CP-conserving if additionally
$\tNd=0,\,\pi/2,\,\pi,\,3\pi/2$ while $\tnud$ is free, with $O\sim P_{\rm rows}R^{(13)}(x_2^\nu)R^{(12)}(iy) P_{\rm columns}$,
or $\tnud = 0,\,\pi/2,\,\pi,\,3\pi/2$ while $\tNd$ is free, with $O\sim P_{\rm rows}R^{(12)}(iy)R^{(23)}(x_2^N) P_{\rm columns}$, both being sub-cases of the CI matrix in Eq.~\eqref{eq:OCP_general}.

%%%%%%%%%%%    CASE 1.4    %%%%%%%%%%%
\item[{\bf 1.4)}] \underline{$\tnuu = 0,\,\pi$ and $\tNt = \pi/2,\,3\pi/2$.}

In this case we have $O\sim P_{\rm rows} R^{(23)}(x_2^\nu)R^{(12)}(iy)R^{(23)}(x_1^N)R^{(13)}(x_2^N)P_{\rm columns}$, which is analogous to the previous case 1.3, but with the re-labelling $\tNu \leftrightarrow \tnuu$, $\tNd \leftrightarrow \tnud$.

\end{enumerate}

%%%%%%%%%%%%%%%%%%%%%%%    CASE 2    %%%%%%%%%%%%%%%%%%%%%%%

\item[{\bf 2)}] \underline{$\Im(O_{22})=0$.}

The reality condition of $O_{22}$, as given in Eq.~\eqref{eq:O22}, for any $y\neq 0$ imposes that
$\cNu \cnuu \sNt = 0$. There are three combinations of the angles allowing for this.
We separately analyse those possibilities in what follows.

\begin{enumerate}

%%%%%%%%%%%    CASE 2.1    %%%%%%%%%%%
\item[{\bf 2.1)}] \underline{$\tNu = \pi/2, \,3\pi/2$.}

In this case we have $O \sim  R^{(13)}(x_2^\nu)R^{(23)}(x_1^\nu) R^{(12)}(iy)R^{(12)}(x_3^N+x_2^N)P_{\rm columns}$. Up to signed permutations, in the second row we have one the element that reads:
\begin{equation}
O_{2\smallerbullet} \sim -\cnuu [\sin(x^N_3+x_2^N) \ch_y-i \cos(x^N_3+x_2^N) \sh_y],
\end{equation}
Evidently, it vanishes if $x_1^\nu = \pi/2,\,3\pi/2$, so that $O\sim  P_{\rm rows} R^{(12)}(iy)R^{(12)}(x_3^N+x_2^N+x_2^\nu)P_{\rm columns}$, but this form of the CI matrix is CP-conserving only if $x_3^N+x_2^N+x_2^\nu = n\pi/2$, implying a CP-conserving form of the CI matrix $O\sim P_{\rm rows} R^{(12)}(iy)P_{\rm columns}$, trivially obtainable from Eq.~\eqref{eq:OCP_general}.

Alternatively, $O_{21}$ is either real or purely imaginary if $x_3^N+x_2^N = n\pi/2$, so that $O\sim P_{\rm rows}R^{(13)}(x_2^\nu)R^{(23)}(x_1^\nu) R^{(12)}(iy)P_{\rm columns}$. Then either $x_1^\nu = 0,\,\pi$ and $O\sim P_{\rm rows}R^{(13)}(x_2^\nu) R^{(12)}(iy)P_{\rm columns}$, or $x_2^\nu = n\pi/2$ and $O\sim R^{(23)}(x_1^\nu) R^{(12)}(iy)P_{\rm columns}$, which are two sub-cases of the form in Eq.~\eqref{eq:OCP_general}.

%%%%%%%%%%%    CASE 2.2    %%%%%%%%%%%
\item[{\bf 2.2)}] \underline{$\tnuu = \pi/2, \,3\pi/2$.} 

Under such condition we have $O \sim P_{\rm rows} R^{(12)}(x_3^N+x_2^\nu) R^{(12)}(iy)R^{(23)}(x_1^N)R^{(13)}(x_2^N)$, and this case is equivalent to the previous one, as they are related by a transposition and redefinition of the angles.

%%%%%%%%%%%    CASE 2.3    %%%%%%%%%%%
\item[{\bf 2.3)}] \underline{$\tNt = 0,\,\pi$.}

In this case we have $O \sim  R^{(13)}(x_2^\nu)R^{(23)}(x_1^\nu) R^{(12)}(iy)R^{(23)}(x_1^N) R^{(13)}(x_2^N)P_{\rm columns}$, and
\begin{equation}
    O_{2\smallerbullet} \sim -\sNd(\cNu\snuu + \cnuu\sNu\ch_y)-i\cNd\cnuu\sh_y.
\end{equation}
If $\tNd = 0,\,\pi$ or $
\pi/2, \,3\pi/2$, so that either $\Re(O_{2\smallerbullet})=0$ or $\Im(O_{2\smallerbullet})=0$, 
we get $O \sim  R^{(13)}(x_2^\nu)R^{(23)}(x_1^\nu) R^{(12)}(iy)R^{(23)}(x_1^N) P_{\rm columns}$ which we have already encountered in case 1.3 and thus will not repeat the analysis. 
If instead we take simultaneously $x_1^N = 0,\,\pi$ and $x_1^\nu = 0,\,\pi$ 
then $\Re(O_{2\smallerbullet})=0$ 
and 
\begin{equation}
    O \sim  P_{\rm rows} R^{(13)}(x_2^\nu)R^{(12)}(iy)R^{(13)}(x_2^N) P_{\rm columns}
    \end{equation}
    which is equivalent to the CP-conserving CI matrix given in Eq.~\eqref{eq:OCP_general}. 
Finally, the last condition to consider is when $x_1^\nu= \pi/2,\,3\pi/2$, so that $\Im(O_{21})=0$, and  $O \sim P_{\rm rows}R^{(12)}(x_2^\nu)R^{(12)}(iy)R^{(23)}(x_1^N)R^{(13)}(x_2^N)P_{\rm columns}$, but the discussion becomes analogous to the previously analysed cases 2.1 and 2.2.

\end{enumerate}

\end{itemize}
 
We have explicitly verified that all the possible restrictions imposed by Eq.~\eqref{eq:CP_O} on the parameters of the CI matrix written as in Eq.~\eqref{eq:alt_O} for any $y \neq 0$ either lead to a matrix that is equivalent to the one in Eq.~\eqref{eq:OCP_general}, or can easily be obtained from the latter. We can thus adopt the parameterisation in Eq.~\eqref{eq:OCP_general} without loss of generality.

%=====================================
\section{Squared-coupling combinations with Dirac-phase CP-violation}\label{app:Theta2_Expr}
We present here the analytical expressions for $\wTheta^2_{1,2,3}$, $\wTheta^2_{e,\mu,\tau}$, and $\wTheta^2$ under the CP-conserving Casas-Ibarra parameterisations given in Eqs.~\eqref{eq:CPconsCI_1}-\eqref{eq:CPconsCI_3}.
\begin{itemize}
    \item{\bf CP1)}
\begin{eqnarray}
 \wMav \wTheta^2_e &=&  m_1 c_{12}^2 c_{13}^2(\sh_y^2+\ch_y^2)+m_2 s_{12}^2c_{13}^2\left[\ch_y^2-  (c_\nu^2-s_\nu^2)\,\sh_y^2\right]+ 
 \\
 \nonumber
 &&+\,m_3 s_{13}^2\left[\ch_y^2+ (c_\nu^2-s_\nu^2)\,\sh_y^2\right]+\\
 \nonumber
 && - \,4\sqrt{m_1}c_{12}c_{13} \left[\sqrt{m_2} s_{12}c_{13}  \sin\left(\frac{\alpha_{21}}{2}\right)\,s_\nu+\sqrt{m_3}  s_{13}  \sin \left(\frac{\alpha_{31}}{2}-\delta \right)\,c_\nu\right]\,\sh_y \ch_y+\\
 \nonumber
 &&+\,4 \sqrt{m_2 m_3}s_{12} s_{13} c_{13}  \cos\left(\frac{\alpha_{23}}{2}+\delta \right)\, s_\nu c_\nu\,\sh_y^2;\\
 %%%%%%%%%%
 \quad\wMav \wTheta^2_\mu &=& m_1(c_{12}^2 s_{23}^2s_{13}^2 +s_{12}^2c_{23}^2 +2 s_{12}c_{12} s_{23}c_{23}  s_{13}  \cos\delta)(\ch_y^2+\sh_y^2) +\\
 \nonumber
 && +\, m_2 (c_{12}^2 c_{23}^2+s_{12}^2 s_{23}^2s_{13}^2-2 s_{12}c_{12} s_{23}c_{23}  s_{13}  \cos \delta)\left[\ch_y^2-(c_\nu^2-s_\nu^2)\,\sh_y^2\right]+\\
 \nonumber
 && +\,m_3s_{23}^2 c_{13}^2 \left[\ch_y^2+ (c_\nu^2-s_\nu^2)\,\sh_y^2\right]+\\
 \nonumber
 && +\,4\sqrt{m_1 m_2}\Big\{\sin\left(\frac{\alpha_{21}}{2}\right)\left[-s_{12}c_{12}s_{23}^2s_{13}^2+s_{12}c_{12}c_{23}^2 + (c_{12}^2-s_{12}^2)s_{23}c_{23}s_{13}\cos\delta\right]+\\
 \nonumber
 && \qquad~~~~~~~~~~~~-\cos\left(\frac{\alpha_{21}}{2}\right)s_{23}c_{23} s_{13}  \sin\delta\Big\}\,s_\nu\ch_y \sh_y+\\
 \nonumber
 &&+ \,4 \sqrt{m_1 m_3}s_{23}c_{13}\Big[\sin\left(\frac{\alpha_{31}}{2}\right)\left(s_{12}c_{23}+c_{12}s_{23}s_{13}\cos\delta\right)+\\
 \nonumber
 &&\qquad~~~~~~~~~~~~~~~~~~-\cos\left(\frac{\alpha_{31}}{2}\right)c_{12}s_{23}s_{13}\sin\delta\Big]\,c_\nu\, \sh_y \ch_y+\\
 \nonumber
 &&-\,4\sqrt{m_2 m_3} s_{23}c_{13}\Big[-\sin\left(\frac{\alpha_{23}}{2}\right) s_{12}s_{23}s_{13}\sin\delta+\\
  \nonumber
  &&\qquad~~~~~~~~~~~~~~~~~~+\cos\left(\frac{\alpha_{23}}{2}\right)(-c_{12}c_{23} +s_{12}s_{23}s_{13}\cos\delta)\Big]\,s_\nu c_\nu\,\sh_y^2;\\
%%%%%%%%%%
 \quad\wMav \wTheta^2_\tau &=& m_1(c_{12}^2 c_{23}^2 s_{13}^2 + s_{12}^2 s_{23}^2-2 s_{12}c_{12} s_{23}c_{23}  s_{13}  \cos\delta)(\ch_y^2+\sh_y^2) +\\
  \nonumber
  && +\, m_2 (c_{12}^2 s_{23}^2 + s_{12}^2c_{23}^2  s_{13}^2  + 2s_{12} c_{12} s_{23}c_{23}  s_{13}  \cos\delta)\left[\ch_y^2-(c_\nu^2-s_\nu^2)\,\sh_y^2\right]+\\
  \nonumber
 && +\,m_3c_{23}^2 c_{13}^2 \left[\ch_y^2+ (c_\nu^2-s_\nu^2)\,\sh_y^2\right]+\\
 \nonumber
 && +\,4\sqrt{m_1 m_2}\Big\{\sin\left(\frac{\alpha_{21}}{2}\right)\left[-s_{12}c_{12} c_{23}^2 s_{13}^2+s_{12} c_{12} s_{23}^2 - (c_{12}^2-s_{12}^2)s_{23}c_{23} s_{13}  \cos\delta\right]+\\
  \nonumber
 && \qquad~~~~~~~~~~~~+\cos\left(\frac{\alpha_{21}}{2}\right)s_{23}c_{23} s_{13}  \sin\delta\Big\}\,s_\nu\ch_y \sh_y+\\
 \nonumber
 &&+ \,4 \sqrt{m_1 m_3}c_{23}c_{13}\Big[\sin\left(\frac{\alpha_{31}}{2}\right)\left(-s_{12}s_{23}+c_{12}c_{23}s_{13}\cos\delta\right)+\\
  \nonumber
 &&\qquad~~~~~~~~~~~~~~~~~~-\cos\left(\frac{\alpha_{31}}{2}\right) c_{12} c_{23} s_{13} \sin \delta\Big]\,c_\nu\, \sh_y \ch_y+\\
  \nonumber
  &&-\,4\sqrt{m_2 m_3} c_{23}c_{13}\Big[-\sin\left(\frac{\alpha_{23}}{2}\right) s_{12}c_{23}s_{13}\sin\delta+\\
  \nonumber
  &&\qquad~~~~~~~~~~~~~~~~~~+\cos\left(\frac{\alpha_{23}}{2}\right)(c_{12}s_{23} + s_{12}c_{23}s_{13}\cos\delta)\Big]\,s_\nu c_\nu\,\sh_y^2;\\
  %%%%%%%%%%%%%%%%%%%%
  \wMav \wTheta^2_1 &=& m_1\, c^2_N \,\sh_y^2+m_2\, (c_N s_\nu \,\ch_y+s_N c_\nu)^2+\\ \nonumber
    && +\, m_3\, (c_N c_\nu \ch_y-s_N s_\nu)^2;\\
    \wMav \wTheta^2_2 &=&m_1\, s^2_N \,\sh_y^2+m_2\, (c_N c_\nu-s_N s_\nu \,\ch_y)^2+\\ \nonumber
    && +\, m_3\, (s_N c_\nu \,\ch_y+c_N s_\nu)^2;\\
    \wMav \wTheta^2_3 &=&m_1\, \ch_y^2+ m_2\, s^2_\nu\, \sh_y^2+m_3\, c^2_\nu\,\sh_y^2;
   \\
    \wMav \wTheta^2 &=& \sum_{a=1}^3 m_a\, \ch^2_y + [m_1+(m_3-m_2) (c_\nu^2-s_\nu^2)] \,\sh^2_y;
\end{eqnarray}

%%%%%%%%%%%%%%%%%%%%%%%%%%%%%%%%%%%
 \item{\bf CP2)}
\begin{eqnarray}
 \wMav \wTheta^2_e &=&  m_1 c_{12}^2 c_{13}^2\left[\ch_y^2-  (c_\nu^2-s_\nu^2)\,\sh_y^2\right]+m_2 s_{12}^2c_{13}^2(\sh_y^2+\ch_y^2)+ 
 \\
 \nonumber
 &&+\,m_3 s_{13}^2\left[\ch_y^2+ (c_\nu^2-s_\nu^2)\,\sh_y^2\right]+\\
 \nonumber
 && + \,4\sqrt{m_1 m_2}s_{12}c_{12}c_{13}^2  \sin\left(\frac{\alpha_{21}}{2}\right)\,s_\nu\,\sh_y\ch_y+\\
 \nonumber
 &&+\,4\sqrt{m_1 m_3}c_{12}s_{13}  c_{13}\cos\left(\frac{\alpha_{31}}{2}-\delta\right)\,s_\nu c_\nu\,\sh_y^2+\\
 \nonumber
 &&+ \,4\sqrt{m_2 m_3}s_{12}s_{13}c_{13}  \sin\left(\frac{\alpha_{23}}{2}+\delta\right)\,c_\nu\,\sh_y\ch_y;\\
 %%%%%%%%%%
 \quad\wMav \wTheta^2_\mu &=& m_1(c_{12}^2 s_{23}^2s_{13}^2 +s_{12}^2c_{23}^2 +2 s_{12}c_{12} s_{23}c_{23}  s_{13}  \cos\delta)\,\left[\ch_y^2-(c_\nu^2-s_\nu^2)\,\sh_y^2\right]+\\
 \nonumber
 && +\, m_2 (c_{12}^2 c_{23}^2+s_{12}^2 s_{23}^2s_{13}^2-2 s_{12}c_{12} s_{23}c_{23}  s_{13}  \cos \delta)(\ch_y^2+\sh_y^2)+\\
 \nonumber
 && +\,m_3s_{23}^2 c_{13}^2 \left[\ch_y^2+ (c_\nu^2-s_\nu^2)\,\sh_y^2\right]+\\
 \nonumber
 && -\,4\sqrt{m_1 m_2}\Big\{\sin\left(\frac{\alpha_{21}}{2}\right)\left[-s_{12}c_{12}s_{23}^2s_{13}^2+s_{12}c_{12}c_{23}^2+ (c_{12}^2-s_{12}^2)s_{23}c_{23}s_{13}\cos\delta\right]+\\
 \nonumber
 && \qquad~~~~~~~~~~~~-\cos\left(\frac{\alpha_{21}}{2}\right)s_{23}c_{23} s_{13}  \sin\delta\Big\}\,s_\nu\ch_y \sh_y+\\
 \nonumber
 &&- \,4 \sqrt{m_1 m_3}s_{23}c_{13}\Big[\sin\left(\frac{\alpha_{31}}{2}\right)c_{12} s_{23}s_{13} \sin\delta+\\
 \nonumber
&&\qquad~~~~~~~~~~~~~~~~~~+\cos\left(\frac{\alpha_{31}}{2}\right)(s_{12}c_{23}  + c_{12} s_{23}s_{13}  \cos\delta)\Big]\,s_\nu c_\nu\, \sh_y^2+\\
 \nonumber
 &&-\,4\sqrt{m_2 m_3} s_{23}c_{13}\Big[\sin\left(\frac{\alpha_{23}}{2}\right) (-c_{12}c_{23}+s_{12}s_{23} s_{13}  \cos\delta)+\\
  \nonumber
&&\qquad~~~~~~~~~~~~~~~~~~+\cos\left(\frac{\alpha_{23}}{2}\right)s_{12}s_{23}s_{13}\sin\delta\Big]\,c_\nu \sh_y \ch_y;\\
 %%%%%%%%%%
 \quad\wMav \wTheta^2_\tau &=& m_1(c_{12}^2 c_{23}^2 s_{13}^2 + s_{12}^2 s_{23}^2-2 s_{12}c_{12} s_{23}c_{23}  s_{13}  \cos\delta)\left[\ch_y^2-(c_\nu^2-s_\nu^2)\,\sh_y^2\right]+\\
  \nonumber
  && +\, m_2 (c_{12}^2 s_{23}^2 + s_{12}^2c_{23}^2  s_{13}^2  + 2s_{12} c_{12} s_{23}c_{23}  s_{13}  \cos\delta)(\ch_y^2+\sh_y^2) +\\
  \nonumber
 && +\,m_3c_{23}^2 c_{13}^2 \left[\ch_y^2+ (c_\nu^2-s_\nu^2)\,\sh_y^2\right]+\\
 \nonumber
 && -\,4\sqrt{m_1 m_2}\Big\{\sin\left(\frac{\alpha_{21}}{2}\right)\left[-s_{12}c_{12} c_{23}^2 s_{13}^2+s_{12} c_{12} s_{23}^2 - (c_{12}^2-s_{12}^2)s_{23}c_{23} s_{13}  \cos\delta\right]+\\
  \nonumber
 && \qquad~~~~~~~~~~~~+\cos\left(\frac{\alpha_{21}}{2}\right)s_{23}c_{23} s_{13}  \sin\delta\Big\}\,s_\nu\ch_y \sh_y+\\
 \nonumber
 &&- \,4 \sqrt{m_1 m_3}c_{23}c_{13}\Big[\sin\left(\frac{\alpha_{31}}{2}\right)c_{12} c_{23}s_{13} \sin\delta +\\
  \nonumber
&&\qquad~~~~~~~~~~~~~~~~~~+\cos\left(\frac{\alpha_{31}}{2}\right) (-s_{12}s_{23} + c_{12} c_{23} s_{13} \cos \delta)\Big]\,s_\nu c_\nu\, \sh_y^2+\\
  \nonumber
  &&-\,4\sqrt{m_2 m_3} c_{23}c_{13}\Big[\sin\left(\frac{\alpha_{23}}{2}\right) (c_{12} s_{23} + s_{12} c_{23} s_{13} \cos\delta)+\\
  \nonumber
  &&\qquad~~~~~~~~~~~~~~~~~~+\cos\left(\frac{\alpha_{23}}{2}\right)s_{12} c_{23} s_{13} \sin\delta\Big]\,c_\nu \,\sh_y \ch_y;\\
  %%%%%%%%%%%%%%%%%%%
  \quad\wMav \wTheta^2_1 &=&m_1\, (c_N s_\nu \,\ch_y+s_N c_\nu)^2 + m_2\, c^2_N \,\sh_y^2+\\ \nonumber
    && +\, m_3\, (c_N c_\nu \ch_y-s_N s_\nu)^2;\\
    \wMav \wTheta^2_2 &=& m_1\, (c_N c_\nu-s_N s_\nu \,\ch_y)^2 + m_2\, s^2_N \,\sh_y^2+\\ \nonumber
    && +\, m_3\, (s_N c_\nu \,\ch_y+c_N s_\nu)^2;\\
    \wMav \wTheta^2_3 &=&m_1\, s^2_\nu\,\sh_y^2 + m_2\, \ch_y^2+m_3\, c^2_\nu\,\sh_y^2;
    \\
    \wMav\wTheta^2 &=& \sum_{a=1}^3 m_a\,\ch^2_y + [m_2 + (m_3-m_1) (c_\nu^2-s_\nu^2)] \,\sh^2_y;
\end{eqnarray}
%%%%%%%%%%%%%%%%%%%%%%%%%%%%%%%%%%%
 \item{\bf CP3)}
\begin{eqnarray}
 \wMav \wTheta^2_e &=&  m_1 c_{12}^2 c_{13}^2\left[\ch_y^2-  (c_\nu^2-s_\nu^2)\,\sh_y^2\right]+m_2 s_{12}^2c_{13}^2\left[\ch_y^2+ (c_\nu^2-s_\nu^2)\,\sh_y^2\right]+ 
 \\
 \nonumber
 &&+\,m_3 s_{13}^2(\sh_y^2+\ch_y^2)+\\
 \nonumber
 && + \,4\sqrt{m_1 m_2}s_{12}c_{12}c_{13}^2  \cos\left(\frac{\alpha_{21}}{2}\right)\,s_\nu c_\nu\,\sh_y^2+\\
 \nonumber
 &&+\,4\sqrt{m_1 m_3}c_{12}s_{13}c_{13}  \sin\left(\frac{\alpha_{31}}{2}-\delta\right)\,s_\nu \,\sh_y \ch_y+\\
 \nonumber
 &&- \,4\sqrt{m_2 m_3}s_{12}s_{13}c_{13}  \sin\left(\frac{\alpha_{23}}{2}+\delta\right)\,c_\nu\,\sh_y\ch_y;\\
 %%%%%%%%%%
 \quad\wMav \wTheta^2_\mu &=& m_1(c_{12}^2 s_{23}^2s_{13}^2 +s_{12}^2c_{23}^2 +2 s_{12}c_{12} s_{23}c_{23}  s_{13}  \cos\delta)\,\left[\ch_y^2-(c_\nu^2-s_\nu^2)\,\sh_y^2\right]+\\
 \nonumber
 && +\, m_2 (c_{12}^2 c_{23}^2+s_{12}^2 s_{23}^2s_{13}^2-2 s_{12}c_{12} s_{23}c_{23}  s_{13}  \cos \delta)\left[\ch_y^2+ (c_\nu^2-s_\nu^2)\,\sh_y^2\right]+\\
 \nonumber
 && +\,m_3s_{23}^2 c_{13}^2(\ch_y^2+\sh_y^2) +\\
 \nonumber
 && -\,4\sqrt{m_1 m_2}\Big\{\cos\left(\frac{\alpha_{21}}{2}\right)\left[- s_{12} c_{12} s_{23}^2 s_{13}^2 + s_{12}c_{12} c_{23}^2 + (c_{12}^2-s_{12}^2)s_{23}c_{23} s_{13}  \cos\delta\right]+\\
 \nonumber
 && \qquad~~~~~~~~~~~~+\sin\left(\frac{\alpha_{21}}{2}\right)s_{23}c_{23} s_{13} \sin\delta\Big\}\,s_\nu c_\nu \sh_y^2+\\
 \nonumber
 &&-\,4 \sqrt{m_1 m_3}s_{23}c_{13}\Big[\sin\left(\frac{\alpha_{31}}{2}\right)(s_{12}c_{23}+ c_{12} s_{23} s_{13} \cos\delta)+\\
 \nonumber
&&\qquad~~~~~~~~~~~~~~~~~~-\cos\left(\frac{\alpha_{31}}{2}\right)c_{12} s_{23} s_{13}\sin\delta\Big]\,s_\nu \sh_y\ch_y+\\
 \nonumber
 &&+\,4\sqrt{m_2 m_3} s_{23}c_{13}\Big[\sin\left(\frac{\alpha_{23}}{2}\right) (-c_{12}c_{23}+s_{12} s_{23} s_{13}  \cos\delta)+\\
  \nonumber
&&\qquad~~~~~~~~~~~~~~~~~~+\cos\left(\frac{\alpha_{23}}{2}\right)s_{12}s_{23}s_{13}\sin\delta\Big]\,c_\nu \sh_y \ch_y;\\
 %%%%%%%%%%
 \quad\wMav \wTheta^2_\tau &=& m_1(c_{12}^2 c_{23}^2 s_{13}^2 + s_{12}^2 s_{23}^2-2 s_{12}c_{12} s_{23}c_{23}  s_{13}  \cos\delta)\left[\ch_y^2-(c_\nu^2-s_\nu^2)\,\sh_y^2\right]+\\
  \nonumber
  && +\, m_2 (c_{12}^2 s_{23}^2 + s_{12}^2c_{23}^2  s_{13}^2  + 2s_{12} c_{12} s_{23}c_{23}  s_{13}  \cos\delta)\left[\ch_y^2+ (c_\nu^2-s_\nu^2)\,\sh_y^2\right] +\\
  \nonumber
 && +\,m_3c_{23}^2 c_{13}^2 (\ch_y^2+\sh_y^2)+\\
 \nonumber
 && -\,4\sqrt{m_1 m_2}\Big\{\cos\left(\frac{\alpha_{21}}{2}\right)\left[-s_{12} c_{12} c_{23}^2 s_{13}^2 + s_{12} c_{12}  s_{23}^2 - (c_{12}^2-s_{12}^2)s_{23}c_{23} s_{13}  \cos\delta\right]+\\
 \nonumber
 && \qquad~~~~~~~~~~~~-\sin\left(\frac{\alpha_{21}}{2}\right)s_{23}c_{23} s_{13} \sin\delta\Big\}\,s_\nu c_\nu \sh_y^2+\\
 \nonumber
 &&-\,4 \sqrt{m_1 m_3}c_{23}c_{13}\Big[\sin\left(\frac{\alpha_{31}}{2}\right)(-s_{12}s_{23}+ c_{12} c_{23} s_{13} \cos\delta)+\\
 \nonumber
&&\qquad~~~~~~~~~~~~~~~~~~-\cos\left(\frac{\alpha_{31}}{2}\right)c_{12} c_{23} s_{13}\sin\delta\Big]\,s_\nu \sh_y\ch_y+\\
 \nonumber
 &&+\,4\sqrt{m_2 m_3} c_{23}c_{13}\Big[\sin\left(\frac{\alpha_{23}}{2}\right) (c_{12}s_{23}+s_{12} c_{23} s_{13}  \cos\delta)+\\
  \nonumber
&&\qquad~~~~~~~~~~~~~~~~~~+\cos\left(\frac{\alpha_{23}}{2}\right)s_{12}c_{23}s_{13}\sin\delta\Big]\,c_\nu \sh_y \ch_y;\\
%%%%%%%%%
    \wMav \wTheta^2_1 &=&m_1\, (c_N s_\nu \,\ch_y+s_N c_\nu)^2 + m_2\, (c_N c_\nu \ch_y-s_N s_\nu)^2+\\ \nonumber
    && +\,m_3\, c^2_N \,\sh_y^2; \\
    \wMav \wTheta^2_2 &=& m_1\, (c_N c_\nu-s_N s_\nu \,\ch_y)^2 + m_2\, (s_N c_\nu \,\ch_y+c_N s_\nu)^2+\\ \nonumber
    && +\,m_3\, s^2_N \,\sh_y^2; \\
    \wMav \wTheta^2_3 &=&m_1\, s^2_\nu\,\sh_y^2 +m_2\, c^2_\nu\,\sh_y^2 +  m_3\, \ch_y^2\\
    \wMav\wTheta^2 &=& \sum_{a=1}^3 m_a \,\ch^2_y + [m_3 + (m_2-m_1) (c_\nu^2-s_\nu^2)] \,\sh^2_y.
\end{eqnarray}
\end{itemize}

\section{More scans of the leptogenesis parameter space}\label{app:otherscans}
\begin{figure}
    \centering
\includegraphics[width=0.45\textwidth]{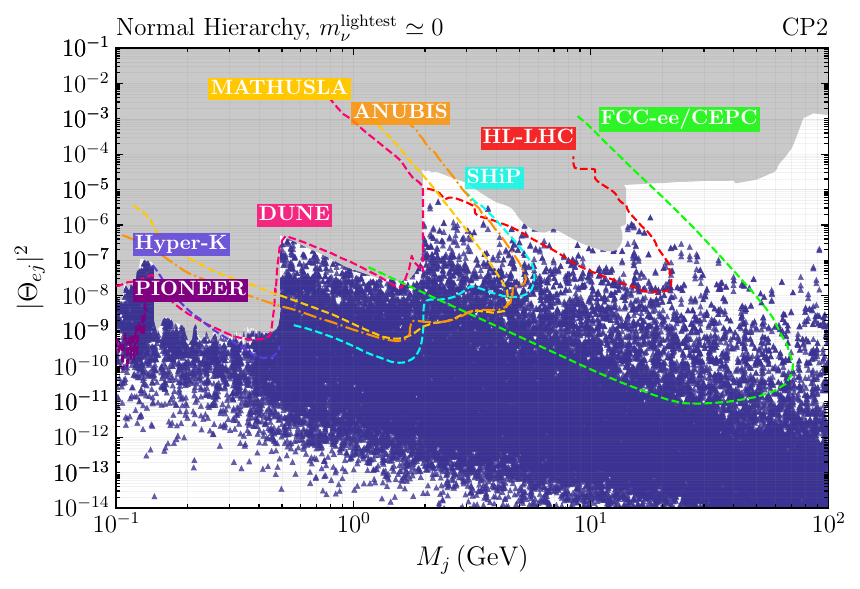}
\includegraphics[width=0.45\textwidth]{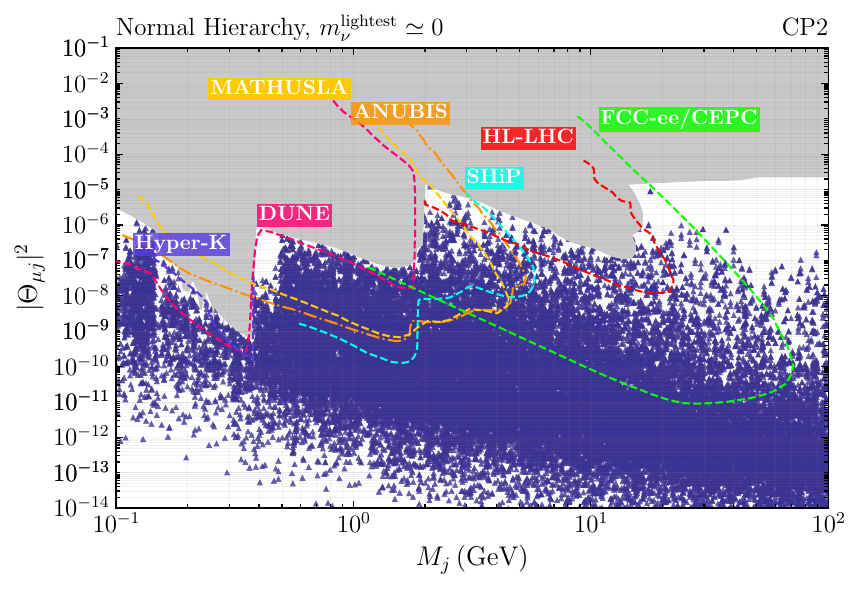}
\includegraphics[width=0.45\textwidth]{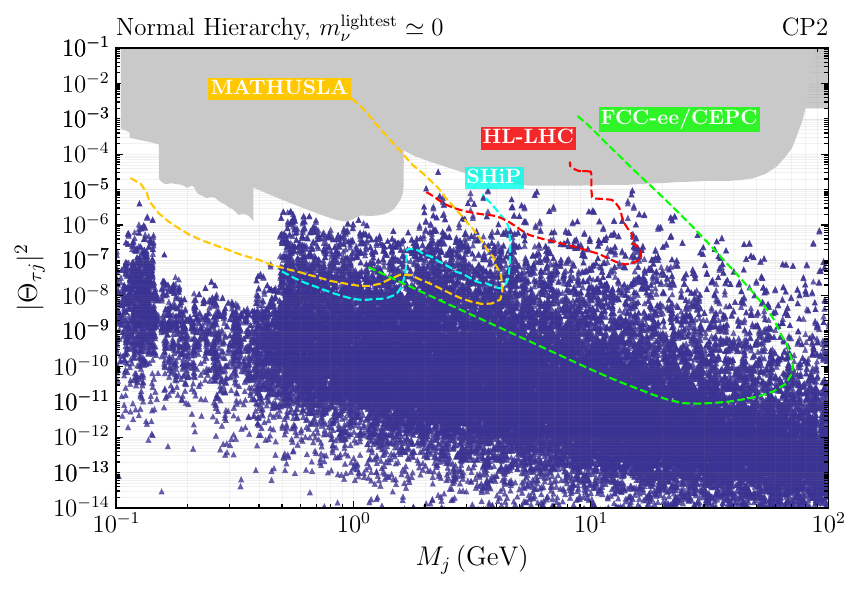}
\hspace{1.2em}
\includegraphics[width=0.4\textwidth]{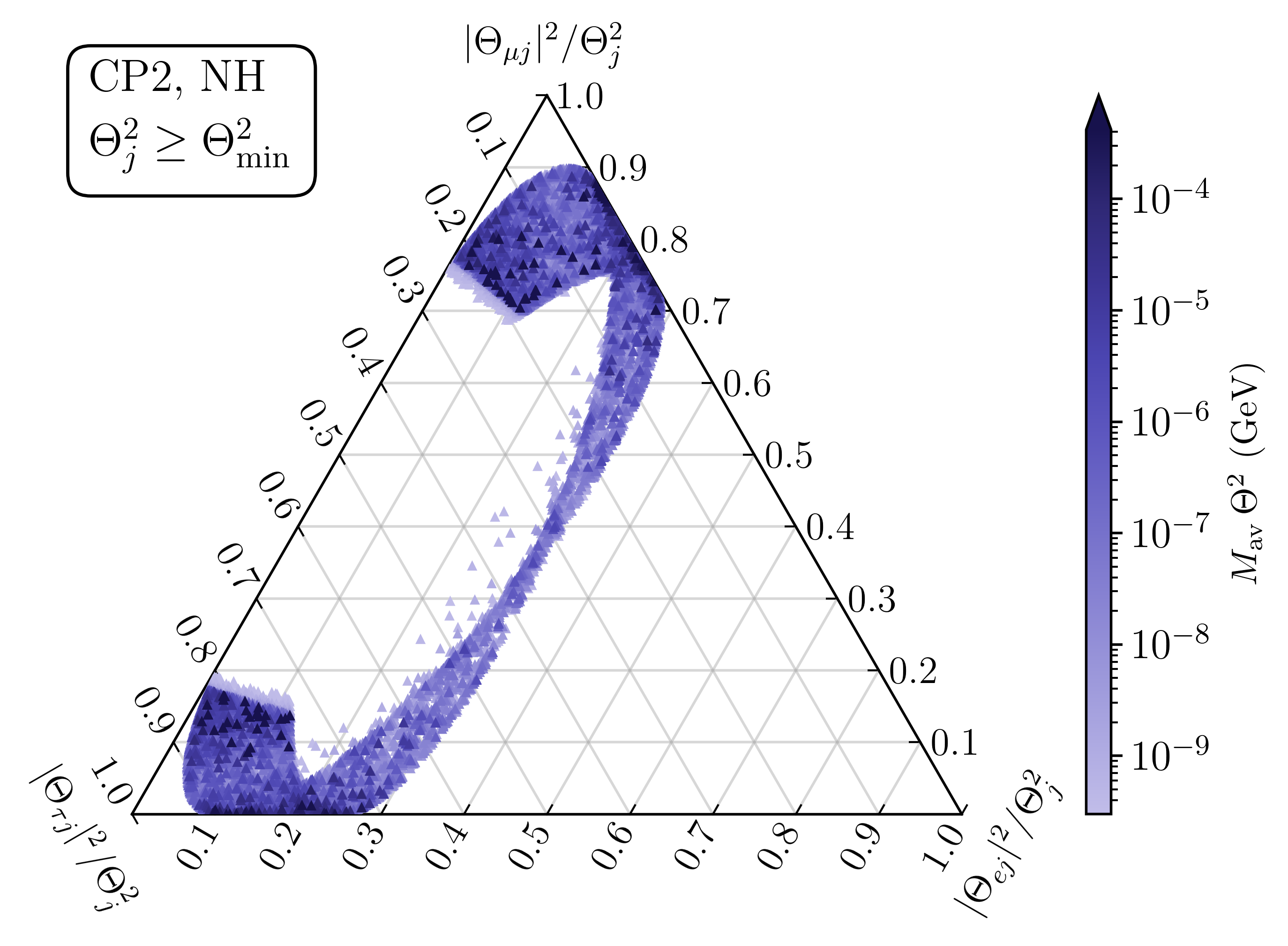}
\includegraphics[width=0.45\textwidth]{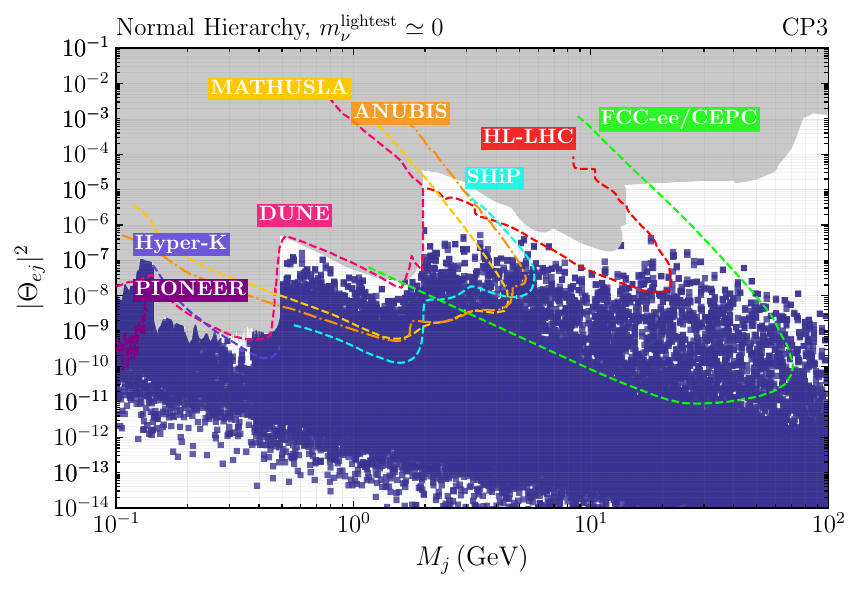}
\includegraphics[width=0.45\textwidth]{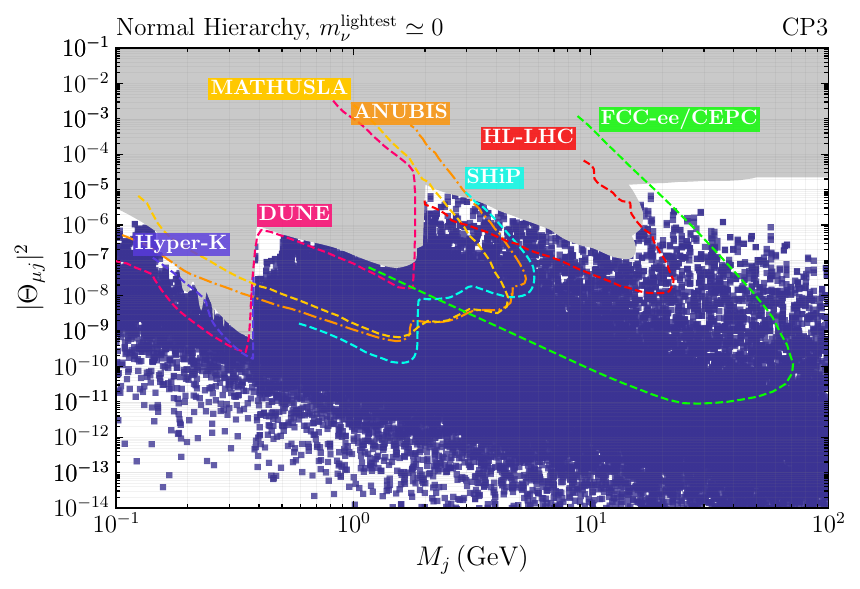}
\includegraphics[width=0.45\textwidth]{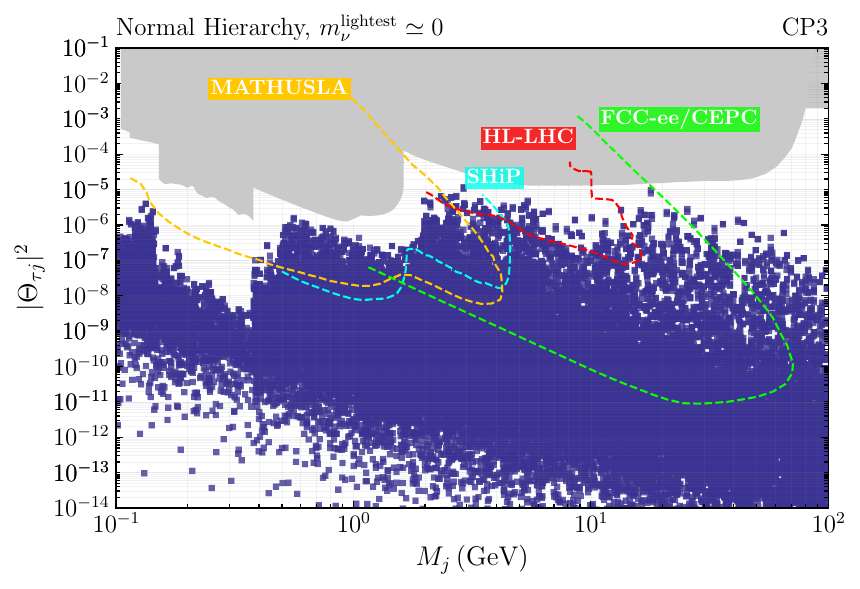}
\hspace{1.2em}
\includegraphics[width=0.4\textwidth]{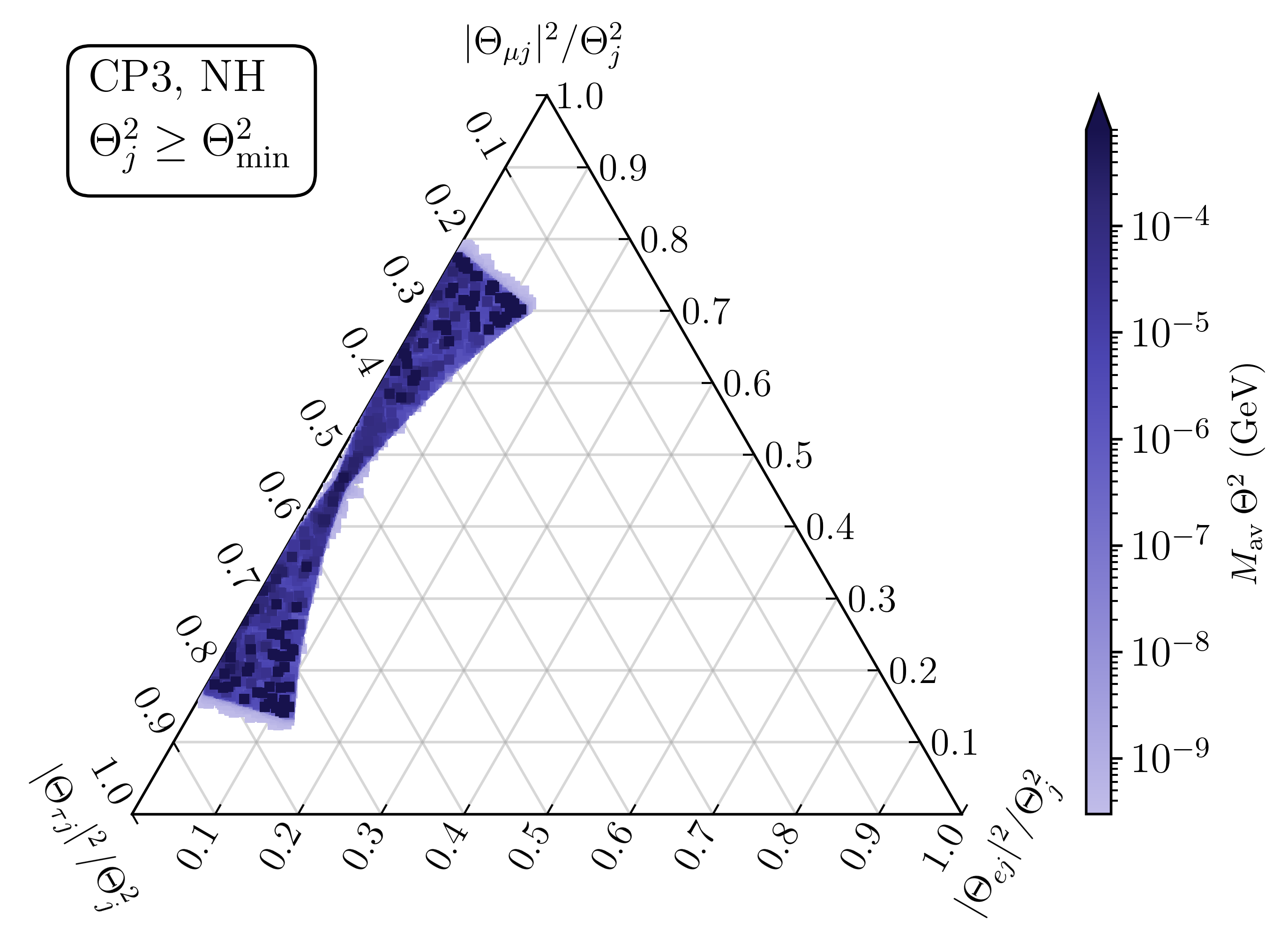}
    \caption{Scan of the parameter space in the CP2 case (first four panels) and CP3 (last four panels) for NH; points are marked with triangles for CP2 and squares for CP3; all other details are as in Fig.~\ref{fig:CP1s_scan_NH}.}
\label{fig:CP2_CP3_scan_NH}
\end{figure}

\begin{figure}
    \centering
\includegraphics[width=0.45\textwidth]{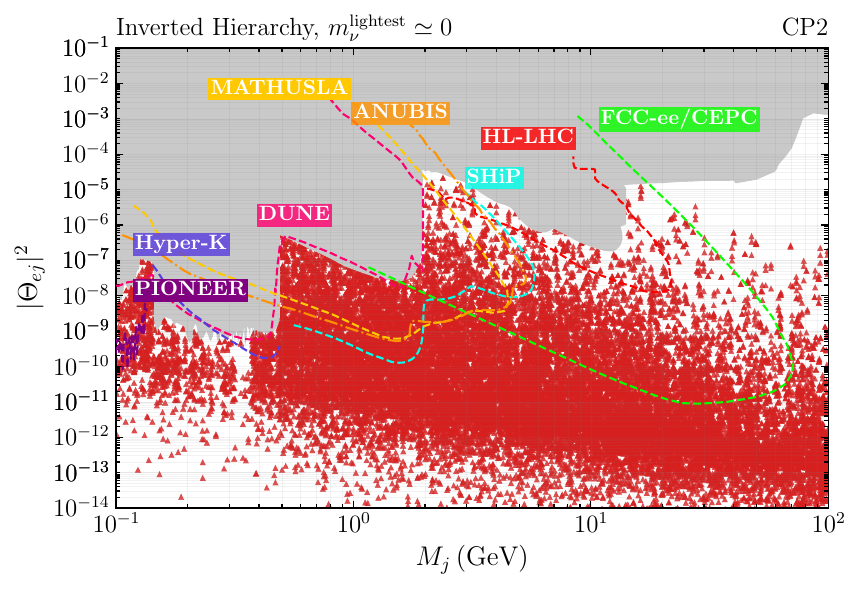}
\includegraphics[width=0.45\textwidth]{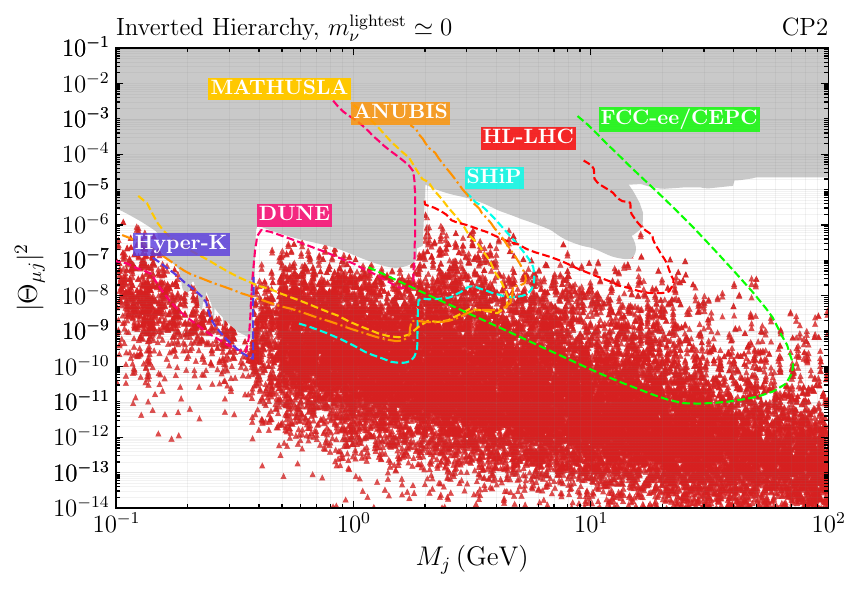}
\includegraphics[width=0.45\textwidth]{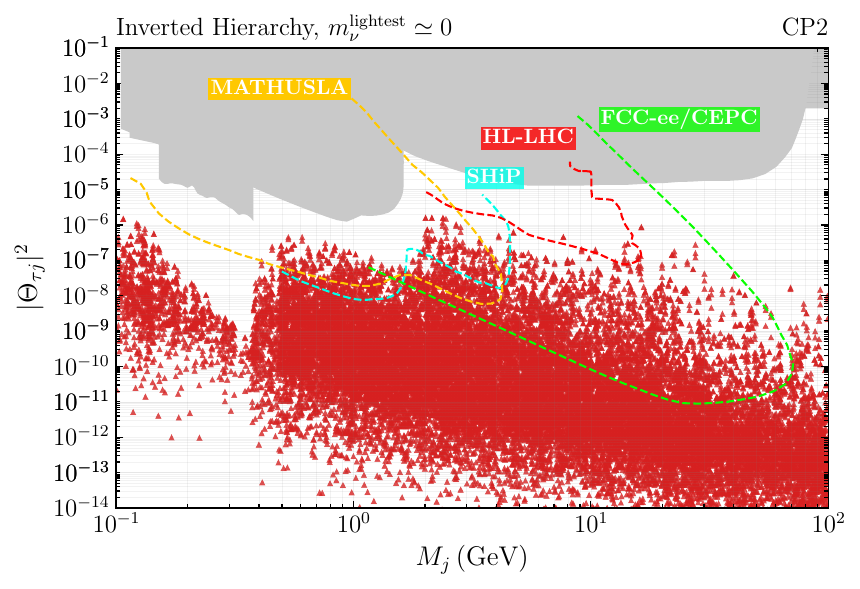}
\hspace{1.2em}
\includegraphics[width=0.4\textwidth]{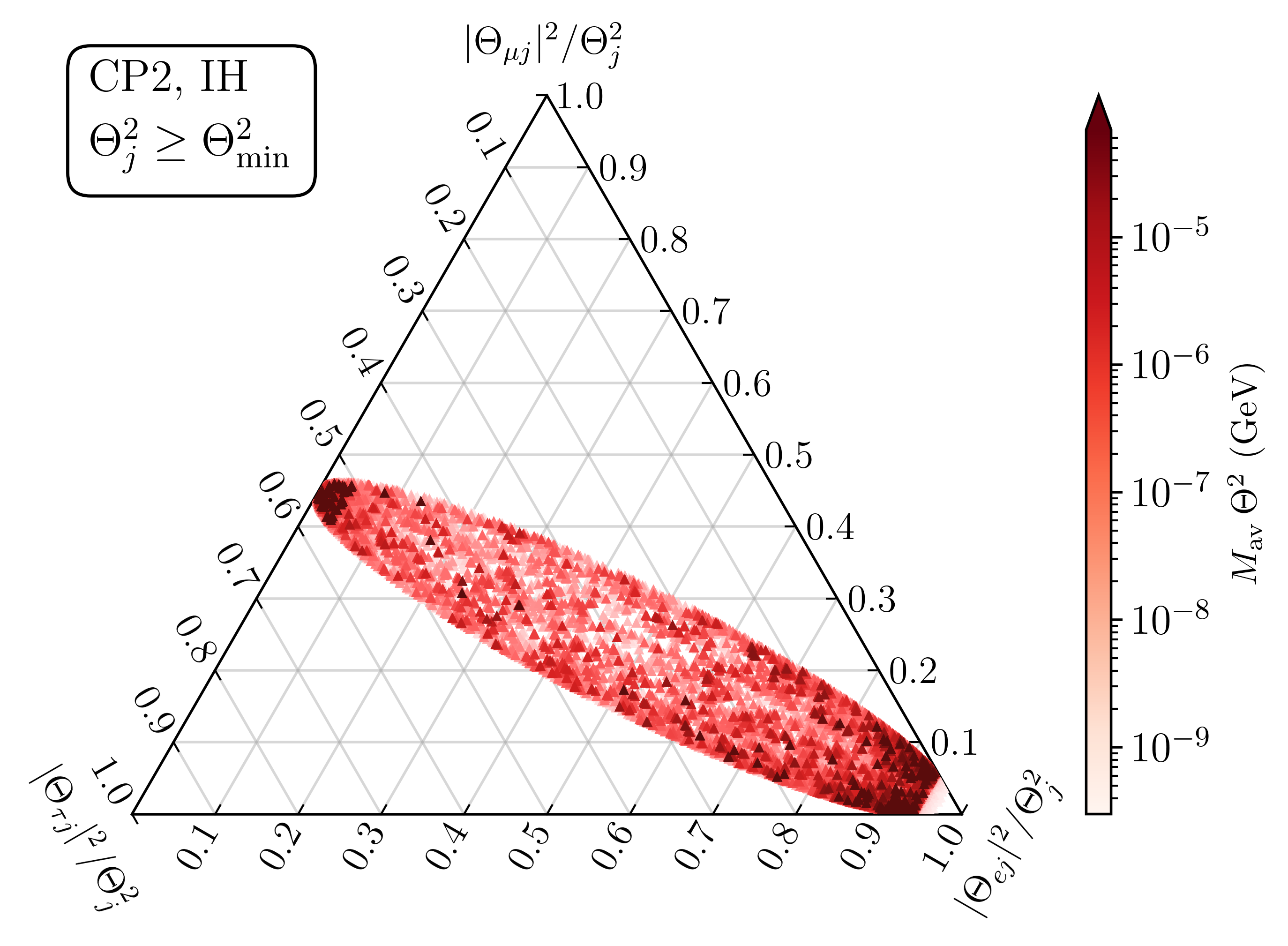}
\includegraphics[width=0.45\textwidth]{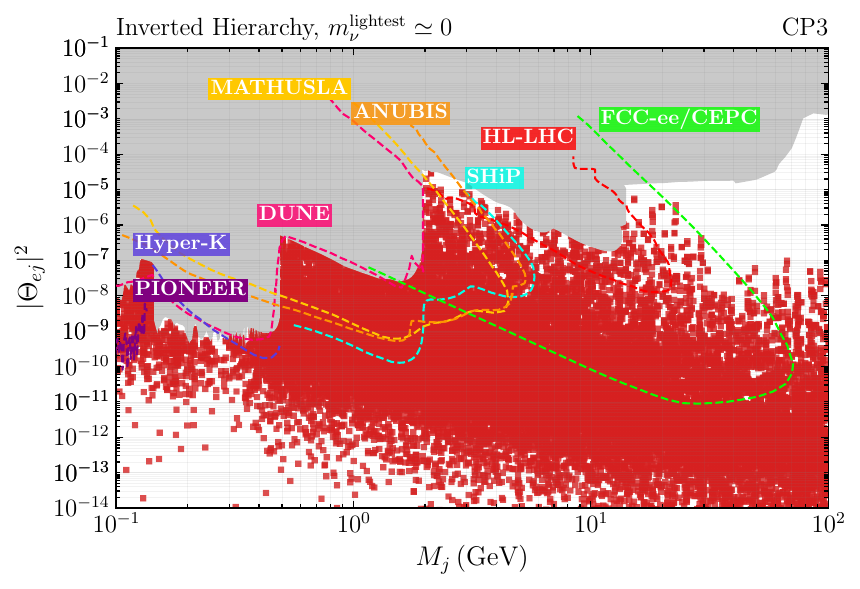}
\includegraphics[width=0.45\textwidth]{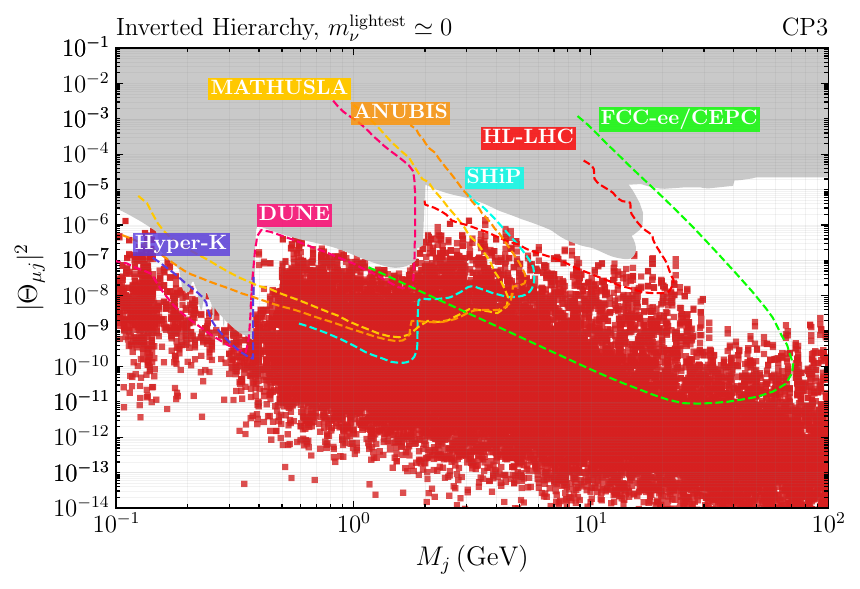}
\includegraphics[width=0.45\textwidth]{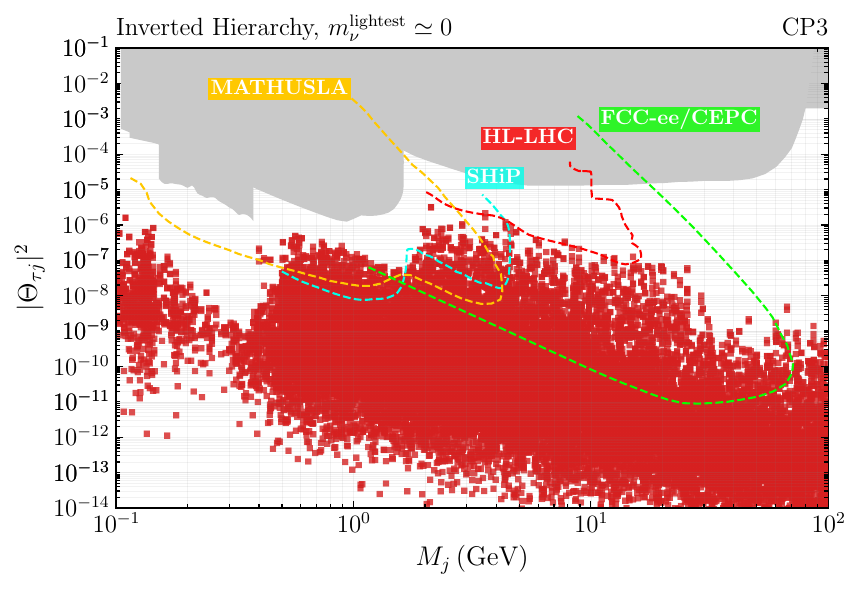}
\hspace{1.2em}
\includegraphics[width=0.4\textwidth]{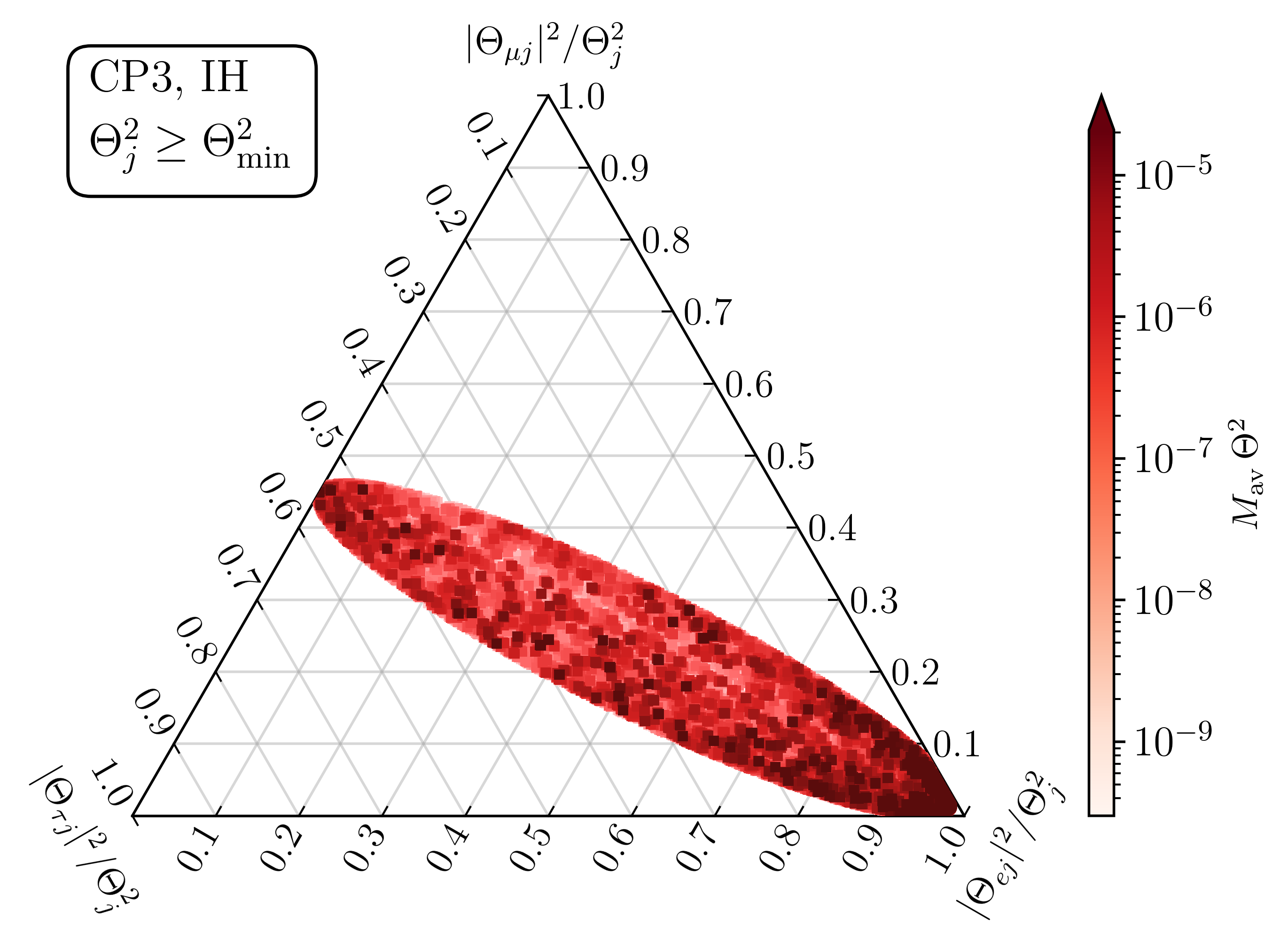}
    \caption{Scan of the parameter space in the CP2 case (first four panels) and CP3 (last four panels) for IH; points are marked with triangles for CP2 and squares for CP3; all other details are as in the four top panels of Fig.~\ref{fig:CP1s_scan_IH_QD}.}
\label{fig:CP2_CP3_scan_IH}
\end{figure}

\begin{figure}
    \centering
\includegraphics[width=0.45\textwidth]{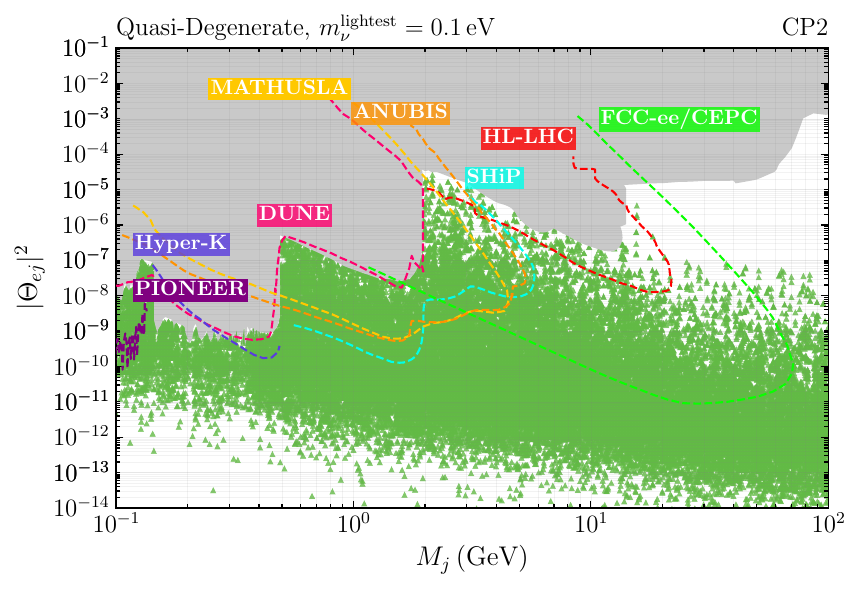}
\includegraphics[width=0.45\textwidth]{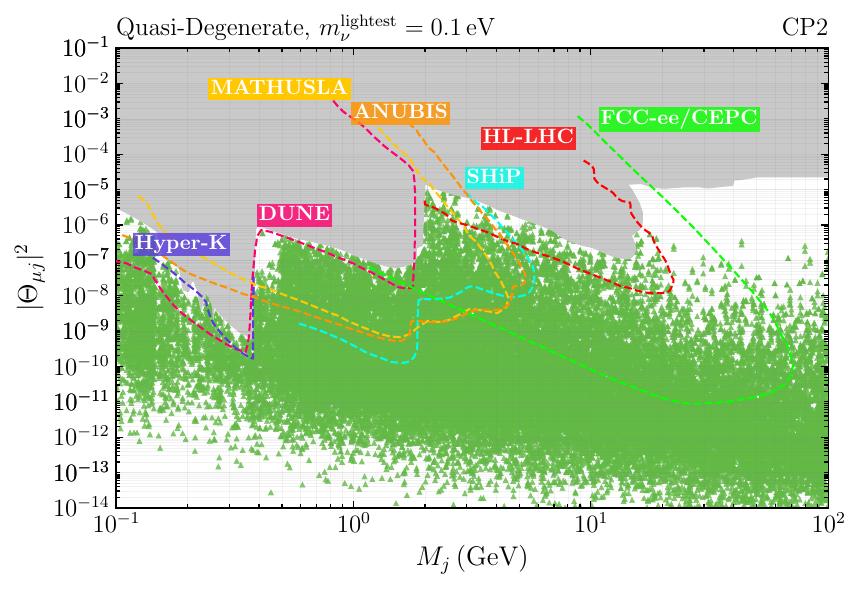}
\includegraphics[width=0.45\textwidth]{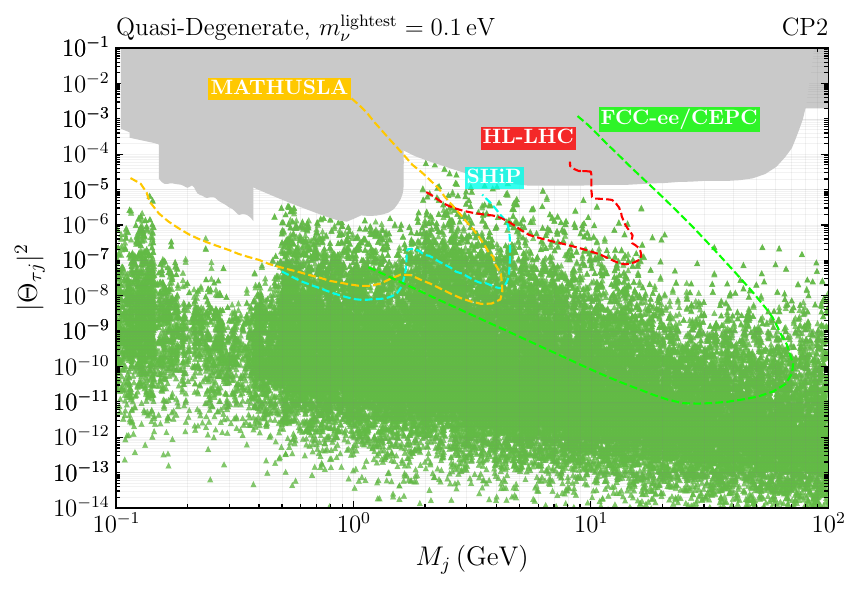}
\hspace{1.2em}
\includegraphics[width=0.4\textwidth]{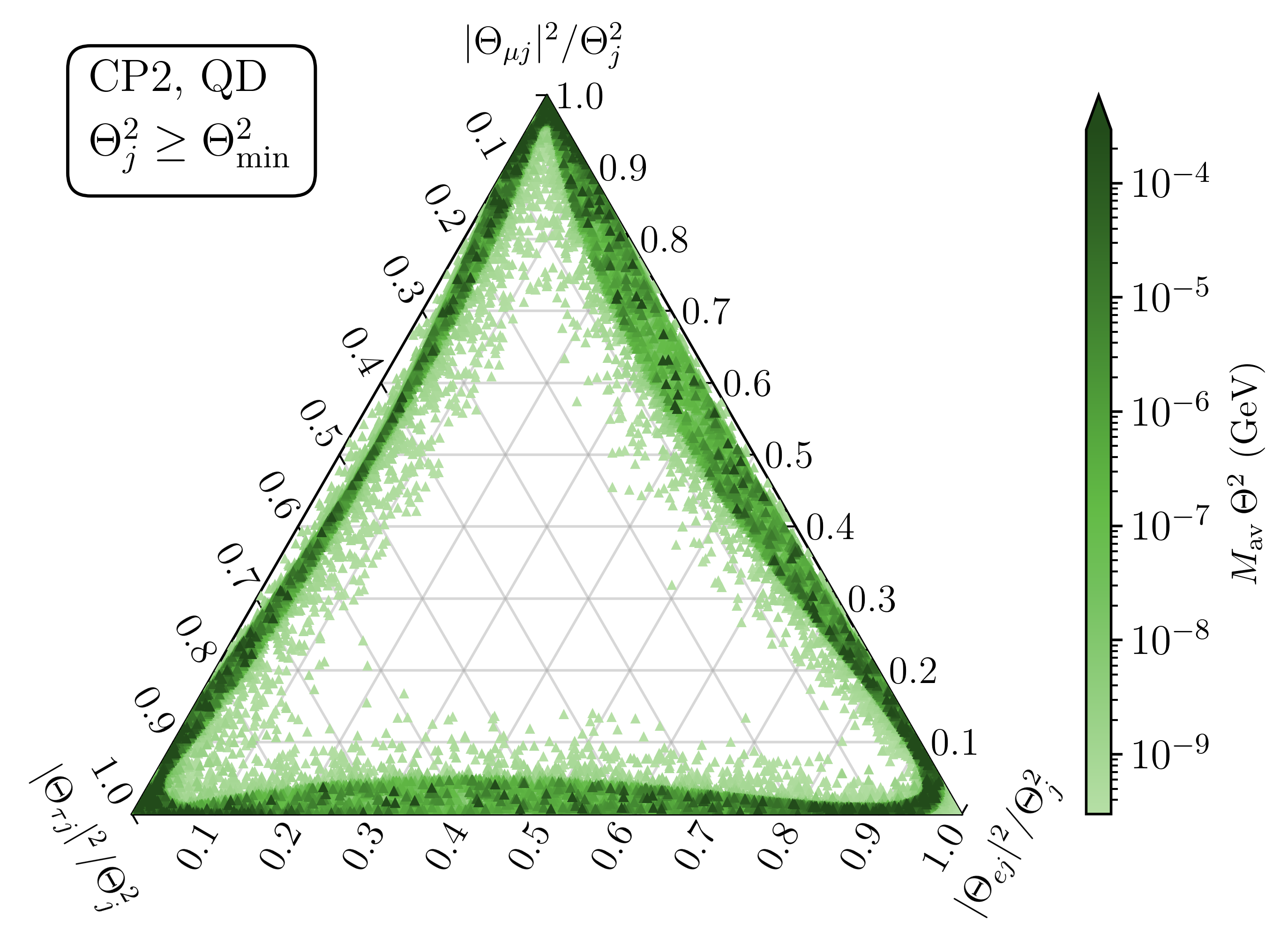}
\includegraphics[width=0.45\textwidth]{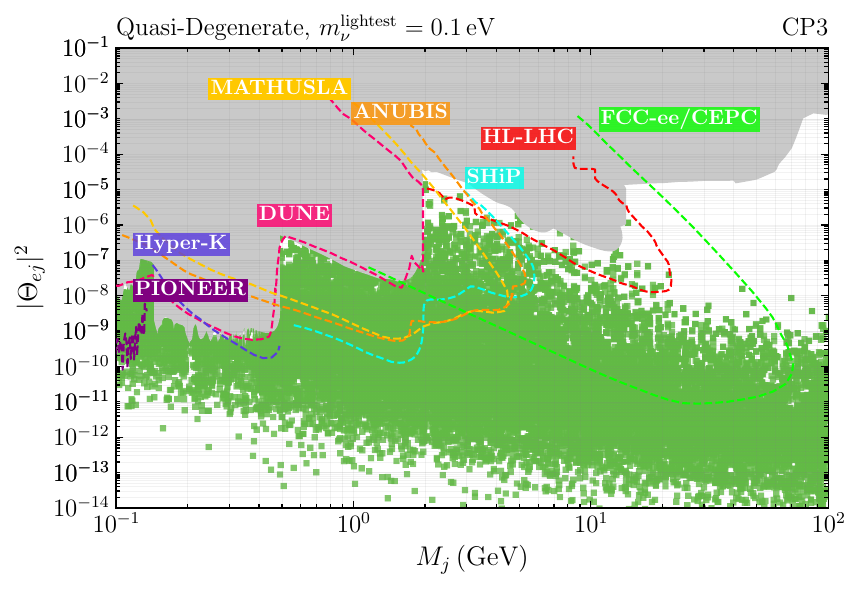}
\includegraphics[width=0.45\textwidth]{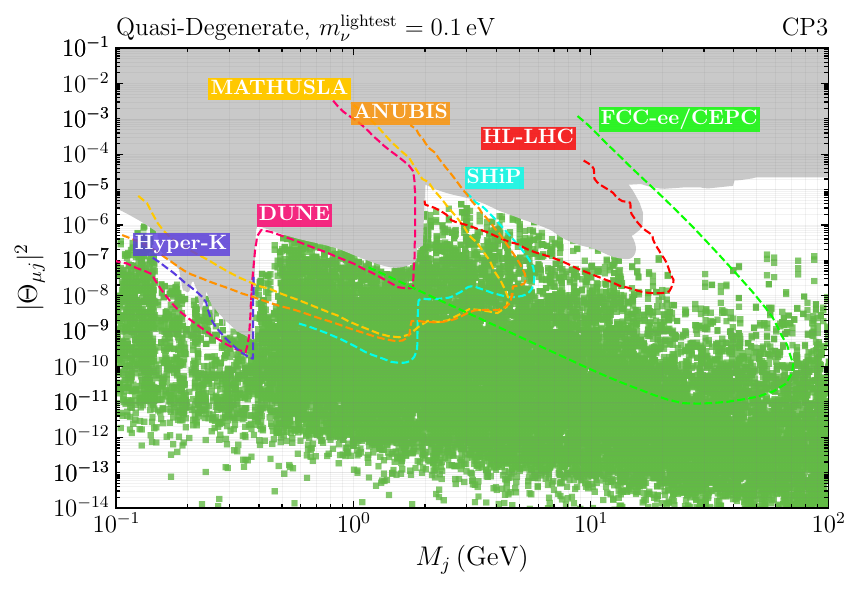}
\includegraphics[width=0.45\textwidth]{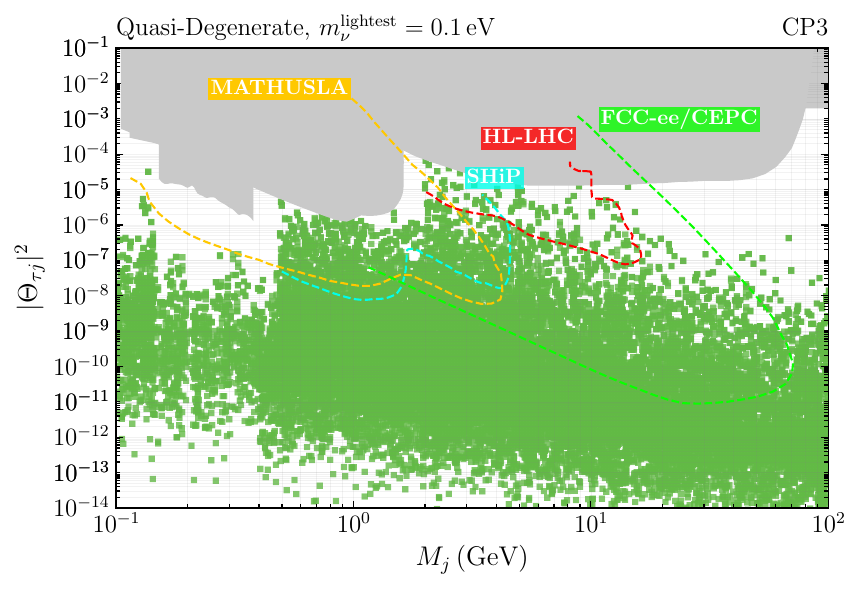}
\hspace{1.2em}
\includegraphics[width=0.4\textwidth]{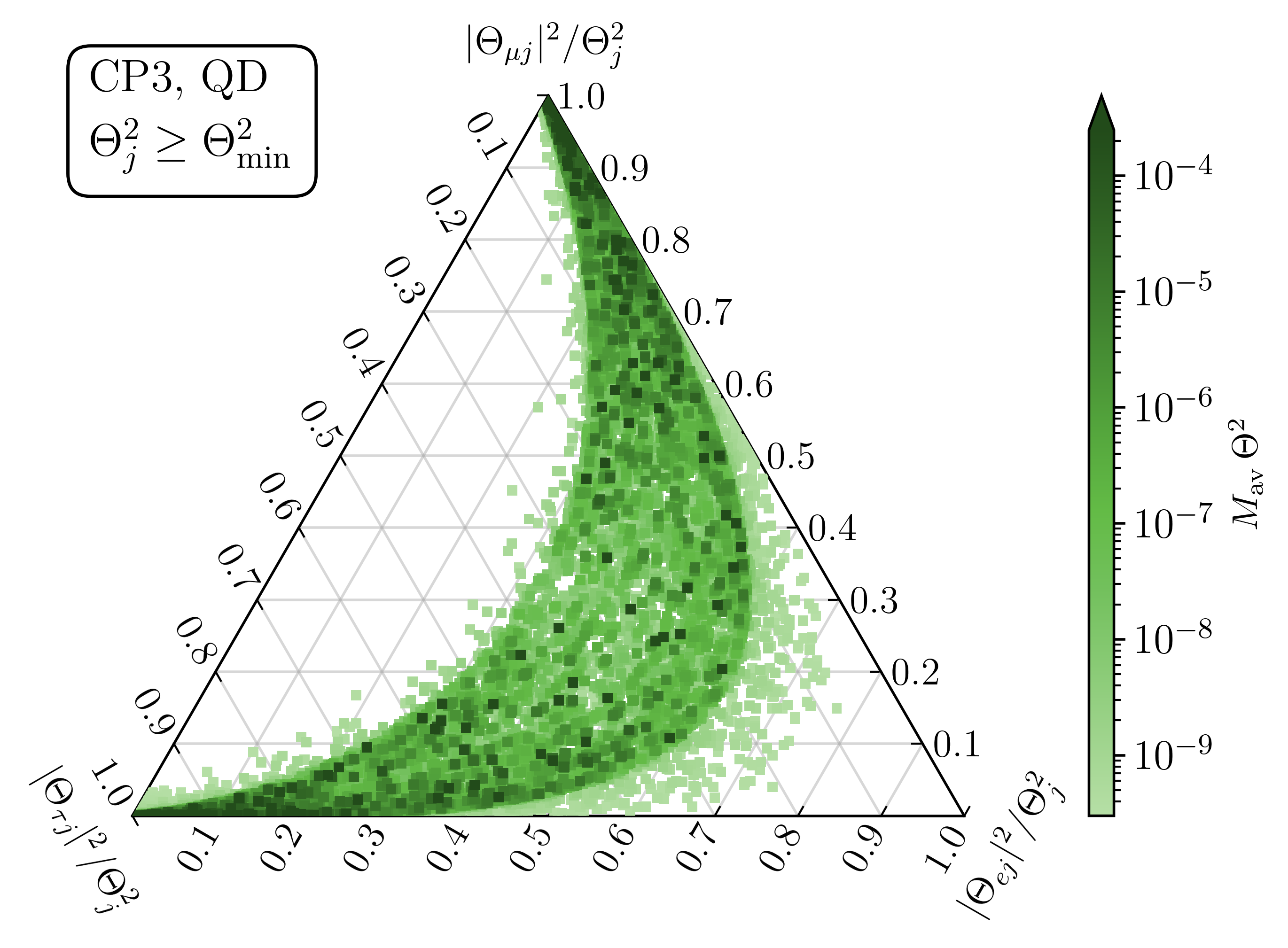}
    \caption{Scan of the parameter space in the CP2 case (first four panels) and CP3 (last four panels) for QD points are marked with triangles for CP2 and squares for CP3; all other details are as in the last four panels of Fig.~\ref{fig:CP1s_scan_IH_QD}.}
\label{fig:CP2_CP3_scan_QD}
\end{figure}

We show here the results of our leptogenesis scans in the CP2 and CP3 cases, for NH in Fig.~\ref{fig:CP2_CP3_scan_NH}, IH in Fig.~\ref{fig:CP2_CP3_scan_IH} and QD in Fig.~\ref{fig:CP2_CP3_scan_QD}.

\newpage
\section{Benchmark points for viable Dirac-phase driven low-scale  leptogenesis}\label{app:BMC}
We provide 
more benchmark points in the CP1 case and investigate around these benchmarks the dependence of the BAU against the Dirac phase $\delta$.

\begin{figure} [t!]
\centering
\includegraphics[width=0.45\linewidth]{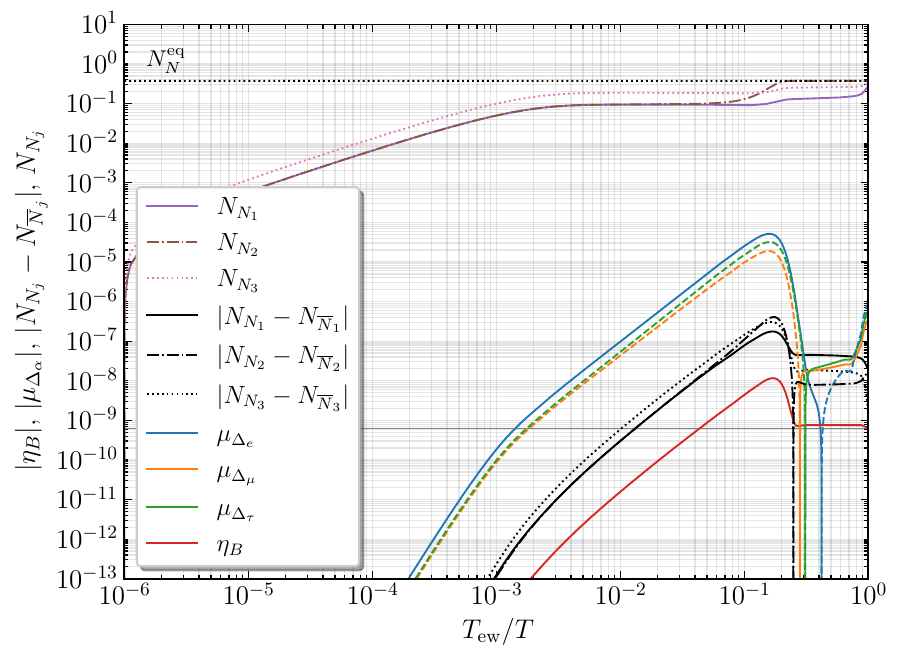}\\
\includegraphics[width=0.45\linewidth]{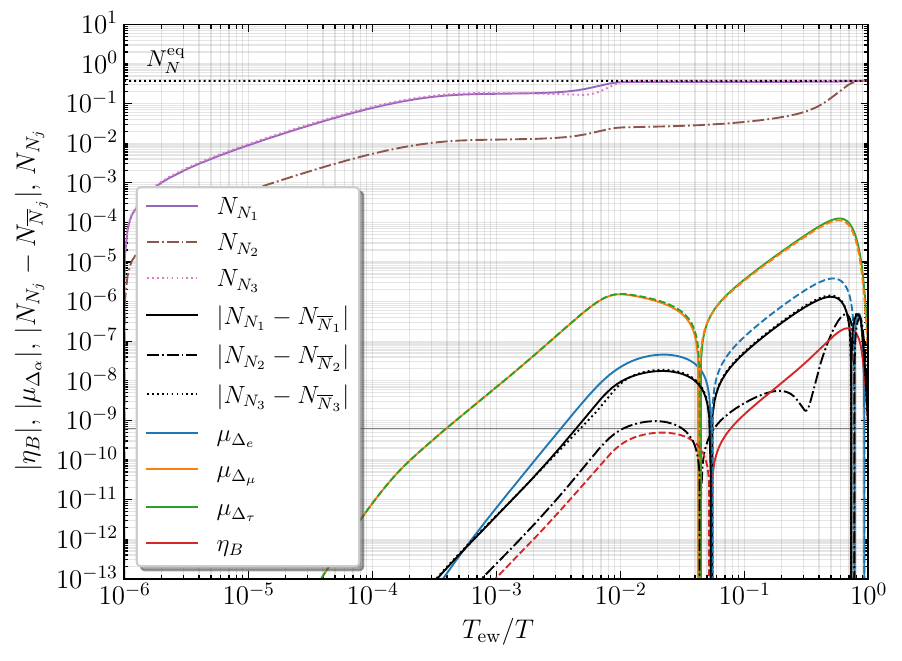}
\includegraphics[width=0.45\linewidth]{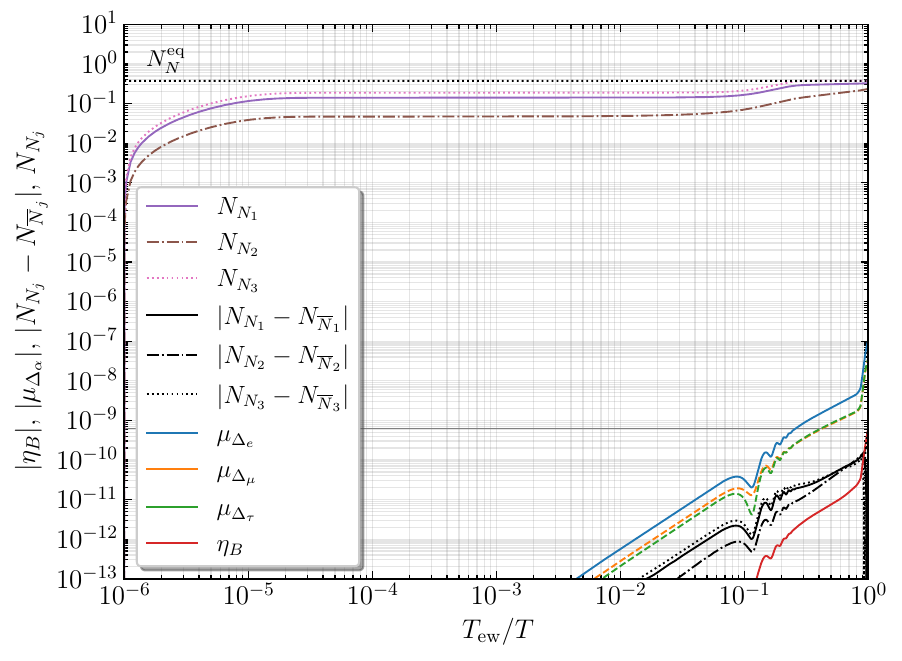}
    \caption{Evolution of the heavy neutrino abundances
$N_{N_j} \equiv (\rho_N)_{jj}$,
their CP-asymmetries $|N_{N_j}-{N}_{\overline{N}_j}|$, $N_{\overline{N}_j}$, the flavour asymmetries $\mu_{\Delta_\alpha}$ with $\alpha=e,\mu,\tau$,
and the baryon-to-photon ratio $\eta_B$,
as functions of $x \equiv T_{\rm ew}/T$, for the benchmarks point in the NH case, Eq.~\eqref{eq:params_BMC_NH} (top), IH, Eq.~\eqref{eq:params_BMC_IH} (bottom-left) and QD, Eq.~\eqref{eq:params_BMC_QD} (bottom-right). All other details are as in Fig.~\ref{fig:LG_Plot_BMC2_NH}.}
    \label{fig:LG_Plot_BMC}
\end{figure}

\paragraph{Second benchmark in the NH case.} Another benchmark point in the NH case is: \begin{equation}\label{eq:params_BMC_NH}
\begin{split}
    \tilde{M}_1 = 2\,\text{GeV}, \quad \Delta \tilde{M}_{21} = 2\times 10^{-7}\,\text{GeV},\quad \Delta \tilde{M}_{31} = 0, \quad \mnulight = 0,\\
    x_\nu = 3\pi/2,\quad x_N = \pi/4,\quad y = 5.44,\\
    \delta = 212^\circ\,\text{(best-fit of } \texttt{NuFit  6.0}\text{)},\quad
    \alpha_{23} = -2\pi.
\end{split}
\end{equation}
Also here $\tilde{N}_1$ and $\tilde{N}_3$ are exactly degenerate in mass (but not $N_1$ and $N_3$).
The other PMNS angles and light neutrino squared mass differences are fixed according to their best-fit values of the \texttt{NuFit 6.0} global analysis in the NO case (with Super-Kamiokande data). This set of parameters fixes the Yukawa $Y$ and the $R$-matrix to be
\begin{equation}
    Y = \begin{pmatrix}
1.06\times10^{-6} + 3.21\times10^{-9}i & 1.07\times10^{-6} - 3.21\times10^{-9}i & -3.56\times10^{-24} - 1.51\times10^{-6}i \\
1.30\times10^{-6} + 5.84\times10^{-8}i & 1.25\times10^{-6} + 5.84\times10^{-8}i & 8.26\times10^{-8} - 1.80\times10^{-6}i \\
-9.83\times10^{-7} + 6.20\times10^{-8}i & -1.04\times10^{-6} + 6.20\times10^{-8}i & 8.77\times10^{-8} + 1.43\times10^{-6}i
\end{pmatrix}
\end{equation}
\begin{equation}
    R = \begin{pmatrix}
9.25\times10^{-5} + 2.79\times10^{-7}i & 9.34\times10^{-5} - 2.79\times10^{-7}i & -3.10\times10^{-22} - 1.31\times10^{-4}i \\
1.13\times10^{-4} + 5.08\times10^{-6}i & 1.08\times10^{-4} + 5.08\times10^{-6}i & 7.19\times10^{-6} - 1.57\times10^{-4}i \\
-8.55\times10^{-5} + 5.40\times10^{-6}i & -9.06\times10^{-5} + 5.40\times10^{-6}i & 7.63\times10^{-6} + 1.25\times10^{-4}i
\end{pmatrix}
\end{equation}
Then, after the procedure outlined around Eq.~\eqref{eq:tMNMN_2}, we obtain $M_1 = (2 + 3.3698\times 10^{-8})\,\text{GeV}$, $\Delta M_{21} = 2.75 \times 10^{-7}\,\text{GeV}$ and $\Delta M_{31} = 5.38 \times 10^{-8} \,\text{GeV}$,
\begin{equation}
V \simeq
\begin{pmatrix}
0.897 & -0.294 & 0.331 \\
-0.193 & -0.933 & -0.304 \\
0.398 & 0.209 & -0.893
\end{pmatrix},
\end{equation}
up to sign changes and permutations of columns, and
\begin{equation}
\Theta \simeq \begin{pmatrix}
6.49\times10^{-5} - 5.20\times10^{-5}i & -1.14\times10^{-4} - 2.72\times10^{-5}i & 2.22\times10^{-6} + 1.18\times10^{-4}i \\
8.34\times10^{-5} - 5.88\times10^{-5}i & -1.33\times10^{-4} - 3.89\times10^{-5}i & -1.92\times10^{-6} + 1.40\times10^{-4}i \\
-5.62\times10^{-5} + 5.34\times10^{-5}i & 1.11\times10^{-4} + 1.94\times10^{-5}i & -7.57\times10^{-6} - 1.11\times10^{-4}i
\end{pmatrix}
\end{equation}
with 
\[
|\Theta_{e1}|^2 = 6.916 \times 10^{-9},\quad 
|\Theta_{e2}|^2 = 1.380 \times 10^{-8},\quad
|\Theta_{e3}|^2 = 1.383 \times 10^{-8}, 
\]
\[
|\Theta_{\mu 1}|^2 = 1.041 \times 10^{-8},\quad 
|\Theta_{\mu 2}|^2 = 1.916 \times 10^{-8},\quad
|\Theta_{\mu 3}|^2 = 1.963 \times 10^{-8},
\]
\[
|\Theta_{\tau 1}|^2 = 6.003 \times 10^{-9},\quad
|\Theta_{\tau 2}|^2 = 1.275 \times 10^{-8},\quad
|\Theta_{\tau 3}|^2 = 1.241 \times 10^{-8},
\]
\[
\Theta_e^2 = 3.455 \times 10^{-8},\quad
\Theta_\mu^2 = 4.920 \times 10^{-8}, \quad
\Theta_\tau^2 = 3.117 \times 10^{-8},
\]
\[
\Theta_1^2 = 2.333 \times 10^{-8},\quad
\Theta_2^2 = 4.572 \times 10^{-8},\quad
\Theta_3^2 = 4.587 \times 10^{-8},
\]
and $\Theta^2 = 1.149 \times 10^{-7}$. The BAU for this benchmark point roughly equals the observed value $\eta_B^{\rm obs}$, i.e.~$\eta_B \simeq 6.1 \times 10^{-10}$. The behaviour of the quantities $N_{N_j}\equiv (\rho_N)_{jj}$ and $|N_{N_j} - \overline{N}_{N_j}|$, $j=1,\,2,\,3$,  $\mu_{\Delta_\alpha}$, $\alpha = e,\,\mu,\,\tau$ and $\eta_B$ against $x = T_{\rm ew}/T$ for this benchmark is shown in the top panel of Fig.~\ref{fig:LG_Plot_BMC}. The effective Majorana mass parameter for this benchmark reads $m_{\beta\beta}^{\rm eff} \simeq 3.23\,\text{meV}$.

\begin{figure}
\centering
\includegraphics[width=0.45\linewidth]{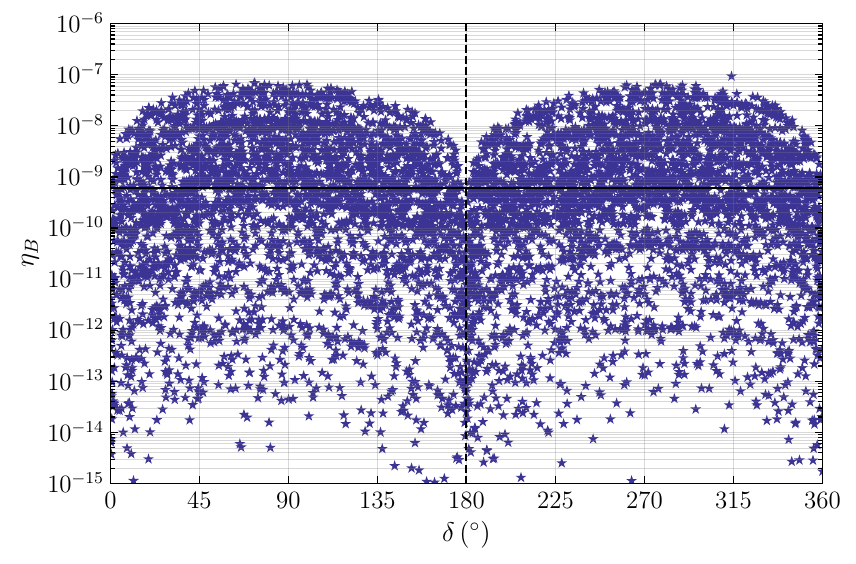}
\includegraphics[width=0.45\linewidth]{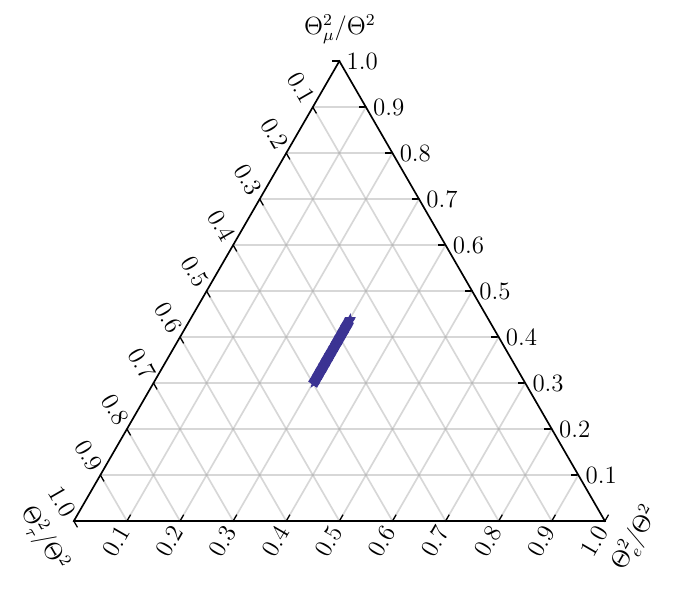}\\
\includegraphics[width=0.45\linewidth]{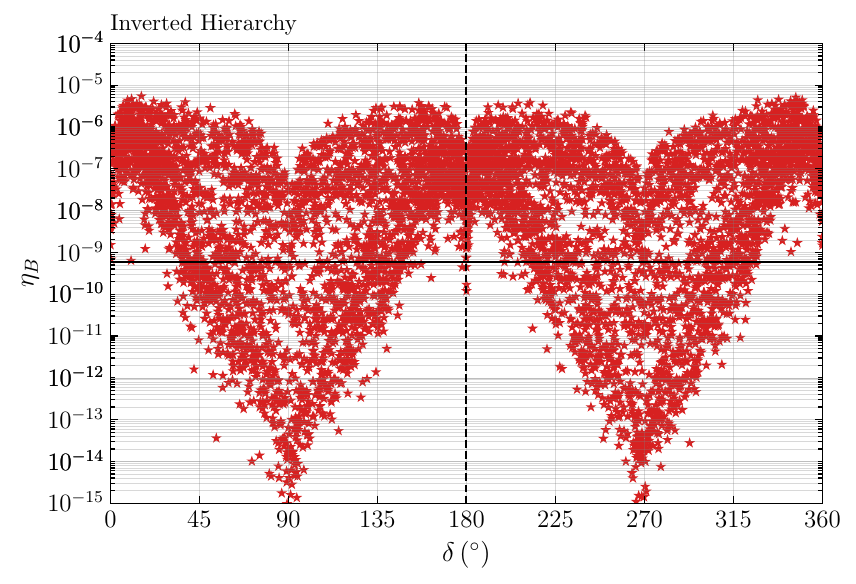}
\includegraphics[width=0.45\linewidth]{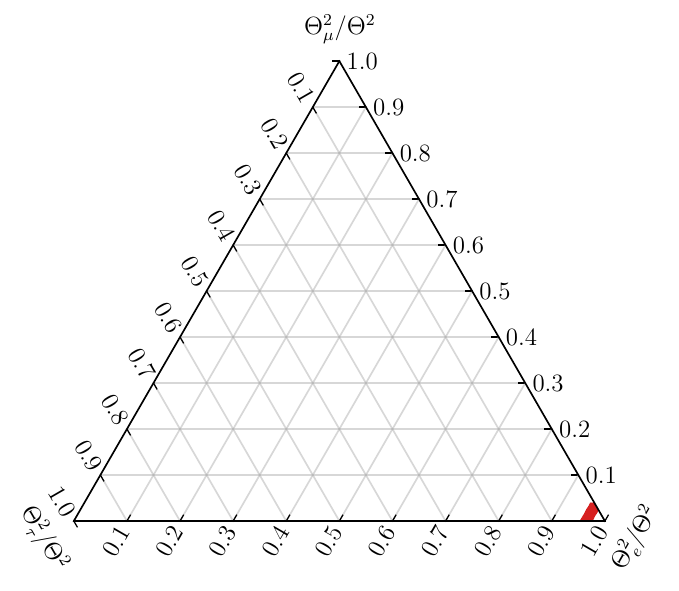}\\
\includegraphics[width=0.45\linewidth]{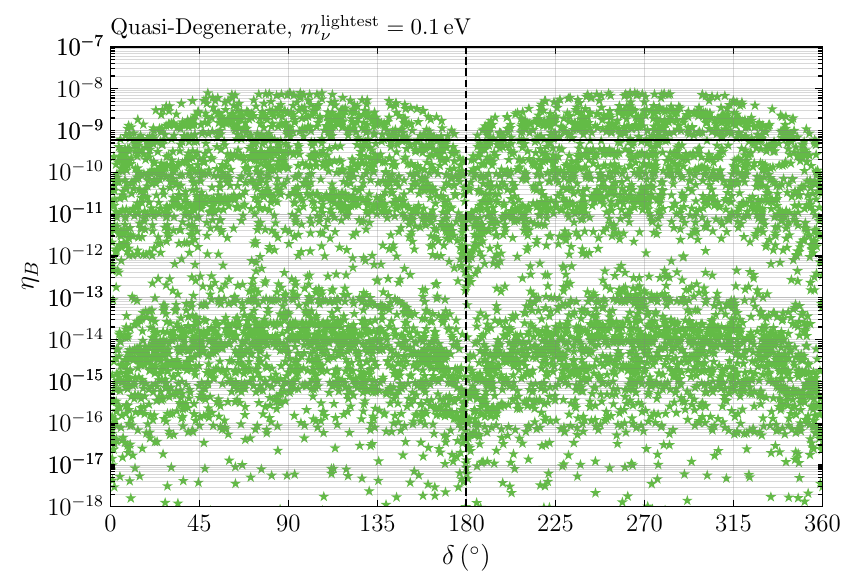}
\includegraphics[width=0.45\linewidth]{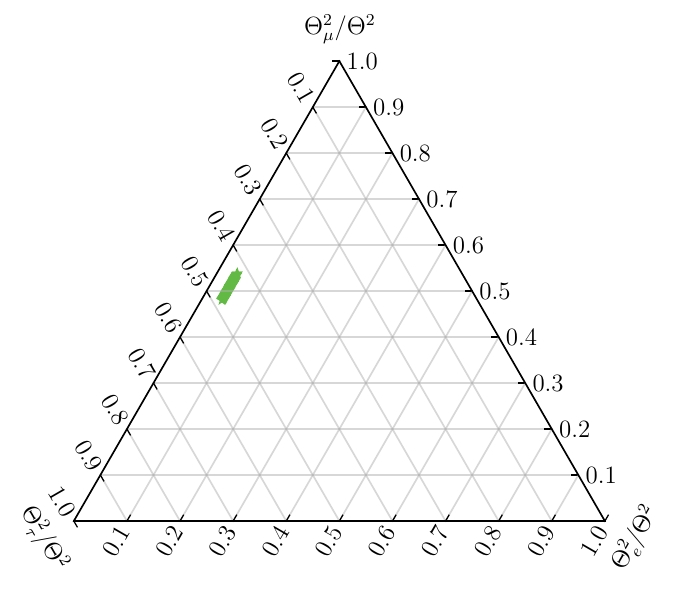}
\caption{On the left we show the scan over the splittings and $x_N$ around the benchmark point in Eqs.~\eqref{eq:params_BMC2_NH} in the top panel, \eqref{eq:params_BMC_IH} middle panel, \eqref{eq:params_BMC_QD} bottom panel. On the right the same points of the scan but in the ternary plane.}
    \label{fig:etaB_vs_delta_BMCs_scans}
\end{figure}

\begin{figure}
\centering
\includegraphics[width=0.8\linewidth]{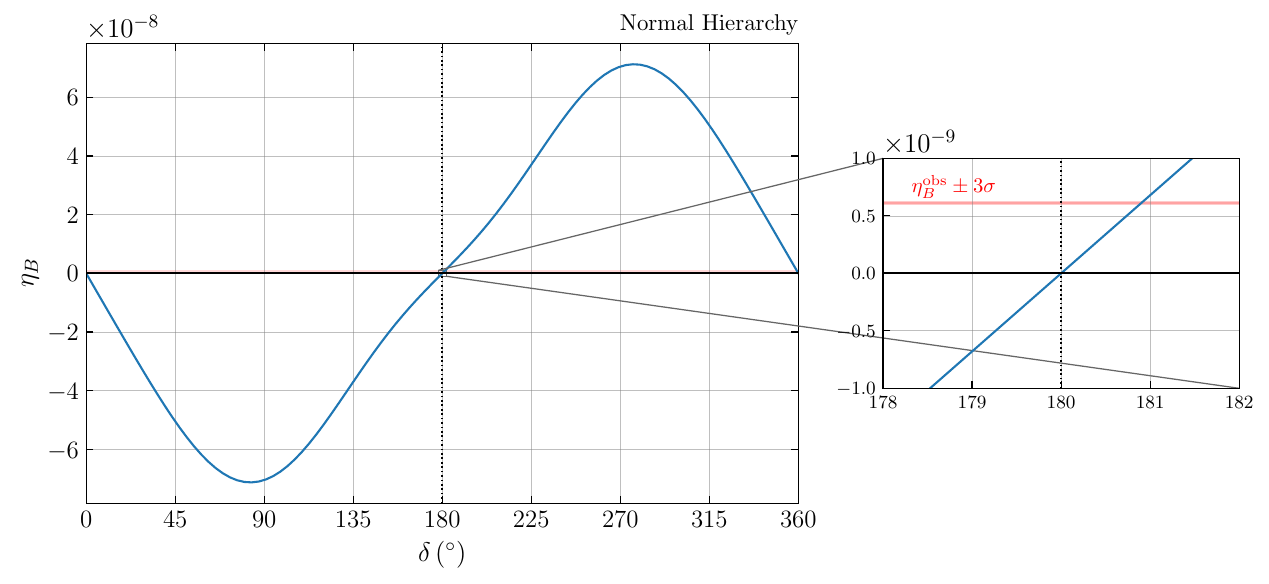}
\includegraphics[width=0.8\linewidth]{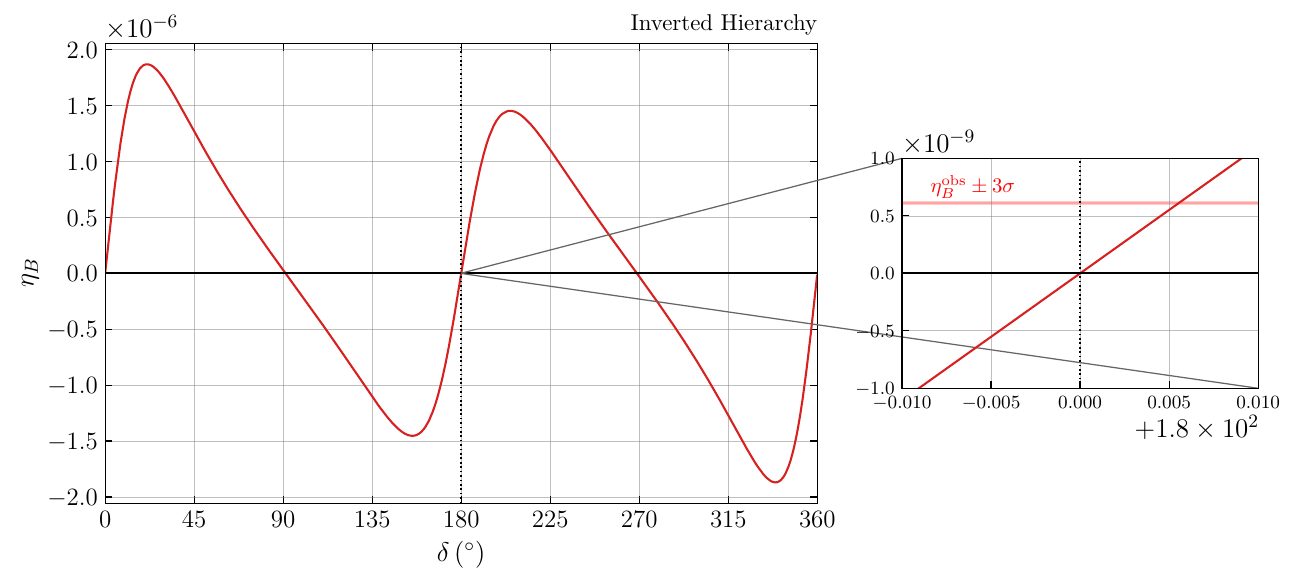}
\includegraphics[width=0.8\linewidth]{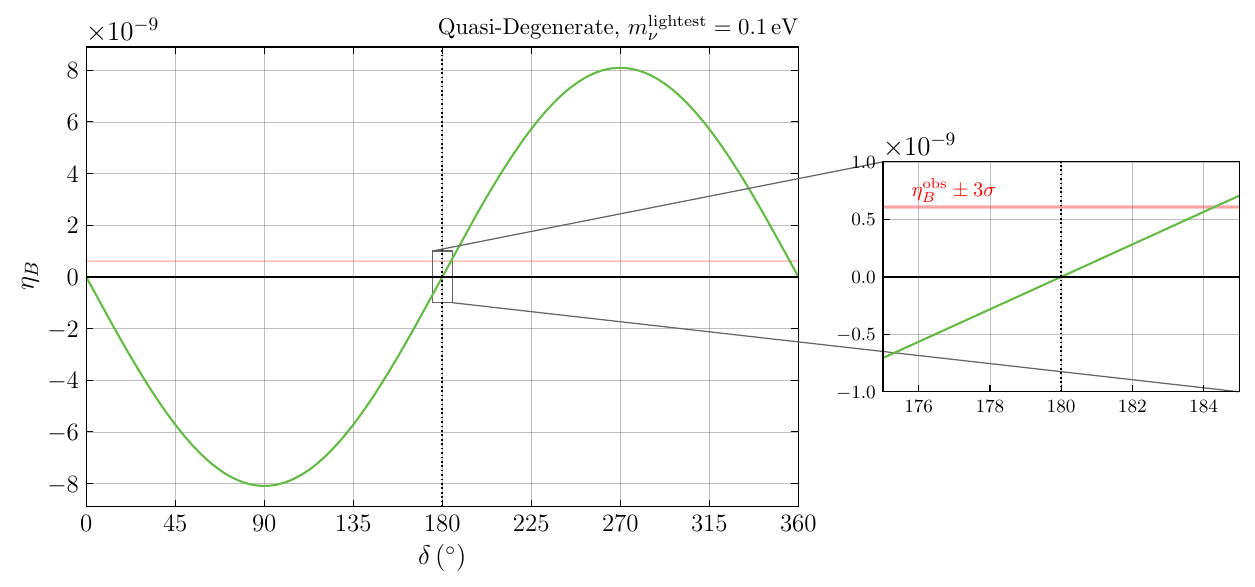}
    \caption{The behaviour of the BAU versus the Dirac phase $\delta$ for the benchmark points discussed in the text in the NH case (top panel), IH case (middle panel) and QD (bottom panel) at which leptogenesis is viable for $\delta$ relatively close to CP-conserving values. See the text for details on the model parameters.}
    \label{fig:etaB_vs_delta_BMCs}
\end{figure}

We also show in the top panel of Fig.~\ref{fig:etaB_vs_delta_BMCs_scans} the scan in the $\eta_B$-$\delta$ plane around this benchmark obtained by varying  $10^{-11} \leq |\Delta \tilde{M}_{21}|/\tilde{M}_1,\,
|\Delta \tilde{M}_{32}|/\tilde{M}_1 \leq 10^{-4}$, 
$0 \leq x_N \leq 2\pi$ and $0 \leq \delta \leq 2\pi$ while keeping the other parameters fixed at the values in Eq.~\eqref{eq:params_BMC_NH}, and in the top panel of Fig.~\ref{fig:etaB_vs_delta_BMCs} the BAU versus $\delta$ for $\Delta \tilde{M}_{12} = 2.45\times 10^{-9}\,\text{GeV}$, $\Delta \tilde{M}_{13} = 2\times 10^{-5}\,\text{GeV}$ and $x_N = 280^\circ$ and the remaining parameters as in Eq.~\eqref{eq:params_BMC_NH}, for which we get successful leptogenesis at $\delta = \pi + 1.57 \times 10^{-2}$.

%%%%%%%%%%%%%%%%%%%%%     BMC IH    %%%%%%%%%%%%%%%%%%%%%%%%%%%

\paragraph{Benchmark in the IH case.} For the point in Fig.~\ref{fig:CP1s_scan_IH_QD} in the IH case,  the model parameters are the following: 
\begin{equation}\label{eq:params_BMC_IH}
\begin{split}
    \tilde{M}_1 = 5.1\,\text{GeV}, \quad \Delta \tilde{M}_{21} = 1.02\times 10^{-8}\,\text{GeV},\quad \Delta \tilde{M}_{13} = 5\times 10^{-4}\,\text{GeV}, \quad \mnulight = 0,\\
    x_\nu = \pi/6,\quad x_N = \pi/12,\quad y = -5.04,\\
    \delta = 274^\circ\,\text{(best-fit of } \texttt{NuFit  6.0}\text{)},\quad
    \alpha_{21} = \pi.
\end{split}
\end{equation}
The other PMNS angles and light neutrino squared mass differences are fixed according to their best-fit values of the \texttt{NuFit 6.0} global analysis in the IO case (with Super-Kamiokande data). This set of parameters fixes the Yukawa $Y$ and the $R$-matrix to be
\begin{equation}
    Y = \begin{pmatrix}
7.47\times10^{-6} & 1.96\times10^{-6} &  7.72\times10^{-6}i \\
-6.62\times10^{-7} + 8.35\times10^{-7}i & -2.23\times10^{-7} + 2.19\times10^{-7}i & -8.63\times10^{-7} - 6.97\times10^{-7}i \\
6.15\times10^{-7} + 7.54\times10^{-7}i & 2.16\times10^{-7} + 1.98\times10^{-7}i & -7.80\times10^{-7} + 6.51\times10^{-7}i
\end{pmatrix}
\end{equation}
\begin{equation}
    R =\begin{pmatrix}
2.55\times10^{-4} & 6.67\times10^{-5}  & 2.63\times10^{-4}i \\
-2.26\times10^{-5} + 2.85\times10^{-5}i & -7.60\times10^{-6} + 7.46\times10^{-6}i & -2.94\times10^{-5} - 2.38\times10^{-5}i \\
2.10\times10^{-5} + 2.57\times10^{-5}i & 7.37\times10^{-6} + 6.74\times10^{-6}i & -2.66\times10^{-5} + 2.22\times10^{-5}i
\end{pmatrix}
\end{equation}
Then, after the procedure outlined around Eq.~\eqref{eq:tMNMN_2}, we obtain $M_1 = (5.1 + 5.2574 \times 10^{-7})\,\text{GeV}$, $\Delta M_{12} = 4.97 \times 10^{-7}\,\text{GeV}$ and $\Delta M_{13} = 5.10 \times 10^{-4} \,\text{GeV}$,
\begin{equation}
V = \begin{pmatrix}
0.9996 & -0.02661 & -5.839\times10^{-5} \\
0.02661& 0.9996 & -2.541\times10^{-6} \\
5.844\times10^{-5} & 9.870\times10^{-7} & 1
\end{pmatrix}
\end{equation}
and 
\begin{equation}
\Theta = \begin{pmatrix}
2.56\times10^{-4}& 5.99\times10^{-5} &  2.63\times10^{-4}i \\
-2.28\times10^{-5} + 2.87\times10^{-5}i & -7.00\times10^{-6} + 6.70\times10^{-6}i & -2.95\times10^{-5} - 2.38\times10^{-5}i \\
2.12\times10^{-5} + 2.59\times10^{-5}i & 6.81\times10^{-6} + 6.05\times10^{-6}i & -2.66\times10^{-5} + 2.22\times10^{-5}i
\end{pmatrix}
\end{equation}
with
\[
|\Theta_{e1}|^2 = 6.576 \times 10^{-8},\quad 
|\Theta_{e2}|^2 = 3.590 \times 10^{-9},\quad
|\Theta_{e3}|^2 = 6.936 \times 10^{-8}, 
\]
\[
|\Theta_{\mu 1}|^2 = 1.341 \times 10^{-9},\quad 
|\Theta_{\mu 2}|^2 = 9.388 \times 10^{-10},\quad
|\Theta_{\mu 3}|^2 = 1.433 \times 10^{-9},
\]
\[
|\Theta_{\tau 1}|^2 = 1.120\times 10^{-9},\quad
|\Theta_{\tau 2}|^2 = 8.295 \times 10^{-11},\quad
|\Theta_{\tau 3}|^2 = 1.201\times 10^{-9},
\]
\[
\Theta_e^2 = 1.387 \times 10^{-7},\quad
\Theta_\mu^2 = 2.868 \times 10^{-9}, \quad
\Theta_\tau^2 = 2.403 \times 10^{-9},
\]
\[
\Theta_1^2 = 6.822  \times 10^{-8},\quad
\Theta_2^2 = 3.767 \times 10^{-9},\quad
\Theta_3^2 = 7.199 \times 10^{-8},
\]
and $\Theta^2 = 1.440 \times 10^{-7}$. The BAU for this benchmark point roughly equals the observed value $\eta_B^{\rm obs}$, i.e.~$\eta_B \simeq 6.1 \times 10^{-10}$. The behaviour of the quantities $N_{N_j}$ and $|N_{N_j} - \overline{N}_{N_j}|$, $j=1,\,2,\,3$,  $\mu_{\Delta_\alpha}$, $\alpha = e,\,\mu,\,\tau$ and $\eta_B$ against $x = T_{\rm ew}/T$ for this benchmark is shown in the bottom-left panel of Fig.~\ref{fig:LG_Plot_BMC}.

We also show in the middle panel of Fig.~\ref{fig:etaB_vs_delta_BMCs_scans} the scan in the $\eta_B - \delta$ plane around this benchmark obtained by varying  $10^{-11} \leq |\Delta \tilde{M}_{21}|/\tilde{M}_1,\,
|\Delta \tilde{M}_{32}|/\tilde{M}_1 \leq 10^{-4}$, 
$0 \leq x_N \leq 2\pi$ and $0 \leq \delta \leq 2\pi$ while keeping the other parameters fixed at the values in Eq.~\eqref{eq:params_BMC_IH}. Although we still find that the BAU can be obtained for basically any sufficiently CP-violating value of the Dirac phase, for this particular benchmark, we find that there is a less fine-tuned range of $\delta$, $\pi/4 \lesssim \delta \lesssim 3\pi/4$ and $5\pi/4 \lesssim \delta \lesssim 7\pi/4$ that leads to viable leptogenesis.

In the middle panel of Fig.~\ref{fig:etaB_vs_delta_BMCs} we depict the BAU versus $\delta$ for $\Delta \tilde{M}_{12} = 5.1\times 10^{-10}\,\text{GeV}$, $\Delta \tilde{M}_{13} = 5.1\times 10^{-5}\,\text{GeV}$ and $x_N = 210^\circ$ and the remaining parameters as in Eq.~\eqref{eq:params_BMC_IH}, for which we get successful leptogenesis at $\delta = \pi + 1 \times 10^{-4}$.
Note that for this particular choices of the parameters the BAU vanishes around $\delta \simeq \pi/2$ and $3\pi/2$ (but not at these exact values), and correspondingly we can get viable leptogenesis around these values as well. We have checked that this is due to a fine-tuned cancellation of the BAU from wash-out processes happening exactly at the sphaleron freeze-out.

%%%%%%%%%%%%%%%%%%%%%     BMC QD     %%%%%%%%%%%%%%%%%%%%%%%%%%%

\paragraph{Benchmark in the QD case.} For the point in Fig.~\ref{fig:CP1s_scan_IH_QD} in the QD case,  the model parameters are the following: 
\begin{equation}\label{eq:params_BMC_QD}
\begin{split}
    \tilde{M}_1 = 3\,\text{GeV}, \quad \Delta \tilde{M}_{12} = 3\times 10^{-10}\,\text{GeV},\quad \Delta \tilde{M}_{13} = 3\times 10^{-8}\,\text{GeV}, \quad \mnulight = 0.1\,\text{eV},\\
    x_\nu = 3\pi/2,\quad x_N = \pi/6,\quad y = 6.47,\\
    \delta = 315^\circ,\quad
    \alpha_{21} = -\alpha_{31} = -\pi.
\end{split}
\end{equation}
The other PMNS angles and light neutrino squared mass differences are fixed according to their best-fit values of the \texttt{NuFit 6.0} global analysis in the NO case (with Super-Kamiokande data). This set of parameters fixes the Yukawa $Y$ and the $R$-matrix to be
\begin{equation}
    Y =\begin{pmatrix}
-7.60\times10^{-6} + 5.54\times10^{-9}i & -4.40\times10^{-6} - 9.60\times10^{-9}i & 8.79\times10^{-6}i \\
2.87\times10^{-5} - 5.55\times10^{-7}i & 1.65\times10^{-5} - 3.21\times10^{-7}i & -6.41\times10^{-7} - 3.31\times10^{-5}i \\
-2.59\times10^{-5} - 5.89\times10^{-7}i & -1.50\times10^{-5} - 3.40\times10^{-7}i & -6.81\times10^{-7} + 2.99\times10^{-5}i
\end{pmatrix}
\end{equation}
\begin{equation}
    R = \begin{pmatrix}
-4.41\times10^{-4} + 3.21\times10^{-7}i & -2.55\times10^{-4} - 5.57\times10^{-7}i &  5.10\times10^{-4}i \\
1.66\times10^{-3} - 3.22\times10^{-5}i & 9.56\times10^{-4} - 1.86\times10^{-5}i & -3.72\times10^{-5} - 1.92\times10^{-3}i \\
-1.50\times10^{-3} - 3.42\times10^{-5}i & -8.71\times10^{-4} - 1.97\times10^{-5}i & -3.95\times10^{-5} + 1.73\times10^{-3}i
\end{pmatrix}
\end{equation}
Then, after the procedure outlined around Eq.~\eqref{eq:tMNMN_2}, we obtain $M_1 = (3 + 4.2539\times 10^{-6})\,\text{GeV}$, $\Delta M_{21} = 6.18 \times 10^{-6}\,\text{GeV}$ and $\Delta M_{31} = 2.28 \times 10^{-5} \,\text{GeV}$,
\begin{equation}
V \simeq
\begin{pmatrix}
0.554 & 0.830 & 0.065 \\
-0.648 & 0.479 & -0.593 \\
-0.523 & 0.287 & 0.803
\end{pmatrix}
\end{equation}
up to sign changes and permutations of columns, and
\begin{equation}
\Theta \simeq \begin{pmatrix}
-7.90\times10^{-5} - 2.66\times10^{-4}i & -4.88\times10^{-4} + 1.46\times10^{-4}i & 1.23\times10^{-4} + 4.09\times10^{-4}i \\
3.23\times10^{-4} + 9.98\times10^{-4}i & 1.83\times10^{-3} - 5.86\times10^{-4}i & -4.89\times10^{-4} - 1.53\times10^{-3}i \\
-2.47\times10^{-4} - 9.13\times10^{-4}i & -1.67\times10^{-3} + 4.59\times10^{-4}i & 3.88\times10^{-4} + 1.40\times10^{-3}i
\end{pmatrix}
\end{equation}
with 
\[
|\Theta_{e1}|^2 = 7.693 \times 10^{-8},\quad 
|\Theta_{e2}|^2 = 2.597 \times 10^{-7},\quad
|\Theta_{e3}|^2 = 1.827 \times 10^{-7}, 
\]
\[
|\Theta_{\mu 1}|^2 = 1.100 \times 10^{-6},\quad 
|\Theta_{\mu 2}|^2 = 3.687 \times 10^{-6},\quad
|\Theta_{\mu 3}|^2 = 2.587 \times 10^{-6},
\]
\[
|\Theta_{\tau 1}|^2 = 8.944 \times 10^{-7},\quad
|\Theta_{\tau 2}|^2 = 3.010 \times 10^{-6},\quad
|\Theta_{\tau 3}|^2 = 2.116 \times 10^{-6},
\]
\[
\Theta_e^2 = 5.194 \times 10^{-7},\quad
\Theta_\mu^2 = 7.374 \times 10^{-6}, \quad
\Theta_\tau^2 = 6.021 \times 10^{-6},
\]
\[
\Theta_1^2 = 2.071 \times 10^{-6},\quad
\Theta_2^2 = 6.957 \times 10^{-6},\quad
\Theta_3^2 = 4.886 \times 10^{-8},
\]
and $\Theta^2 = 1.391 \times 10^{-5}$. The BAU for this benchmark point roughly equals the observed value $\eta_B^{\rm obs}$, i.e.~$\eta_B \simeq 6.1 \times 10^{-10}$. The behaviour of the quantities $N_{N_j}$ and $|N_{N_j} - \overline{N}_{N_j}|$, $j=1,\,2,\,3$,  $\mu_{\Delta_\alpha}$, $\alpha = e,\,\mu,\,\tau$ and $\eta_B$ against $x = T_{\rm ew}/T$ for this benchmark is shown in the bottom-right panel of Fig.~\ref{fig:LG_Plot_BMC}.

We also show in the bottom panel of Fig.~\ref{fig:etaB_vs_delta_BMCs_scans} the scan in the $\eta_B - \delta$ plane around this benchmark obtained by varying  $10^{-11} \leq |\Delta \tilde{M}_{21}|/\tilde{M}_1,\,
|\Delta \tilde{M}_{32}|/\tilde{M}_1 \leq 10^{-4}$, 
$0 \leq x_N \leq 2\pi$ and $0 \leq \delta \leq 2\pi$ while keeping the other parameters fixed at the values in Eq.~\eqref{eq:params_BMC_QD}, and in the bottom panel of Fig.~\ref{fig:etaB_vs_delta_BMCs} the BAU versus $\delta$ for $\Delta \tilde{M}_{12} = 2.274\times 10^{-8}\,\text{GeV}$, $\Delta \tilde{M}_{13} = 3\times 10^{-4}\,\text{GeV}$ and $x_N = 5\pi/12$ and the remaining parameters as in Eq.~\eqref{eq:params_BMC_QD}, for which we get successful leptogenesis at $\delta = \pi + 7.6 \times 10^{-2}$.
%=====================================

\bibliography{Biblio}
\bibliographystyle{JHEP}
\addcontentsline{toc}{section}{References}

\end{document}